\documentclass[acmsmall]{acmart} %anonymous

\AtBeginDocument{%
  \providecommand\BibTeX{{%
    \normalfont B\kern-0.5em{\scshape i\kern-0.25em b}\kern-0.8em\TeX}}}

\setcopyright{acmcopyright}
\copyrightyear{2021}
\acmYear{2021}
\acmJournal{TELO}
\acmArticle{xxx}

\usepackage{amsmath}
\usepackage{amsbsy}
\usepackage{amsfonts}
\usepackage{textcomp}

\usepackage{appendix}
\usepackage{enumitem}

\usepackage[T1]{fontenc}

\usepackage{color}
\usepackage{url}
\usepackage{soul}
\usepackage{hyperref}
\usepackage{pdflscape}

\usepackage{tikz}
\usepackage{multirow}

\usepackage{graphicx}

\usepackage{comment}
\usepackage{algorithm}
\usepackage{algpseudocode}

\usepackage{caption}
\usepackage{subcaption}
\definecolor{pblue}{rgb}{0.13,0.13,1}
\definecolor{pgreen}{rgb}{0,0.5,0}
\definecolor{pred}{rgb}{0.9,0,0}
\definecolor{pgrey}{rgb}{0.46,0.45,0.48}
\usepackage{listings}

\makeatletter
\newcommand{\highlight}[3]{
    \begingroup
    \lst@basicstyle
    \color{white}
    \ifnum\value{lstnumber}>#1
        \ifnum\value{lstnumber}<#2
            \color{#3}
        \fi
    \fi
    \rlap{\hspace*{\lst@numbersep}
    \color@block{\linewidth}{\ht\strutbox}{\dp\strutbox}
    }
    \endgroup
}
\makeatother

\usepackage{etoolbox}
\global\providetoggle{found}
\global\settoggle{found}{false}

\makeatletter
\newcommand{\highlightlist}[2]{%
    \begingroup
    \lst@basicstyle
    \color{white}%
    \global\settoggle{found}{false}
    \foreach \x in {#1} {
        \ifnum\value{lstnumber}=\x
            \global\settoggle{found}{true}
        \fi
    }
    \global\iftoggle{found}{\color{#2}}{\color{white}}
    \rlap{\hspace*{\lst@numbersep}%
    \color@block{\linewidth}{\ht\strutbox}{\dp\strutbox}%
    }%
    \endgroup
}
\makeatother

\algnewcommand\algorithmicforeach{\textbf{for each}}
\algdef{S}[FOR]{ForEach}[1]{\algorithmicforeach\ #1\ \algorithmicdo}
\usepackage{algpseudocode}
\makeatletter
\renewcommand{\ALG@beginalgorithmic}{\scriptsize}
\algrenewcommand\alglinenumber[1]{\scriptsize #1:}
\makeatother
\algrenewcommand\textproc{}

\let\oldReturn\Return
\renewcommand{\Return}{\State\oldReturn}

\newcommand{\NUMRUNS}{10}

\begin{document}

\title{Genetic Improvement of Routing Protocols for Delay Tolerant Networks}

\author{Michela Lorandi}
\email{michela.lorandi@studenti.unitn.it}

\author{Leonardo Lucio Custode}
\email{leonardo.custode@unitn.it}

\author{Giovanni Iacca}
\email{giovanni.iacca@unitn.it}
\orcid{0000-0001-9723-1830}

%\authornote{}

\affiliation{%
  \institution{\\Department of Information Engineering and Computer Science, University of Trento}
  \streetaddress{Via Sommarive 9}
  \city{Povo}
  \state{Italy}
  \postcode{38123}
}

\renewcommand{\shortauthors}{Lorandi, et al.}

%The abstract should be approximately 250 words and should be a concise statement of the main contributions of the paper. Abstracts should not be in first person, should not present mathematical formulas, and do not include citations.
\begin{abstract}
Routing plays a fundamental role in network applications, but it is especially challenging in Delay Tolerant Networks (DTNs). These are a kind of mobile ad hoc networks made of e.g. (possibly, unmanned) vehicles and humans where, despite a lack of continuous connectivity, data must be transmitted while the network conditions change due to the nodes' mobility. In these contexts, routing is NP-hard and is usually solved by heuristic ``store and forward'' replication-based approaches, where multiple copies of the same message are moved and stored across nodes in the hope that at least one will reach its destination. Still, the existing routing protocols produce relatively low delivery probabilities. Here, we genetically improve two routing protocols widely adopted in DTNs, namely Epidemic and PRoPHET, in the attempt to optimize their delivery probability. First, we dissect them into their fundamental components, i.e., functionalities such as checking if a node can transfer data, or sending messages to all connections. Then, we apply Genetic Improvement (GI) to manipulate these components as terminal nodes of evolving trees. We apply this methodology, in silico, to six test cases of urban networks made of hundreds of nodes, and find that GI produces consistent gains in delivery probability in four cases. We then verify if this improvement entails a worsening of other relevant network metrics, such as latency and buffer time. Finally, we compare the logics of the best evolved protocols with those of the baseline protocols, and we discuss the generalizability of the results across test cases.
\end{abstract}




\maketitle

% ----------------------

%\hl{NOTE: I enabled the anonymous flag in the documentclass because the ACM TELO review process is double-blind, i.e., the authors' names must be omitted.}
%Guidelines: https://dlnext.acm.org/journal/telo/author-guidelines
%Submission: https://mc.manuscriptcentral.com/telo

%Source code: https://github.com/michiL96/protocol_evolution

% -------------------------------------------------------------------------
% -------------------------------------------------------------------------
% -------------------------------------------------------------------------

\section{Introduction}
\label{sec:intro}

%communication protocols basic enabling technology for IoT
A fundamental element in many modern applications is the use of networked systems: be it environment monitoring, smart industries, smart cities, or distribution systems, networks of various scales and complexity are employed today practically everywhere. One of the most important aspects in networking is the concept of \emph{network protocol}, i.e., a set of well-defined data format and rules that allow nodes in a network to communicate with each other \cite{holzmann1991design}. Typically, a physical network relies on multiple protocols, which are arranged as a \emph{protocol stack} where protocols at lower layers provide basic functionalities which are progressively enriched by the higher layer protocols. Among those basic functionalities, routing is a crucial one, as it allows data to flow across the network and reach their destination. While well-established efficient routing protocols exist for IP networks, routing in mobile ad hoc networks (MANETs) and other Internet of Things (IoT) instances, such as networks of cars, unmanned vehicles, or collectives of vehicles and humans, is still a very active area of research. One particularly challenging kind of MANETs is represented by the \emph{delay tolerant networks} (DTNs), also known as \emph{disruption tolerant networks}, \emph{opportunistic networks} or \emph{intermittently connected wireless networks}. Originally designed in the 1970s for space communications, but applied since the early 2000s also to terrestrial applications such as urban mobile networks, DTNs are (typically, heterogeneous) decentralized networks that lack continuous connectivity due to their node sparsity, limited wireless radio access, and limited energy resources.%attack, and noise.

In these contexts, routing is NP-hard \cite{balasubramanian2007dtn} and as such it is usually solved by heuristic (best-effort) ``store and forward'' replication-based approaches, where multiple copies of the same message are moved and stored across nodes in the hope that at least one will eventually reach its destination. Still, the sparsity and mobility of the nodes causes unpredictable meeting patterns and frequent disconnections \cite{alouf2010fitting}, which result in relatively low data delivery probabilities (also called delivery rates, or delivery ratios), even with well-established protocols such as Epidemic \cite{vahdat2000epidemic}, PRoPHET \cite{doria2003prophet}, and their variants. As shown in \cite{hong2002scalable,abolhasan2004review,boukerche2011routing}, the node density and their mobility affect the delivery probability in most kinds of MANETs (not only DTNs), although the problem is further exacerbated in DTNs: while in dense MANETs composed of slow-moving nodes (e.g. pedestrians) the delivery probability obtained by state-of-the-art routing protocols can be higher than 90\% \cite{johansson1999scenario}, in sparse MANETs made of fast-moving nodes (e.g. vehicles) the delivery probability can be as low as 15-18\%, \cite{saudi2019mobile,clausen:inria-00071448}. Furthermore, other factors such as the number of sources (i.e., the nodes transmitting data) can also affect the delivery probability: the lower the number of sources, the higher the delivery probability \cite{clausen:inria-00071448,perkins2001performance}.
%NOTE: hiranandani2012benchmarking, mwanza2008performance and thong2004performance are master theses

Traditionally, network protocols are modelled as a \emph{reactive system}, i.e., a two-player game (agent vs environment) where an agent (a node in the network) \emph{reacts}, by performing a certain action, to pre-defined conditions in the environment (the rest of the network): for instance, the agent retries a message transmission if it does not receive an acknowledgment. As such, a protocol can be described with an automaton, for which formal specifications can be logically expressed and verified. For that, one usually needs to have complete knowledge about (and strict assumptions on) the environment. This approach, rooted in the theory of Temporal Logic and infinite (B{\"u}chi) automata \cite{buchi1990solving}, has been the gold standard in protocol design and verification for decades. %Since the late 1960s, an impressive number of theoretical and practical results have been obtained in this area, gearing towards the automatic synthesis of protocol from service specifications \cite{saleh1991automatic,probert1991synthesis,carchiolo1992formal,saleh1996synthesis} and the development of automatic model checker tools, such as SPIN \cite{holzmann1997model}. Despite these many successes, this approach to protocol design has also limitations.
However, there are limitations. First of all, this way of designing protocols assumes in general the environment, and all its states, to be known: this is usually not the case for DTNs, where the environment conditions can be unpredictable due to the nodes' mobility. Secondly, the numerical (time and space) complexity of these design methods makes them impractical when the number of states of protocol and environment grows \cite{vardi2018siren}. This is the case of DTNs, where the number of states of the environment can be very large, depending on the number of nodes and their state. For these reasons, finding a way to design better routing protocols for DTNs, or at least improving the performance of the existing ones, is still an open research question. 

Here, we consider the \emph{Genetic Improvement} (GI) \cite{langdon2015genetically} of two of the main routing protocols used in DTNs, namely Epidemic and PRoPHET. Our methodology consists in the following: first, we dissect the two protocols into their fundamental components, i.e., basic network functionalities such as checking if a node can start transferring data, or sending messages to all connections; then, we apply Genetic Programming (GP) to rearrange these components into evolving trees, in the attempt to maximize the data delivery probability. It is worth stressing that, in principle, this methodology can be easily generalized to other protocols, also at different layers of the protocol stack, and to different kinds of networks.

To evaluate the proposed methodology, we perform a broad \emph{in silico} experimentation where we improve Epidemic and PRoPHET on six test cases of urban networks made of three different kinds of mobile nodes (pedestrians, cars, and trams). Overall, we find that GI consistently produces a gain in data delivery probability. We then verify if this improvement in the delivery probability entails a worsening of other relevant network metrics, and find some counterintuitive results obtained as ``byproducts'' of GI: in fact, we find that the best evolved protocols not only increase the delivery probability, but also reduce the overall network overhead (i.e., the number of retransmissions). However, they trade these improvements for a higher latency. We also investigate the generalizability of the best evolved protocols across test cases, and find that apart from one specific map (Manhattan), the evolved protocols are able to generalize to unseen test cases, producing results that are still better than the baseline protocols. Finally, we compare the working logics of the best evolved protocols with those of the baseline protocols, and identify some common aspects which underlie their improved performance.

\subsubsection*{Novel aspects of this work} Compared to the existing works on GP applied to protocol evolution, our work presents various elements of novelty which are worth highlighting. In particular, among the literature we will briefly survey in the next Section, the closer works on the application of GP to the evolution of protocols act either on the application layer, focusing in particular on aggregation protocols (i.e., protocols that calculate an aggregation function of distributed data, such as their mean) \cite{weise2007genetic,weise2008evolving,weise2011evolving}, protocol adaptors, i.e., interfaces between incompatible application protocols \cite{van2003using}, or application logics in Wireless Sensor Networks \cite{johnson2005genetic,valencia2010distributed}; or, they use GP to optimize only a very specific equation used in existing protocols, such as the formula used to variate the contention window size in IEEE802.11 DCF \cite{lewis2006enhancing}, or the update rule of the routing table \cite{roohitavaf2018synthesizing} (the latter work being based on rather simplified network simulations). Besides these works, three other papers have addressed, although with various limitations, very related research questions, namely
\cite{yamamoto2005experiments,yamamoto2005genetic,alouf2010fitting}. Among these, the first two works apply GP to perform the online distributed adaptation of an extremely simplified delivery protocol (which allows only direct message exchanges, i.e., without routing): in this case GP automatically selects different combinations of modules in order to adapt the resulting protocol to the network conditions. This approach is quite different from the one we propose here, which instead is based on offline genetic improvement of \emph{realistic} routing protocols for DTNs. The last one, on the other hand, does focus on the same protocols we study in this paper, i.e., Epidemic and PRoPHET, however it uses a GA to adjust the protocols' parameters (such as the number of copies) in response to the network dynamics, in order optimize the overall delivery probability. Again, this approach is quite different from our proposal. Thus, to the best of our knowledge no prior work has used GP to genetically improve the inner logic of existing routing protocols, yet alone those used in DTNs.
% \cite{tekken2017derivation} Derivation of Network Reprogramming Protocol with Z3: automated synthesis via SMT solvers from network specifications (GP mentioned as future work!)

\subsubsection*{Structure of the paper} The remaining of this paper is organized as follows. In the next Section, we present the related work. Then, in Section \ref{sec:methods} we introduce the method, in particular the GP configuration and the simulation setup. In Section \ref{sec:setup}, we describe the details of the experimentation, while the numerical results are discussed in Section \ref{sec:results}. Finally, in Section \ref{sec:conclusions} we draw the conclusions and hint at possible extensions of this work.

% -------------------------------------------------------------------------
% -------------------------------------------------------------------------
% -------------------------------------------------------------------------

\section{Related work}
\label{sec:related}

In the following we briefly summarize the related works on Genetic Improvement and the applications of Evolutionary Learning to networked systems.

\subsection{Genetic Improvement}

While the application of GP for 
bug fixing, sometimes referred to as \emph{automatic repair} of programs, dates back to the early 2000s \cite{weimer2010automatic,orlov2011flight,le_goues_systematic_2012}, the term ``Genetic Improvement'' was originally coined by Langdon and collaborators in some seminal works from 2014 \cite{langdon2014genetic,langdon2014genetically,langdon2014improving,langdon2014optimizing,petke2014using} and finally formalized in \cite{langdon2015genetically}. The broad definition of GI is the application of optimization techniques, particularly evolutionary search algorithms such as GP \cite{woodward2016gp}, to improve existing software w.r.t. either functional requirements (and in particular bug fixing), or non-functional requirements, such as speed or memory. 

In the past few years, several works have shown the potential of GI. Some of these works have focused on speed improvement, e.g. in the case of CUDA code \cite{langdon2014genetically,langdon2014improving} and CUDA-based sequencing tools such as BarraCUDA \cite{langdon2017genetically}, complex C/C++ applications such as Bowtie2 \cite{langdon2014optimizing}, the OpenCV library \cite{langdon2016api}, or the MiniSAT solver \cite{petke2014using}. A particular case of GI aimed at speed improvement is discussed in \cite{lopez2019applying}, where the authors applied GI to a C++ GP library and noted a speedup due to the deletion of some operators (crossover, point mutation and others). Another area of GI research is the accuracy improvement of low-level implementations of various mathematical functions, such as $sqrt$ \cite{langdon2019genetic-b}, $log_2$ \cite{langdon2019genetic-a,langdon2018evolving} and other functions based on lookup tables \cite{krauss2020automatically}. In other works, energy consumption has been considered as primary goal of GI \cite{bokhari2016optimising,bruce2015reducing}.

As for bug fixing, successful applications of GI have been reported in \cite{le_goues_systematic_2012}, \cite{weimer2010automatic}, and \cite{schulte2015repairing}. This latter work is especially interesting since it uses GI to repair a bug in a MIPS binary without accessing the source code. Beyond bug fixing, automatic code generation has also shown promising results e.g. in the porting of existing C code to CUDA \cite{langdon2014genetic}, or in the automatic generation of equivalent Java bytecode starting from existing programs \cite{orlov2011flight}, 

%In \cite{le_goues_systematic_2012} the authors use GenProg, a Genetic Improvement technique, to fix bugs in software totalling 5.1 Millions lines of code. The results show that the proposed approach is able to fix about a third of the known bugs for a cost of about 8\$/bug. In \cite{weimer2010automatic}, Genetic Improvement is applied to several legacy C programs. The results show that the proposed approach was able several type of bugs: infinite loops, segmentation faults and others. The authors note that in some of the proposed test cases, i.e., the ones in which the fault is easy to localize, random search can be better or equal to GP. Instead, in the opposite case, GP shows better performance. Schulte et al., in \cite{schulte2015repairing}, propose a GI approach to repair a bug in a MIPS binary without accessing the source code.

%Langdon, in \cite{langdon2014genetic}, uses GI to port a functionality of the gzip utility (written in C) to CUDA, in order to have better parallelization of the execution. In \cite{orlov2011flight}, the authors propose a system that is able to evolve Java bytecode starting from existing programs. They show that the program is able to generate good solutions in a wide variety of settings.

Finally, it is worth mentioning two recent surveys of the GI literature, \cite{petke2017genetic} and \cite{langdon2016genetic}. The latter work, in particular, mentions the application to programming languages different from C/C++ (which has attracted so far most of the attention of the GI literature) as one of the key challenges for GI. We find that our work is loosely related to this issue. 

%----------------------------------------------------------------

\subsection{Evolutionary Learning applied to networked systems}

In the past two decades, various Machine Learning and bio-inspired technique, especially Evolutionary Algorithms (EAs), have been applied to network problems and particularly protocol optimization. For instance, some researchers have proposed various solutions based on collective intelligence \cite{wolpert1999using} and Reinforcement Learning (RL) \cite{tao2001multi,peshkin2002reinforcement,stampa2017deep} to optimize routing protocols for Wireless Sensor Networks \cite{kulkarni2010computational,forster2010machine,alsheikh2014machine}. Albeit quite powerful, the main limitation of most of these approaches is that they often require a large amount of data collected from the network, in order to train a model of the protocol, to be used later at runtime for further optimization. 

As for EAs, there is a very large body of research in the general area of networked systems, as surveyed for instance in \cite{nakano2010biologically,dressler2010survey}\footnote{Another link between network engineering and evolutionary theory also exists: recent evidence has shown that the typical hourglass-shaped protocol stacks are the result of an implicit evolutionary process that led to a minimal complexity, maximal robustness architecture \cite{siyari2017emergence,dovrolis2008would}.}, with several works focusing especially on protocol optimization (not only for routing). In this context, a seminal paper is represented by the work from late 1990s by El-Fakih et al. \cite{el1999method}, who formulated the protocol design problem in the form of a 0-1 integer programming message exchange model, optimized by means of a Genetic Algorithm in order to minimize the number of messages to be exchanged while meeting a given specification of network services. As for GP, it has been successfully used to optimize protocol adaptors \cite{van2003using}, aggregation protocols \cite{weise2007genetic,weise2008evolving,weise2011evolving}, or MAC access protocols \cite{lewis2006enhancing,roohitavaf2018synthesizing}. The latter have been synthesized also by means of probabilistic Finite State Machines (FSMs) whose transition probabilities are optimized by means of a Genetic Algorithm \cite{sharples2000protocol,hajiaghajani2015feasibility,hajiaghajani2015mac}.
% one FSM for all nodes
% hajiaghajani2015mac: GA achieves a performance close to pure-ALOHA protocol.
% hajiaghajani2015feasibility: extension to Slotted ALOHA and CSMA logic.

While the methods above are meant for centralized offline protocol optimization, some works have also considered distributed online protocol optimization, although these usually focus on higher network layers (e.g. application) rather than routing. For instance, a bio-inspired distributed learning approach was introduced in \cite{su2010dynamic}, where each node observes the other nodes' behavior and forms internal conjectures on how they would react to its actions, in order to choose the action that maximizes a local utility function: the authors demonstrated, analytically and through numerical simulations, that this method reaches Nash equilibria corresponding to optimal traffic fairness and throughput. Other works have investigated distributed EAs \cite{iacca2013distributed} and distributed GP \cite{johnson2005genetic,valencia2010distributed} to evolve the nodes' parameters and functioning logics (at the application layer) of Wireless Sensor Networks. Finally, two notable online methods are STEM-Net \cite{aloi2014stem} and Fraglets \cite{tschudin2003fraglets,yamamoto2007self}. The first one is a wireless network where each node uses an EA ``to reconfigure itself at multiple layers of the protocol stack, depending on environmental conditions, on the required service and on the interaction with other analogous device''. The latter is based on the concept of ``autocatalytic software'' \cite{tschudin2005self}, or chemical computing \cite{miorandi2008evolutionary}: essentially, protocols emerge automatically as collections of ``fraglets'', i.e., combinations of code segments and parameters which are evolved, respectively, by distributed GP \cite{yamamoto2005experiments,yamamoto2005genetic} and distributed EAs \cite{alouf2010fitting}, and spread over the network through opportunistic (epidemic) propagation \cite{alouf2007embedding} regulated by interactions with the environment. On top of this, another EA optimizes the combination of protocols, i.e., the protocol stack \cite{miorandi2006service,baude2010mixing,imai2010practical}.

Finally, it is worth mentioning recent works on the application of GA to DTNs which are somehow related to this paper in a broader sense, namely \cite{bucur2015black}, where a GA is used to find specific DTN conditions characterized by abnormally low delivery rates, and \cite{bucur2016optimizing,bucur2017improved}, where the parameters of groups of heterogeneous malicious nodes attacking the network are optimized in the attempt to \emph{reduce} the delivery probability. Indeed, security is a major concern in this kind of networks: their inherent intermittent (and open) nature makes it difficult to apply encryption and authentication techniques which are common in other kinds of networks \cite{kate2007anonymity,farrell2006security}, although some possible countermeasures have been proposed.

% -------------------------------------------------------------------------
% -------------------------------------------------------------------------
% -------------------------------------------------------------------------

\section{Methods}
\label{sec:methods}

For the numerical experiments, we used \textbf{The ONE} (Opportunistic Network Environment) \cite{keranen2009one}, which allows us to simulate a given urban environment composed of a map and a number of moving agents of different types, producing a network traffic with user-defined characteristics. As for the GP algorithm, we used the implementation of the strongly typed GP \cite{montana1995strongly} provided by the \textbf{Jenetics} library \cite{wilhelmstotter2017jenetics}. At each generation, the GP algorithm generates a set of Java classes that implement candidate instances of the routing protocol, differing only for the instructions contained in the \texttt{update()} method defined in the abstract Java class \texttt{ActiveRouter} implemented in The ONE. We also include in our scheme a validity check with repair mechanism, in order to ensure the validity of the code generated by GP. The dynamically generated classes are then compiled at runtime and fed to the simulator, in order to extract the relevant information needed for the fitness evaluation. In the following, we detail the GP configuration, the validity check and repair mechanism, and the way the code generated by GP is evaluated in The ONE. %before feeding it to The ONE simulator. 

\subsection{Genetic Programming configuration}
\label{sec:gp}

The candidate GP individuals, represented as tree structures, are obtained by composing the elements in the non-terminal and terminal sets specified in Table \ref{tab:nonterminals} and Table \ref{tab:terminals} respectively (see Section \ref{sec:repair}). The non-terminals include basic Boolean operators as well as the inequality test, the \texttt{if} and the \texttt{sequence} operators. Note that \texttt{if} is considered as a 2-argument operator (i.e., without \texttt{else}) since \texttt{if}-\texttt{else} statements can be obtained as a concatenation of \texttt{if} and \texttt{sequence}. The terminals, instead, are obtained by ``dissecting'' the \texttt{update()} method of the baseline protocols into its \emph{main functional components}, which are then rearranged by GP. This is an important aspect of our proposal: rather than evolving from scratch the entire protocol's logic, which would entail an excessively large, hard-to-explore protocol space, we use available knowledge in the form of protocol basic components, for which we then try to identify a better rearrangement by means of GP. We believe that this form of Genetic Improvement is particularly interesting in that it couples the advantages of using pre-existing code (in this case, in the form of Java functions), representing necessary functional elements, with the power of the genetic search. In our case, we identify as basic components six functions, including the superclass' \texttt{update()} method, that implement basic operations which are at the base of the two selected protocols considered in our experimentation. The terminals also include the \texttt{return} keyword to exit from the \texttt{update()} method.

The implementation of the GP algorithm follows the logics of Jenetics \cite{wilhelmstotter2017jenetics} and the genetic operators available therein. At the first generation, the individuals are randomly generated using the aforementioned non-terminals and terminals (note that the baseline protocols are not used as seeds in the initial GP population), with given maximum tree depth and maximum number of nodes (``grow'' initialization method). Then, the evolutionary loop starts, where at each cycle:
\begin{enumerate}[leftmargin=*]
    \item The individuals in the current population are verified and, if needed, repaired (see Section \ref{sec:repair}), before being evaluated into The ONE (see Section \ref{sec:evaluation}).
    \item The new population is obtained by applying tournament selection to the current population. A predefined offspring fraction is set, which determines how many selected individuals (the offspring) will be altered by crossover and mutation; the remaining individuals (the survivors) are kept unaltered in the new population.
    \item The offspring are altered by applying \emph{single-node crossover}: with a given crossover probability, every individual among the offspring is selected to undergo crossover; the selected individuals are randomly paired (for a total number of pairs equal to half the number of offspring), and for every pair two new individuals are obtained by swapping two nodes (randomly chosen from the two original individuals in the pair), with the corresponding subtrees. The two new individuals replace the two original ones in the pair.
    \item The individuals obtained by crossover are then selected to undergo also mutation, with a given individual mutation probability (otherwise individuals are not mutated). Each individual selected to undergo mutation is altered by \emph{swap-mutation}: with a given node mutation probability, every node in the GP individual is selected and swapped with another randomly chosen node from the same individual. 
\end{enumerate}
This loop (steps 1 to 4) goes on until one of two stop criteria is met: a) the maximum number of generations is reached, or b) the maximum number of generations with steady fitness is reached. The latter criterion refers to the maximum number of generations in which the best fitness in the population does not improve. %(note that no elitism is used) 
No specific anti-bloat mechanism (apart from the limited depth and number of nodes in the initial GP population) or history of previously evaluated solutions is used. See Table \ref{tab:gp_parameters} for the detailed parameters (default values in Jenetics) used in the GP algorithm.

%NOTE: see file GPelements.java
\begin{table}[ht!]
\centering
\caption{Non-terminals used, their corresponding Java code, argument types and return types.}
\label{tab:nonterminals}
\begin{tabular}{llll}
\toprule
\textbf{Non-terminal} & \textbf{Java code} & \textbf{Argument types} & \textbf{Return type} \\ \midrule
\texttt{or} & \texttt{( arg1 $||$ arg2 )} & \emph{condition, condition} & \emph{condition}\\
\texttt{not} & \texttt{!arg1} & \emph{condition} & \emph{condition}\\
\texttt{notEqual} & \texttt{( arg1 != arg2 )} & \emph{condition}, \emph{condition} & \emph{condition}\\
\texttt{if} & \texttt{if ( arg1 )\{ arg2 \};} & \emph{condition}, \emph{body} & \emph{body} \\
\texttt{sequence} & \texttt{arg1; arg2;} & \emph{body}, \emph{body} & \emph{body} \\ \bottomrule
\end{tabular}
\end{table}

\begin{table}[ht!]
\centering
\caption{Terminals used, their corresponding Java code and type. The terminal \texttt{tryOtherMessages} is used only for improving the PRoPHET protocol. All the other terminals are used for both Epidemic and PRoPHET. Note that the method \texttt{tryAllMessagesToAllConnections()} is called on the child class (\texttt{this}) in order to override the method of the parent class \texttt{ActiveRouter}.}
\label{tab:terminals}
\resizebox{\textwidth}{!}{
\begin{tabular}{llc}
\toprule
\textbf{Terminal} & \textbf{Java code} & \textbf{Type} \\ \midrule
\texttt{isTransferring} & \texttt{isTransferring()} & \emph{condition} \\
\texttt{canStartTransfer} & \texttt{canStartTransfer()} & \emph{condition} \\
\texttt{update} & \texttt{super.update()}; & \emph{body} \\
\texttt{exchangeDeliverableMessages} & \texttt{exchangeDeliverableMessages();} & \emph{body} \\
\texttt{tryAllMessagesToAllConnections} & \texttt{this.tryAllMessagesToAllConnections();} & \emph{body} \\
\texttt{tryOtherMessages} & \texttt{tryOtherMessages();} & \emph{body} \\
\texttt{return} & \texttt{return;} & \emph{body} \\\bottomrule
\end{tabular}
}
\end{table}

\begin{table}[ht!]
\centering
\caption{Parameter setting (Koza-style tableau) of the Genetic Programming algorithm.}
\label{tab:gp_parameters}
\begin{tabular}{ll}
\toprule
\textbf{Parameter} & \textbf{Value} \\ \midrule
Objective & Delivery probability \\
Non-terminal set & See Table \ref{tab:nonterminals} \\
Terminal set & See Table \ref{tab:terminals} \\
Population size & 150 \\
Initialization & Grow, max depth 5, max nodes 50 \\
Offspring fraction & 0.6 \\
Crossover & Single-node crossover, prob. 0.1 \\
Mutation & Swap-mutation, individual prob. $0.1^{2/3}$, node prob. $0.1^{1/3}$ \\ %$\sqrt[3]{0.1}$
Selection & Tournament selection, size 3 \\
Steady fitness generations & 50 \\
Max number of generations & 100 \\ \bottomrule
\end{tabular}
\end{table}

%------------------------------------------------

\subsection{Validity check and repair method}
\label{sec:repair}

\begin{figure}[ht!]
\begin{minipage}{0.49\textwidth}
    \centering
    \begin{subfigure}[b]{\textwidth}
    \centering
    \includegraphics[width=\textwidth]{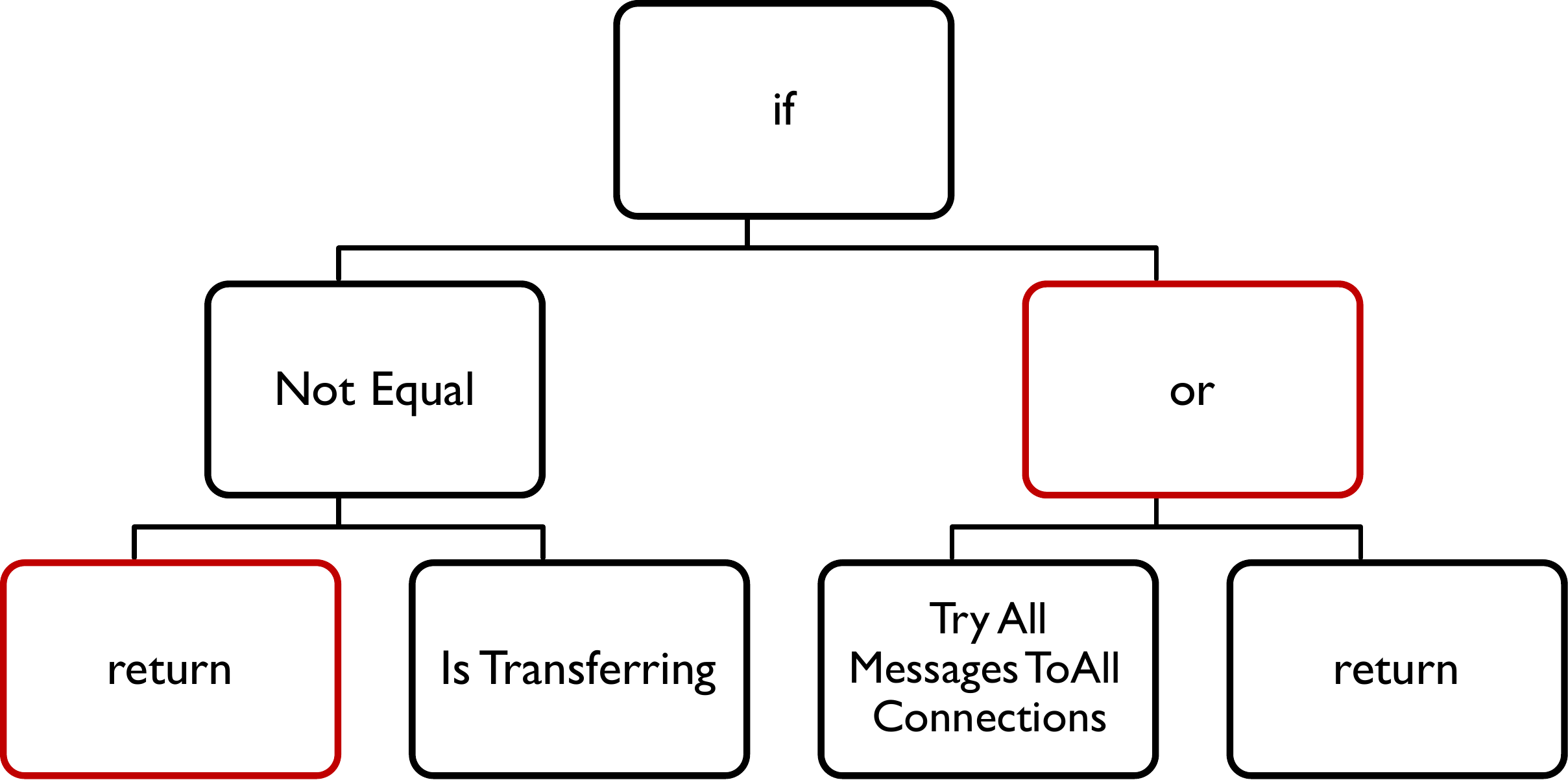}
    \caption{Invalid tree}%\label{}
    \vspace{0.5cm}
    \end{subfigure}
    \begin{subfigure}[b]{\textwidth}
    \centering
    \includegraphics[width=\textwidth]{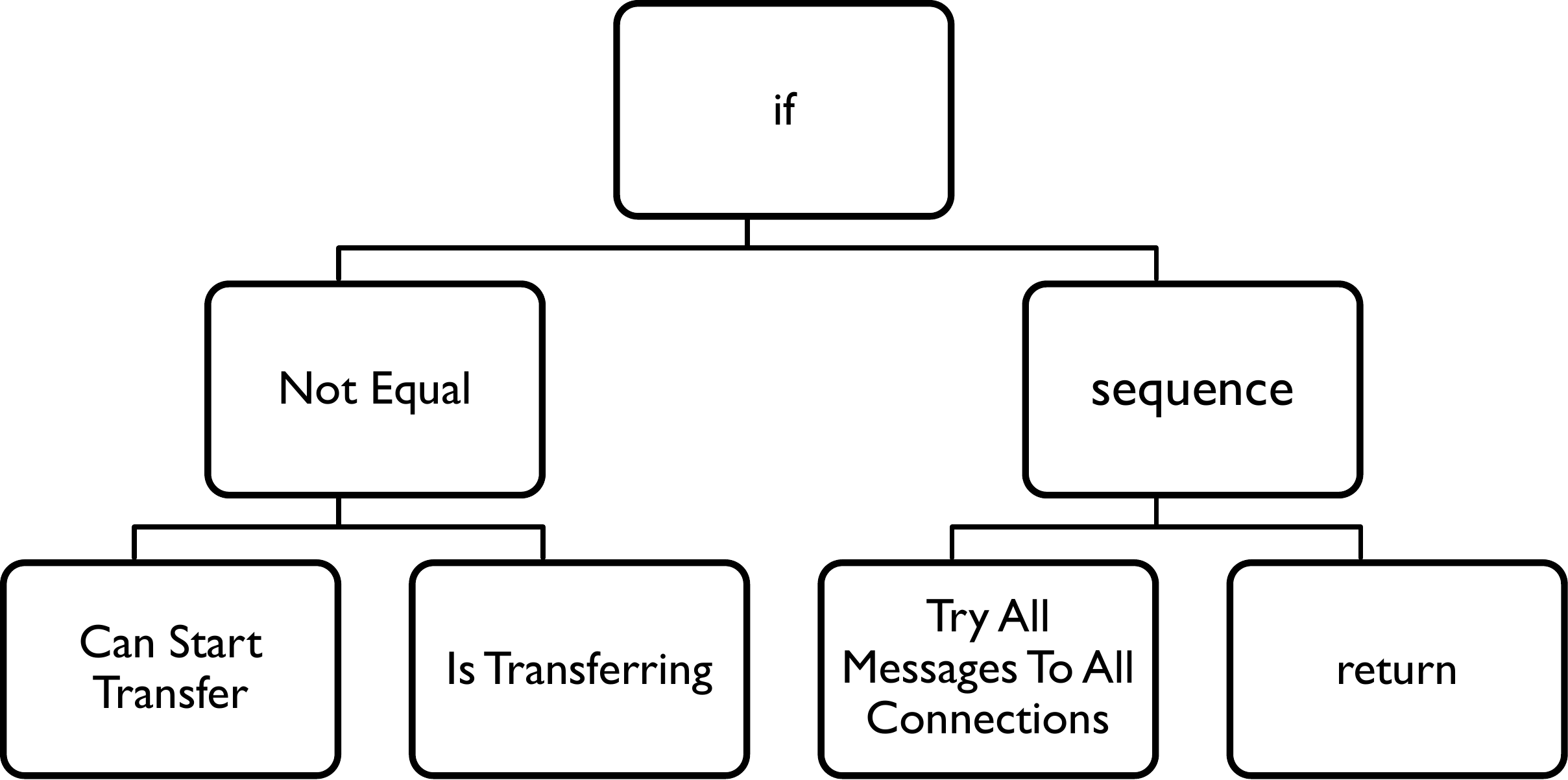}
    \caption{Correct tree}%\label{}
    \end{subfigure}
    \caption{Example of the repair mechanism on an invalid tree in which the \texttt{if} condition contains a \texttt{return} instruction and the \texttt{if} body contains an \texttt{or} condition.}
    \label{fig:repair}
\end{minipage}
\hfill
\begin{minipage}{0.49\textwidth}
    \includegraphics[width=\textwidth]{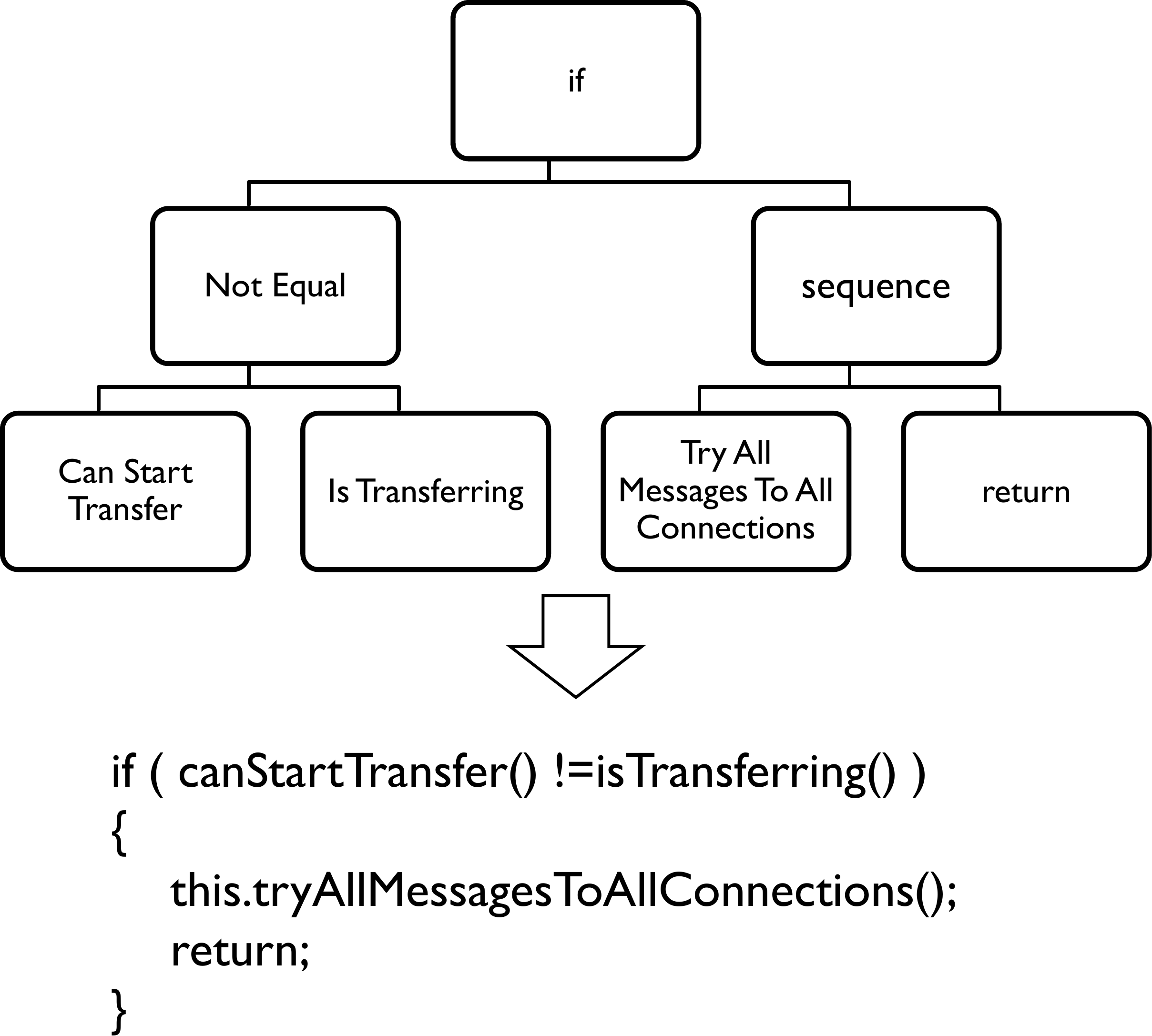}
    \caption{Example of translation from tree to Java code.}
    \label{fig:tree2class}
\end{minipage}
\end{figure}

Not all the elements in the non-terminal and terminal sets defined in Tables \ref{tab:nonterminals}-\ref{tab:terminals} can be an argument of every non-terminal: for example, a \texttt{return} instruction cannot be inside an \texttt{if} condition and an \texttt{or} condition cannot be inside an \texttt{if} body, as shown in Figure \ref{fig:repair} (top). To address this issue, we have developed two methods: the first one checks the validity of each individual tree structure, the latter tries to repair it, if needed. In order to check the validity of the trees and repair them, an additional structure is built specifying, for each non-terminal, the number of arguments and their types. In our case, we defined two possible types, namely \emph{body} and \emph{condition}. 

The validity check starts from the root node and checks if it has the correct number and types of arguments. For example, if the root node is an \texttt{if} node, it must have two arguments, respectively of type \emph{condition} and \emph{body}, as shown in Table \ref{tab:nonterminals}. If this check fails, the current tree is considered not correct. Otherwise, the method recursively checks the arguments, until it reaches the terminals, whose type must comply with those of the arguments of their corresponding parent node. If this recursive check does not fail, the tree is correct.

The repair method is invoked whenever the validity check fails and, following a similar logic, it visits recursively all the nodes in the tree (see Figure \ref{fig:repair} for an example). More specifically: 1) in the case of the non-terminals, if it finds that a node does not have the correct number of arguments, or these are not of the correct type, the method tries to randomly replace that node with another non-terminal that has that number of arguments of those types; 2) in the case of terminals, these are replaced, if needed, to comply with the type of the arguments of their parent node. The method also ensures that the tree contains at least one \texttt{return} terminal. If the tree cannot be repaired, the tree is not evaluated in the simulator and it is assigned a delivery probability of zero. %Figure \ref{fig:repair} shows an example of the repair mechanism: on the top, the tree is not correct and the repair method acts on it, generating the tree on the bottom. %The pseudocode of the repair method is shown in Algorithm \ref{alg:repair}.

% \begin{algorithm}[ht!]
% \caption{Repair method}
% \label{alg:repair}
% \begin{algorithmic}[1]
% \Procedure{repair}{$currentNode, expectedType$}
% \State $currentType \gets$ type of $currentNode$
% \State $numberArguments \gets$ number of arguments of $currentNode$
% \If{$currentType \not= expectedType$}
% \State $elements \gets$ set of nodes with $expectedType$ and $numberArguments$
% \If{$elements \not= empty$}
% \State $currentNode \gets$ random node from $elements$
% \EndIf
% \EndIf
% \For{$argument \gets$ arguments of $currentNode$}
% \State $expectedType \gets$ expected type of $argument$
% \State \Call{repair}{$argument, expectedType$} \Comment{recursive repair}
% \EndFor
% \EndProcedure
% \end{algorithmic}
% \end{algorithm}

\subsection{Individual evaluation}
\label{sec:evaluation}

Each individual (i.e., a candidate implementation of the \texttt{update()} method, and its corresponding Java class that is dynamically generated and compiled) is evaluated by measuring the delivery probability in The ONE simulations. This is calculated as the fraction of messages that reach their destination over the total number of messages generated during the simulation. In particular, the individual evaluation proceeds according to the following steps:

\begin{enumerate}[leftmargin=*]
    \item Class name generation: A \emph{unique id} is appended to the class name of each GP individual, in order to avoid having GP classes with the same name (the dynamically generated classes are placed in the same folder, and each class is loaded by The ONE at the beginning of a simulation).
    \item Node evaluation: Each node composing the current individual tree is translated into the corresponding Java instructions (see Tables \ref{tab:nonterminals}-\ref{tab:terminals}), generating the code that will be placed in the \texttt{update()} method of the dynamically generated Java class. %For example, a node \texttt{or} with arguments \texttt{isTransferring} and \texttt{canStartTransfer} is translated into the Java code: \texttt{isTransferring()} $||$ \texttt{canStartTransfer()}. 
    In Figure \ref{fig:tree2class}, we show an example of translation from a GP tree to the corresponding Java code inserted into the \texttt{update()} of the Java class implementing the routing protocol to be evaluated.
    \item Class generation: The generation of the Java class is done by using a template class, which contains the structure of a routing protocol that extends the \texttt{ActiveRouter} class present in The ONE simulator. The templates of the Epidemic and PRoPHET routing protocols are presented in Appendix \ref{sec:appendix_templates}, where it can be seen that each template contains the necessary imports, the class definition, its constructors, public and private fields, the \texttt{update()} method and other additional methods contained in the baseline protocol. It should be noted that apart from the \texttt{update()} method, all the other parts of the class are not modified by GP. For each individual, the template is modified by replacing the code of the baseline protocol inside the \texttt{update()} method with the code generated at the previous step.
    \item Class compiling: The dynamically generated class is compiled by calling \texttt{javac}, and placed in the correct classpath, in order to allow the simulator to execute it. If the class fails to compile, the corresponding tree is assigned a delivery probability of zero.
    \item Settings file configuration: The ONE uses a settings file to store all the relevant parameters of the simulation. An example of this file is given in Appendix \ref{sec:theone_config}, where we highlight the parameters that are relevant to our experimentation. The main parameters of interest (see Table \ref{tab:one_settings}) are:
    \begin{itemize}
    \item \texttt{Scenario.name}: name of the simulation (i.e., the Java class name generated in the first step);
    \item \texttt{Group.router}: name of the routing protocol (same as \texttt{Scenario.name});
    \item \texttt{Group.nrofHosts}: number of nodes (a.k.a. hosts) per group (pedestrians and cars only);
    \item \texttt{Report.nrofReports}: number of reports created as output of the simulation;
    \item \texttt{Report.report1}: type of report created as output of the simulation;
    \item \texttt{Events1.hosts}: range of message source/destination addresses;
    \item \texttt{MovementModel.worldSize}: size of the map;
    \item \texttt{MapBasedMovement.nrofMapFiles}: number of map files used in the simulation;
    \item \texttt{MapBasedMovement.mapFile[1-4]}: maps used in the simulation;
    \item \texttt{Group[4-5-6].routeFile}: routes followed by the first/second/third group of trams;
    \item \texttt{ProphetRouter.secondsInTimeUnit}: specific setting for the PRoPHET routing protocol.
    \end{itemize}
    Before each individual simulation, a specific settings file is created starting from the template given in Appendix \ref{sec:theone_config}, where the relevant settings are set accordingly to the specific test case.
    \item Simulation execution: The DTN simulation uses the automatically generated Java class, producing a report containing the delivery probability of the messages sent in the network.
    \item Result retrieval: The report file \texttt{MessageStatsReport} generated by The ONE is parsed in order to extract the delivery probability, which is then used as fitness of the GP individual.
\end{enumerate}

In order to automate these steps, we have implemented a procedure that creates a script to automatically compile the generated Java class starting from the code generated by GP, prepare the settings file, and run The ONE simulation (in batch mode). This script is executed by using the Java Process Builder. When the simulation is terminated, the procedure accesses the report file, and extracts the delivery probability. If the report does not exist, e.g. because the generated class is invalid or the simulation fails, the current individual is assigned a delivery probability of zero.%and the other metrics of interest. 

It should be noted that the bottleneck in the process described above is the simulation execution (step 6): in our experiments, the wall-clock time of a single simulation ranged between 1 minute in the simplest test case to 10 minutes in the most complex one. On the other hand, the steps 1 to 5 and 7 are executed overall in approximately 3 seconds per GP individual.
%(Default map, 40 hosts per group)
%(Manhattan map, 100 hosts per group)

\begin{table}[ht!]
    \centering
    \caption{The ONE settings used in the different test cases.}
    \label{tab:one_settings}
    \resizebox{\textwidth}{!}{
    \begin{tabular}{lrl}
        \toprule
        \multirow{5}{*}{\textbf{Common settings}} & Scenario.name: & $\langle$Name of the Java class generated by Jenetics$\rangle$ \\
         & Group.router: & $\langle$Name of the Java class generated by Jenetics$\rangle$ \\ 
         & Group.nrofHosts: & 40 or 100 \\
         & Report.nrofReports: & 1 \\ 
         & Report.report1: & MessageStatsReport \\
         & Events1.hosts: & 0,126 or 0,306 \\
         \midrule
        \multirow{4}{*}{\textbf{Default settings}}          & MovementModel.worldSize: & 4500 $\times$ 3400 meters\\
         & MapBasedMovement.nrofMapFiles: & 4 \\
         & MapBasedMovement.mapFile[1-4]: & Roads, Main Roads, Pedestrian Paths, Shops \\ 
         & Group[4-5-6].routeFileGroups: & Tram 3-4-10 \\ \midrule
        \multirow{4}{*}{\textbf{Helsinki settings}} & MovementModel.worldSize: & 100000 $\times$ 100000 meters\\
         & MapBasedMovement.nrofMapFiles: & 1 \\
         & MapBasedMovement.mapFile1: & Helsinki Medium - Roads \\ 
         & Group[4-5-6].routeFileGroups : & Helsinki Medium - Bus A. Bus B, Bus C \\ \midrule
        \multirow{4}{*}{\textbf{Manhattan settings}} & MovementModel.worldSize: & 100000 $\times$ 100000 meters\\
         & MapBasedMovement.nrofMapFiles: & 1 \\
         & MapBasedMovement.mapFile1: & Manhattan - roads \\ 
         & Group[4-5-6].routeFileGroups: & Manhattan - Bus \\ \midrule
        \multirow{1}{*}{\textbf{PRoPHET settings}} & ProphetRouter.secondsInTimeUnit: & 30 \\
        \bottomrule
    \end{tabular}
    }
\end{table}

% -------------------------------------------------------------------------
% -------------------------------------------------------------------------
% -------------------------------------------------------------------------

\section{Experimental setup}
\label{sec:setup}
\begin{figure}[ht!]
    \centering
    \begin{subfigure}[b]{0.32\textwidth}
    \centering
    \includegraphics[height=0.81\textwidth]{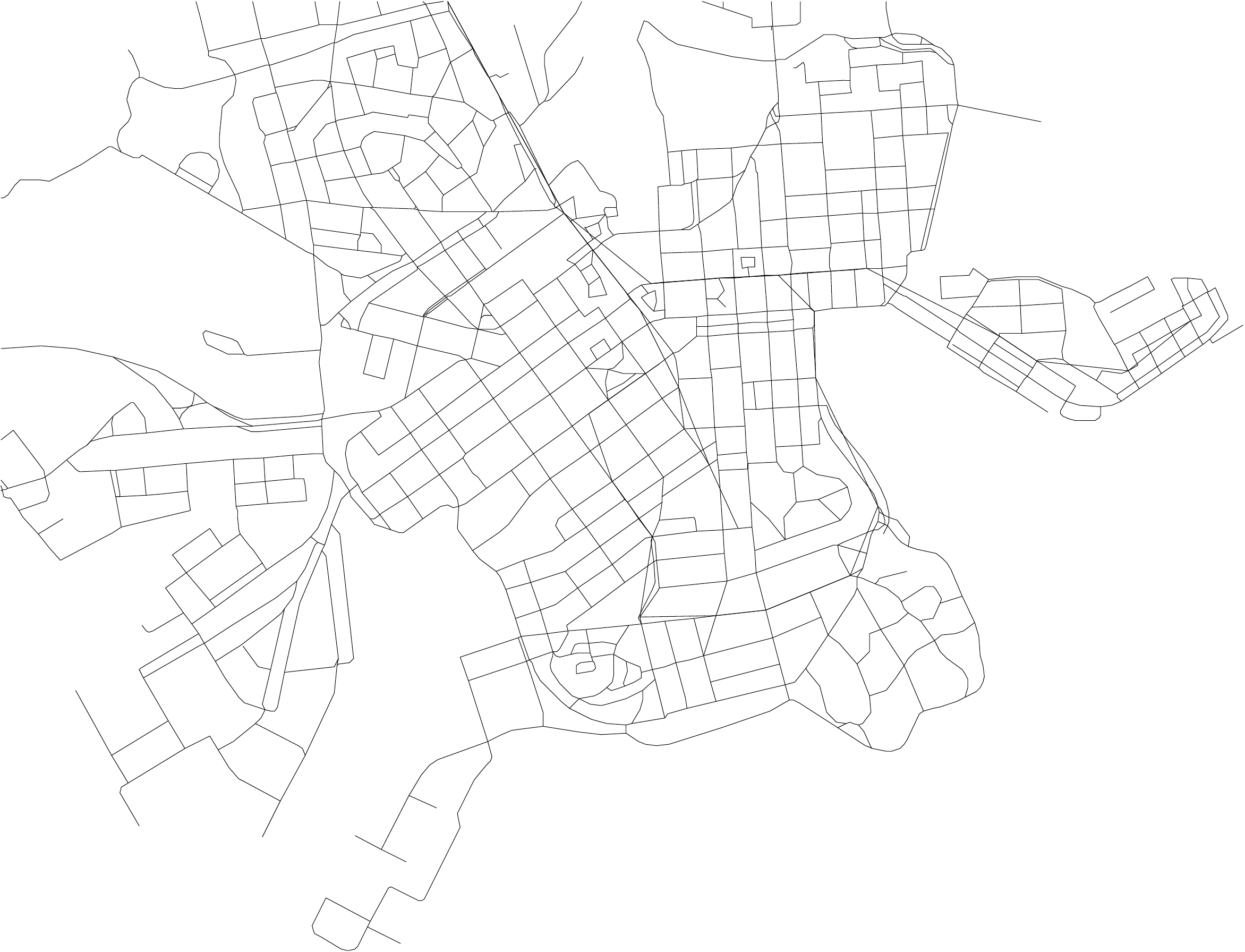}
    \caption{Default}%\label{}
    \end{subfigure}
    \begin{subfigure}[b]{0.32\textwidth}
    \centering
    \includegraphics[height=0.81\textwidth]{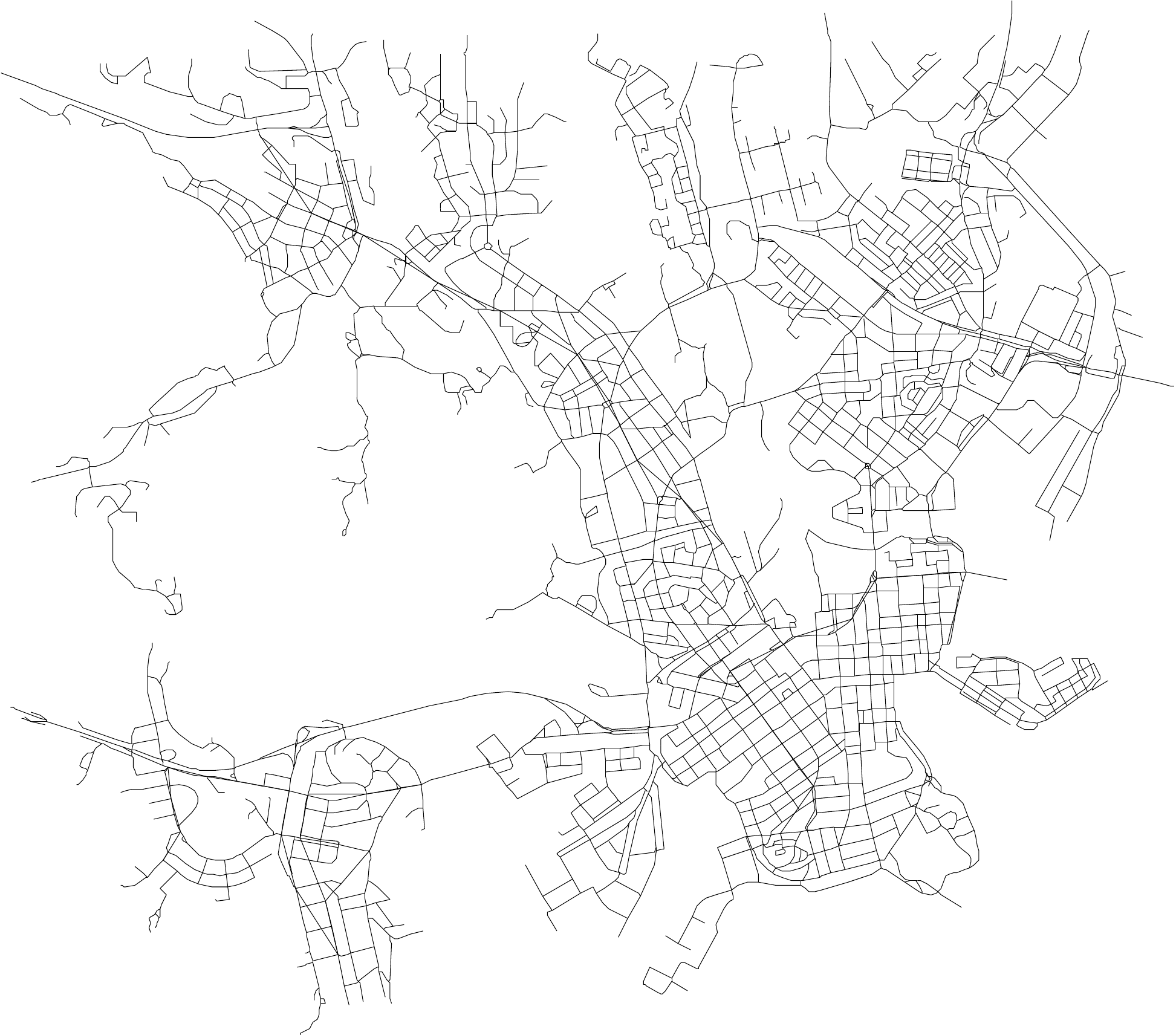}
    \caption{Helsinki}%\label{}
    \end{subfigure}
    \begin{subfigure}[b]{0.32\textwidth}
    \centering
    \includegraphics[height=0.81\textwidth]{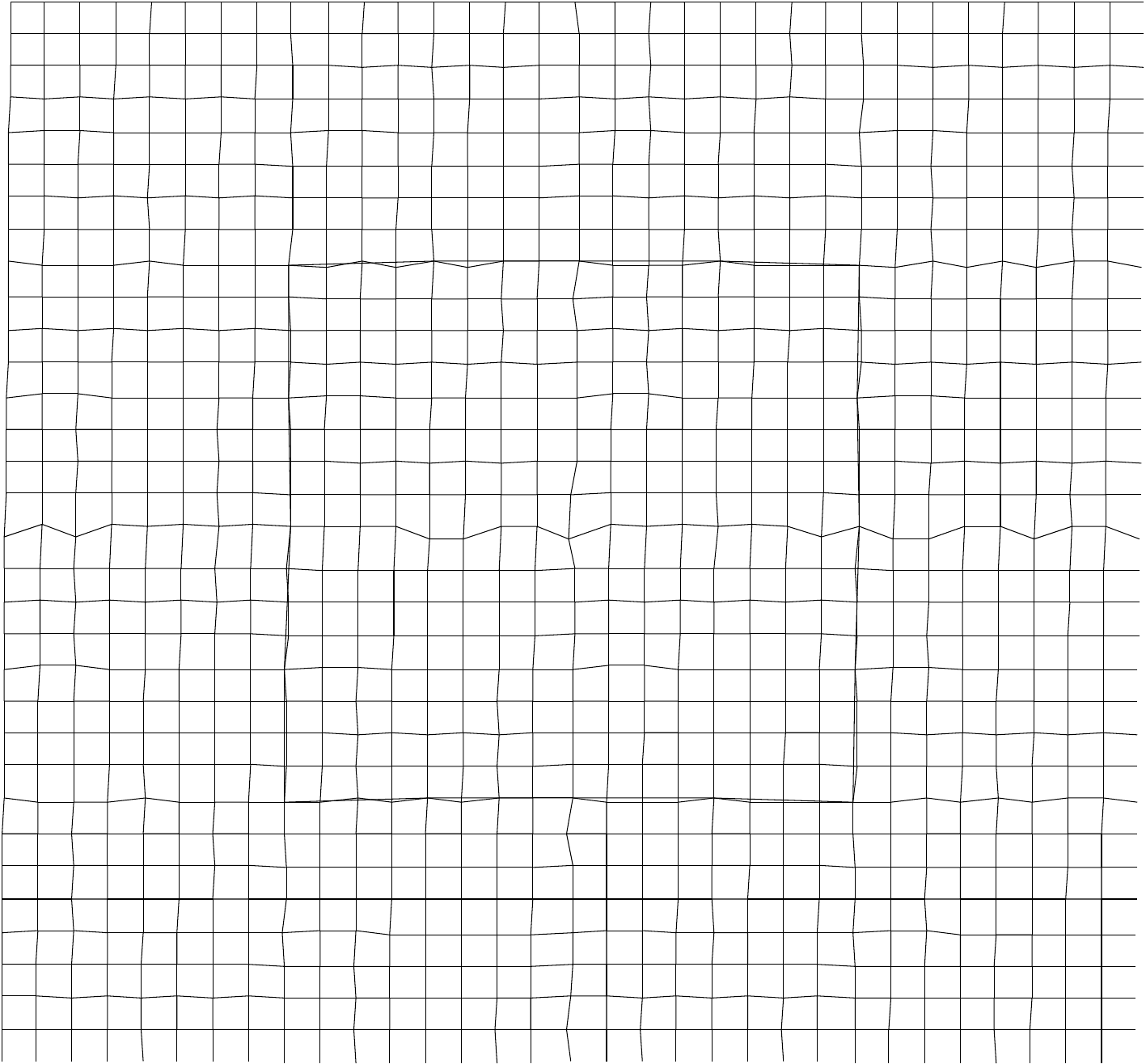}
    \caption{Manhattan}%\label{}
    \end{subfigure}
    \caption{City maps used in the experimentation.}
    \label{fig:maps}
\end{figure}

We considered six different test cases, based on three different maps and two different numbers of agents. The maps used are those available in The ONE, namely the city center of Helsinki (that in The ONE is identified as the ``default'' map), the metropolitan area of Helsinki (a.k.a. Greater Helsinki), and a Manhattan-like map, see Figure \ref{fig:maps}. In the following, we will refer to these three maps as \textbf{Default}, \textbf{Helsinki} and \textbf{Manhattan} respectively. 

%------------------------------------------------

\subsection{Simulation configuration}
\label{sec:simulation}

The simulation time of each DTN simulation is 12 hours (with an update interval of 0.1 seconds), starting after a warm-up period of 1000 seconds of simulation time needed to allow the node mobility to reach steady state conditions. The simulated DTNs are open networks (i.e., without authentication) made of three types of mobile nodes (referred to as \emph{hosts}): pedestrians, cars and trams. These heterogeneous networks represent realistic scenarios of hosts moving at different velocity so that establishing a cabled network infrastructure is either obviously impossible (for pedestrian and cars) or too costly to set up and maintain (for trams). In our experiments, hosts are further divided into 6 groups: two groups of pedestrians, one group of cars and three groups of trams. The groups of pedestrians and cars are composed of 40 or 100 hosts each (depending on the experiments), while in all the experiments the groups of trams are composed of 2 hosts each. Thus, the simulations with 40 hosts per group have a total number of $40\times2 + 40 + 3\times2=126$ hosts. The simulations with 100 hosts per group have a total number of $100\times2 + 100 + 3\times2=306$ hosts. During the simulations, a new message of size 500KB-1MB is generated every 25-35 seconds (both the message size and the interval are uniformly sampled in these ranges), with source and destination randomly chosen among all the hosts in the network.

Each group has different networking parameters and mobility behaviors, which are specified in the settings file and have been set according to the default parameters of The ONE (see Appendix \ref{sec:theone_config}). Concerning the networking parameters, we considered two network interfaces, namely: low-speed short-range Bluetooth (transmit speed: 250kBps, message buffer: 5MB, range: 10m), and a high-speed connection (transmit: 10MBps, message buffer: 10MB, range: 1km). Bluetooth is available to all groups, while the high-speed connection is available only to the first group of trams.

As for the mobility behavior, the hosts are randomly placed on the map at the beginning of the simulation, and the destination of each host is chosen randomly between a set of available target points. The hosts then move according to one of the following mobility models:
\begin{itemize}[leftmargin=*]
    \item Pedestrians and cars use the \texttt{ShortestPathMapBasedMovement}, in which each host moves towards its destination following the shortest valid path on the map. Once the destination is reached, the host waits for a given period of time before choosing (randomly) the next destination. Pedestrians move at 0.5-1.5 m/s, while cars move at 2.7-13.9 m/s. Both kinds of hosts have a waiting time of 0-120 seconds.
    \item Trams use the \texttt{MapRouteMovement}, in which the hosts move between adjacent points on a predefined set of routes of 10-100 points. Once a point is reached, each host waits for 10-30 seconds and then moves to the next randomly selected adjacent point, trying to avoid the point where it came from. Trams move at 7-10 m/s.
\end{itemize}

%------------------------------------------------

\subsection{Routing protocols}
\label{sec:routing_protocols}

%NOTE: ad hoc routing is different from DTN routing
%https://en.wikipedia.org/wiki/List_of_ad_hoc_routing_protocols
%https://en.wikipedia.org/wiki/Routing_in_delay-tolerant_networking
As discussed in Section \ref{sec:intro}, routing in DTNs is challenging (it is a NP-hard problem \cite{balasubramanian2007dtn}) due to their highly dynamic conditions caused by the sparsity and mobility of the nodes. In fact, while in other kinds of ad hoc networks (denser and/or less mobile) nodes can build explicit routes and forward data via established paths (this is the case, for instance, of routing protocols like Ad hoc On-demand Distance Vector (AODV) \cite{perkins1999ad} or Dynamic Source Routing (DSR) \cite{johnson1996dynamic}) this is not possible in DTNs due to their lack of continuous connectivity and the consequent absence of instantaneous (and stable) end-to-end paths. Rather than building explicit routes, routing in these cases must take a ``store and forward'' approach, leveraging the nodes' mobility itself to allow nodes to exchange and carry messages as they move, hoping that this \emph{epidemic-like} propagation will eventually allow messages to reach their intended destination. In the simplest form (called \emph{forwarding-based}) of this kind of routing, only one copy of each message exists at a time in the network: however, this usually does not provide sufficiently high delivery rates. On the other hand, \emph{replication-based} epidemic protocols allow multiple copies of each message to exist in the network at the same time. While this introduces some obvious overhead and hinders scalability, it is the only way to ensure a satisfactory (yet, sub-optimal) delivery probability. This second class of protocols is the focus of our work. In particular, we consider two of the main replication-based protocols, namely Epidemic and PRoPHET, whose functioning can be summarized as below:
\begin{itemize}[leftmargin=*]
    \item The Epidemic routing protocol \cite{vahdat2000epidemic} is a flooding-based protocol in which each node transmits its messages to every other node met that does not have a copy of them. The only limitation is the maximum number of hops for each message, or alternatively its time-to-live (TTL), i.e. the predefined maximum lifetime of a message over the network.%NOTE: >5000 citations
    \item The PRoPHET (Probabilistic Routing Protocol using the History of Encounters and Transitivity) protocol \cite{doria2003prophet} is a variant of the Epidemic routing protocol. This protocol defines a delivery predictability between any two nodes based on the history of contacts between them. A high delivery predictability means a high probability of future contacts between the two nodes. Instead of copying all messages, a message is copied only if the destination node's delivery predictability is higher than the transmitting node's delivery predictability. With respect to Epidemic, this mechanism allows PRoPHET to obtain comparable delivery probability yet with a lower overhead.
    %NOTE: >3000 citations
\end{itemize}
%
% \noindent{}We should note that several other epidemic protocols exist, such as RAPID \cite{balasubramanian2007dtn}, MaxProp \cite{burgess2006maxprop}, or Spray \& Wait \cite{spyropoulos2005spray}, which will be in the scope of future works.
%
In all the experimental test cases, we used the original implementation of Epidemic and PRoPHET available in The ONE, see Listings \ref{lst:epidemic} and \ref{lst:prophet} reported in Appendix \ref{sec:appendix_templates}.
%as baseline templates, and used GP to optimize the \texttt{update()} method, 

% -------------------------------------------------------------------------

\subsection{Computing environment}
\label{sec:computing_environment}

% The experiments related to the Default and Helsinki test cases have been performed on a Linux workstation with a 28-core CPU Intel i9-7940x @3.10GHz and 64 GB RAM. Since the simulations of the Manhattan test case resulted computationally more expensive, we performed the related experiments on 
The experiments have been performed on the High Performance Computing (HPC) facility available at our host institution. We used the JDK v1.8.0\_201 64bit, with Jenetics v5.2.0 and The ONE v1.6.0, with thread parallelization at the level of GP individuals (i.e., each thread handles the code generation of a single GP individual and its related simulation in The ONE)\footnote{Our code is publicly available at \url{https://github.com/michiL96/evolution_routing_protocol}.}. The total HPC runtime of our experiments changed based on the size of the map and the number of hosts, ranging from $\sim$3-5 hours (Default map, 40 hosts per group) to $\sim$9-11 hours (Helsinki map or Manhattan map, 40 hosts per group), $\sim$13-14 hours (Default map, 100 hosts per group), and $\sim$1 day (Helsinki map or Manhattan map, 100 hosts per group). Each simulation has been executed with 12 cores.
%https://sites.google.com/unitn.it/hpc/architecture?authuser=0

% -------------------------------------------------------------------------
% -------------------------------------------------------------------------
% -------------------------------------------------------------------------

\section{Experimental results}
\label{sec:results}

\begin{figure}[ht!]
\begin{minipage}{0.49\textwidth}
    \centering
    \includegraphics[width=\textwidth]{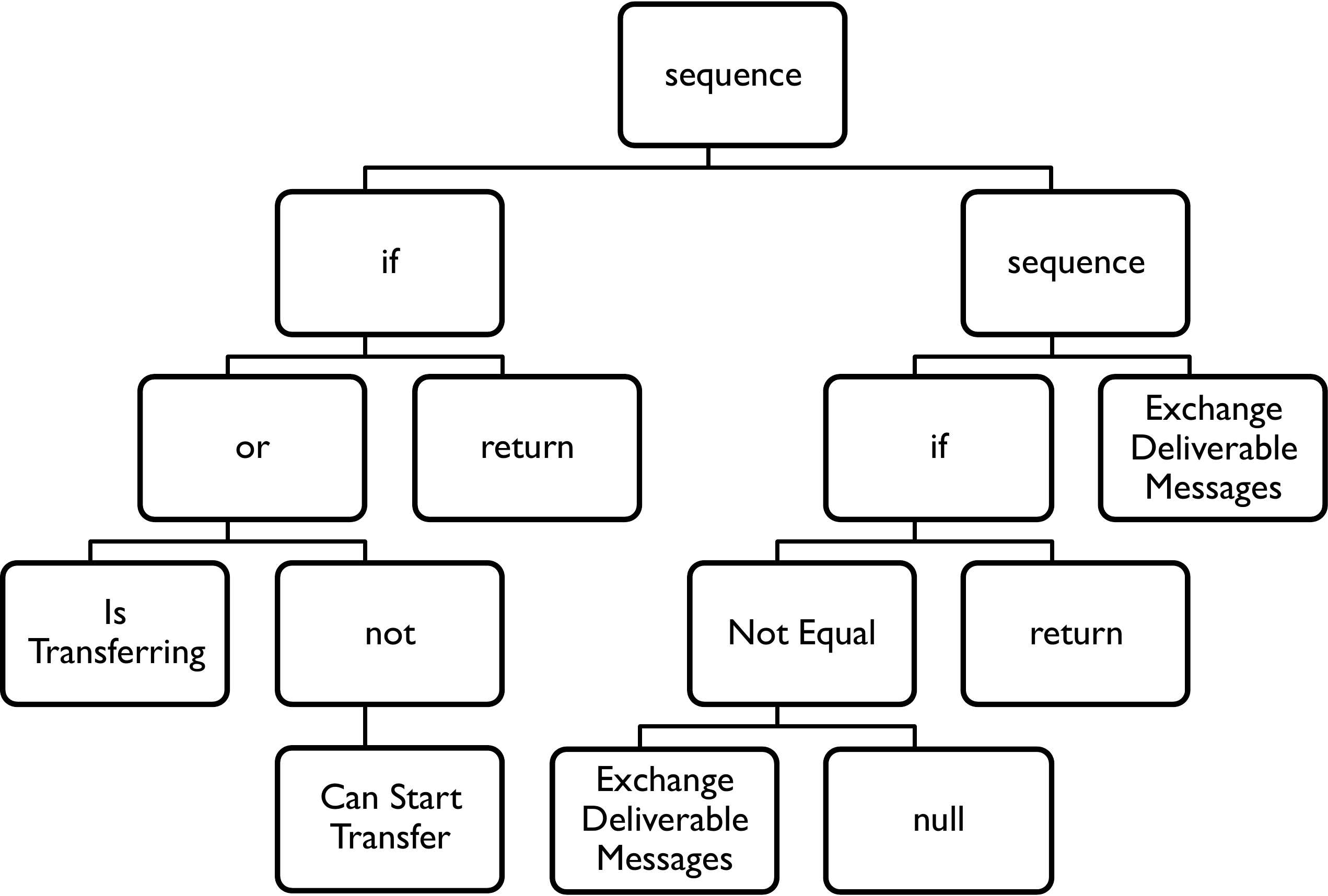}
    \caption{Tree of the baseline Epidemic routing protocol (note that this tree is not produced by GP).}
    \label{fig:tree_baseline_epidemic}
\end{minipage}
\hfill
\begin{minipage}{0.49\textwidth}
    \centering
    \includegraphics[width=.885\textwidth]{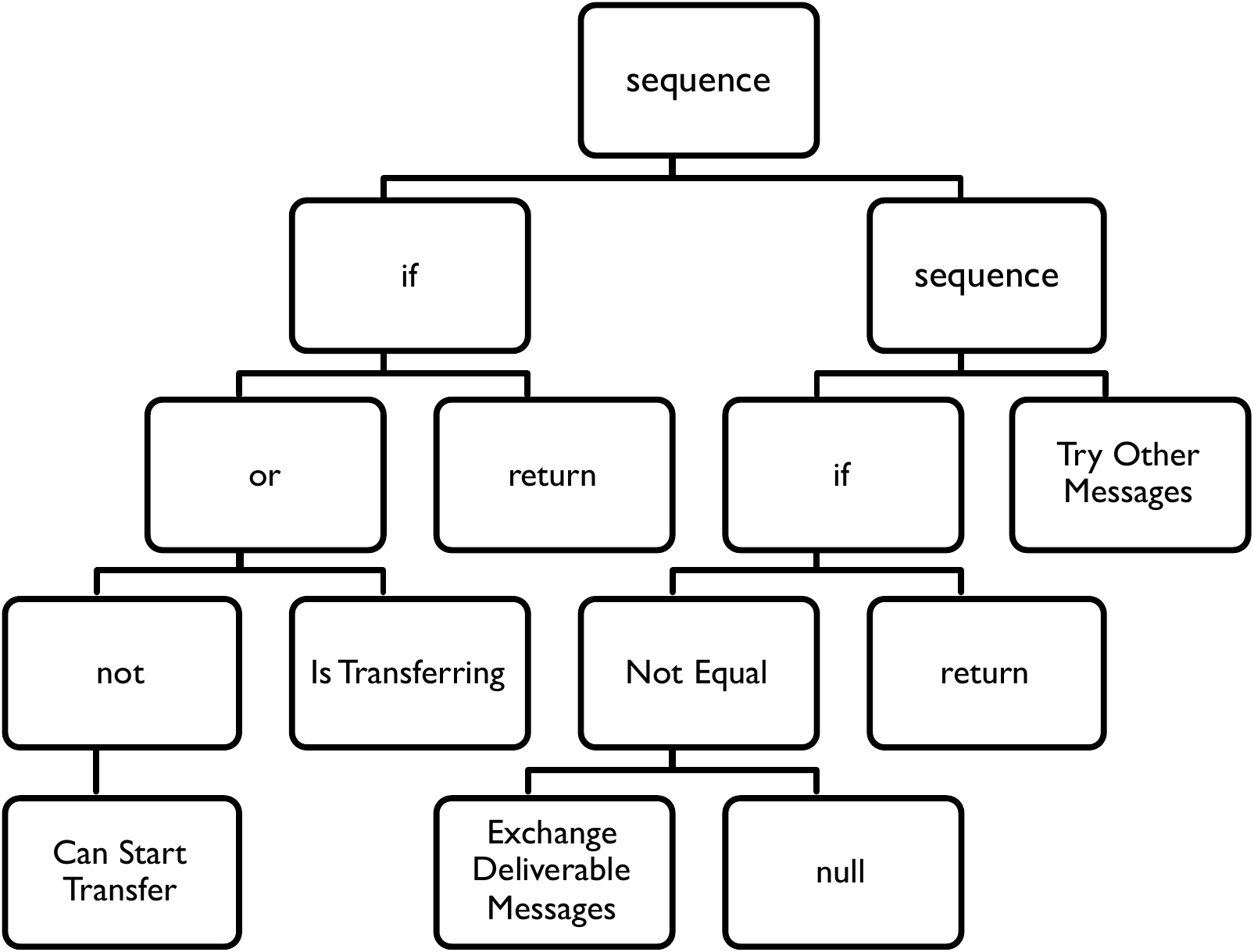}
    \caption{Tree of the baseline PRoPHET routing protocol (note that this tree is not produced by GP).}
    \label{fig:tree_baseline_prophet}
\end{minipage}
\end{figure}

In order to test the proposed method for genetically improving routing protocols, we considered the three urban maps discussed above, with two numbers of hosts per group (40 or 100), thus for a total of six test cases. In each test case, we used as baseline Epidemic and PRoPHET. For reference, we show in Figures \ref{fig:tree_baseline_epidemic}-\ref{fig:tree_baseline_prophet} the functioning logic (in the form of GP tree) corresponding to the \texttt{update()} method of the two baseline protocols. We conducted our experimental campaign as follows:
\begin{itemize}[leftmargin=*]
    \item First, we compared the delivery probability obtained by each of the two baseline protocols against the corresponding genetically improved protocols obtained by GP, in each test case. We considered all the combinations of $\langle$protocol, map, number of hosts per group$\rangle$ in: \{Epidemic, PRoPHET\} $\times$ \{Default, Helsinki, Manhattan\} $\times$ \{40 hosts, 100 hosts\}. For each test case, we executed the baseline protocol for \NUMRUNS~simulations (with different seeds). Similarly, we executed \NUMRUNS~runs (with different seeds) of the GP algorithm on each test case. We then compared the delivery probability of baseline vs best evolved protocols, and performed a statistical analysis based on the Wilcoxon rank-sum test. To further validate our results, we compared our improved PRoPHET protocols against three PRoPHET variants recently proposed in the literature (see Section \ref{sec:comparison_delivery}).
    \item Then, we performed a trade-off analysis aimed at understanding if optimizing for the delivery probability produces a degradation of other relevant network metrics  (see Section \ref{sec:other_metrics}).
\end{itemize}
In addition to this, we assessed the generalizability of the evolved protocols between one test case and another, in order to understand if the improved protocols are optimized for a specific test case or rather they can be used in different test cases (see Appendix \ref{sec:generalization}). Furthermore, we analyzed the functioning logics of the evolved test cases and identified some common patterns exploited by evolution to improve the existing protocols, and compared these logics with the ones underlying the baseline protocols (see Appendix \ref{sec:logics}). %We refer the interested reader to the Appendix for this additional analysis.

%----------------------------------------------------------------

\vspace{-0.5cm}
\subsection{Evolved vs baseline protocols: comparison on the data delivery probability}\label{sec:comparison_delivery}

\begin{figure}[ht!]
    \begin{subfigure}[b]{0.32\textwidth}
    \centering
    \includegraphics[width=\textwidth]{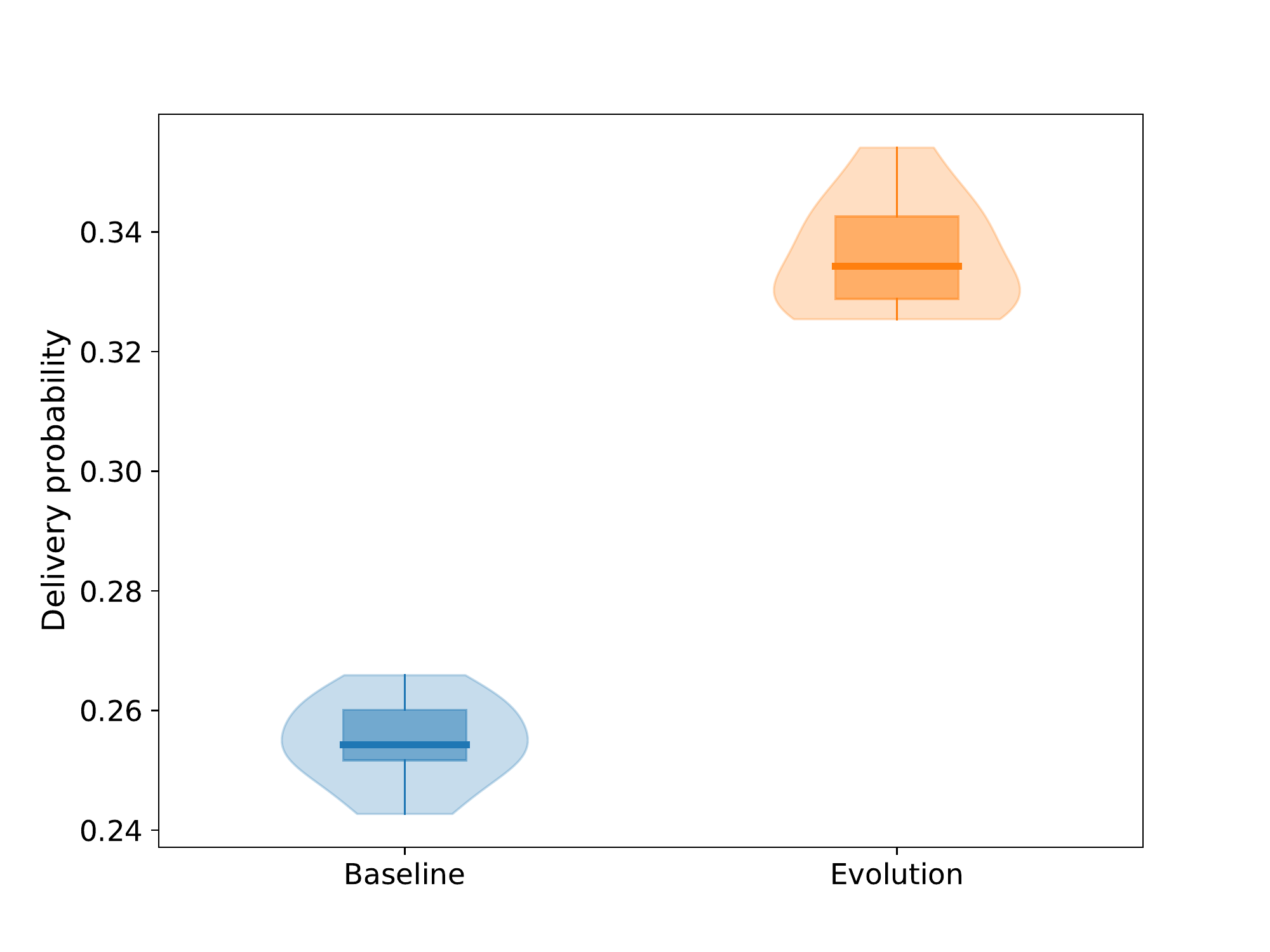}
    \caption{Default (40 hosts)}%\label{}
    \end{subfigure}
    \begin{subfigure}[b]{0.32\textwidth}
    \centering
    \includegraphics[width=\textwidth]{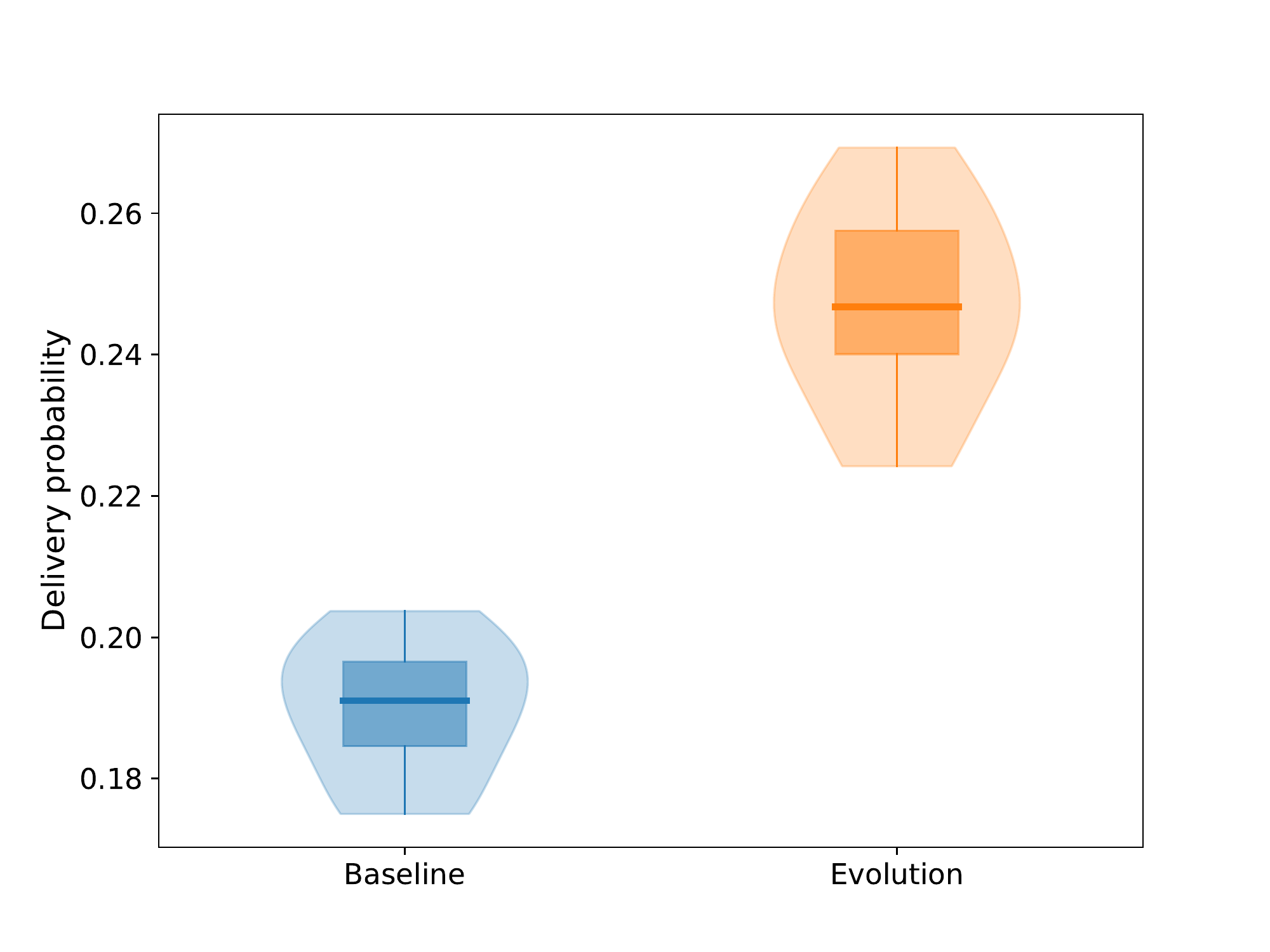}
    \caption{Helsinki (40 hosts)}%\label{}
    \end{subfigure}
    \begin{subfigure}[b]{0.32\textwidth}
    \centering
    \includegraphics[width=\textwidth]{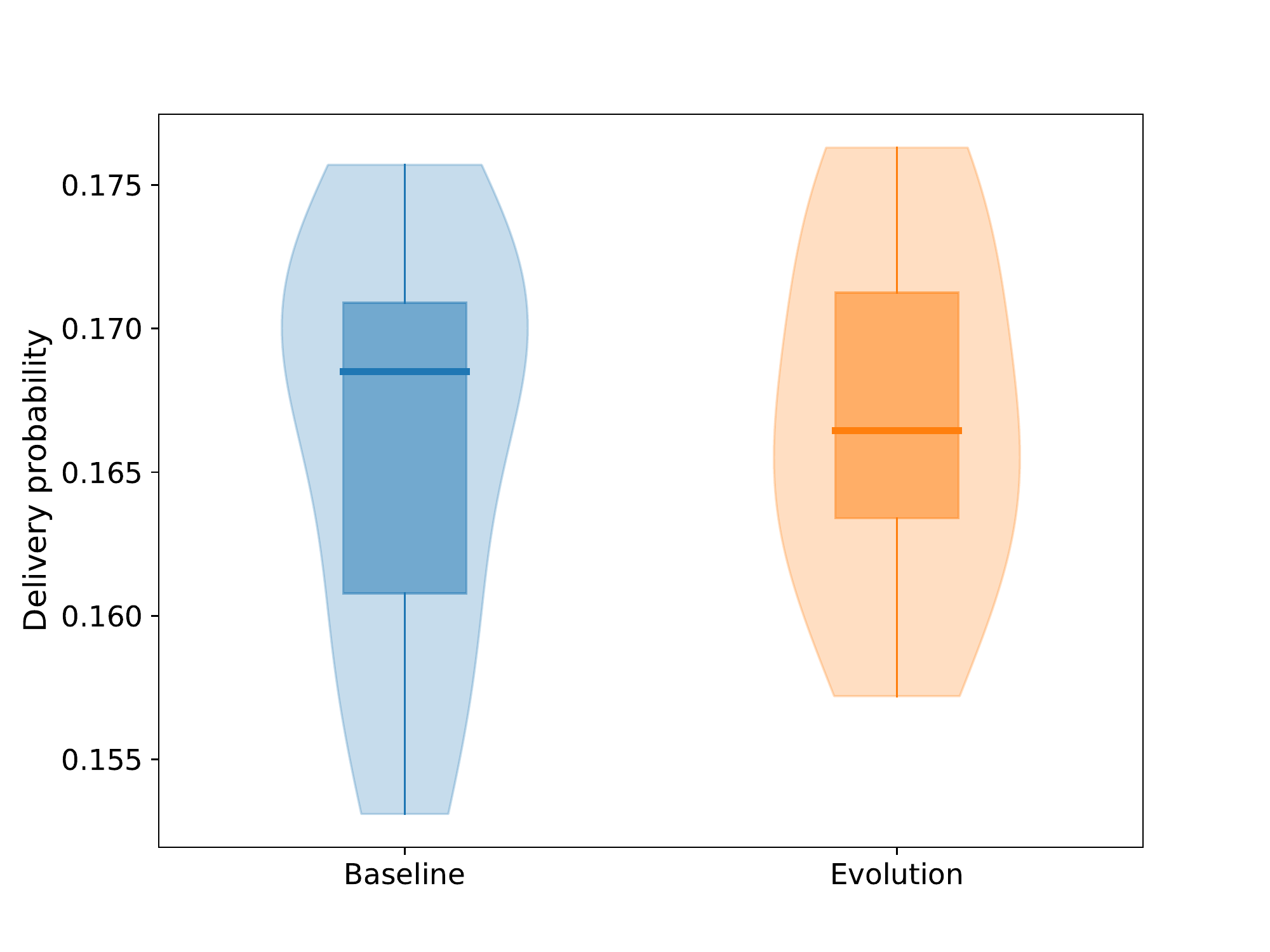}
    \caption{Manhattan (40 hosts)}%\label{}
    \end{subfigure}
    \begin{subfigure}[b]{0.32\textwidth}
    \centering
    \includegraphics[width=\textwidth]{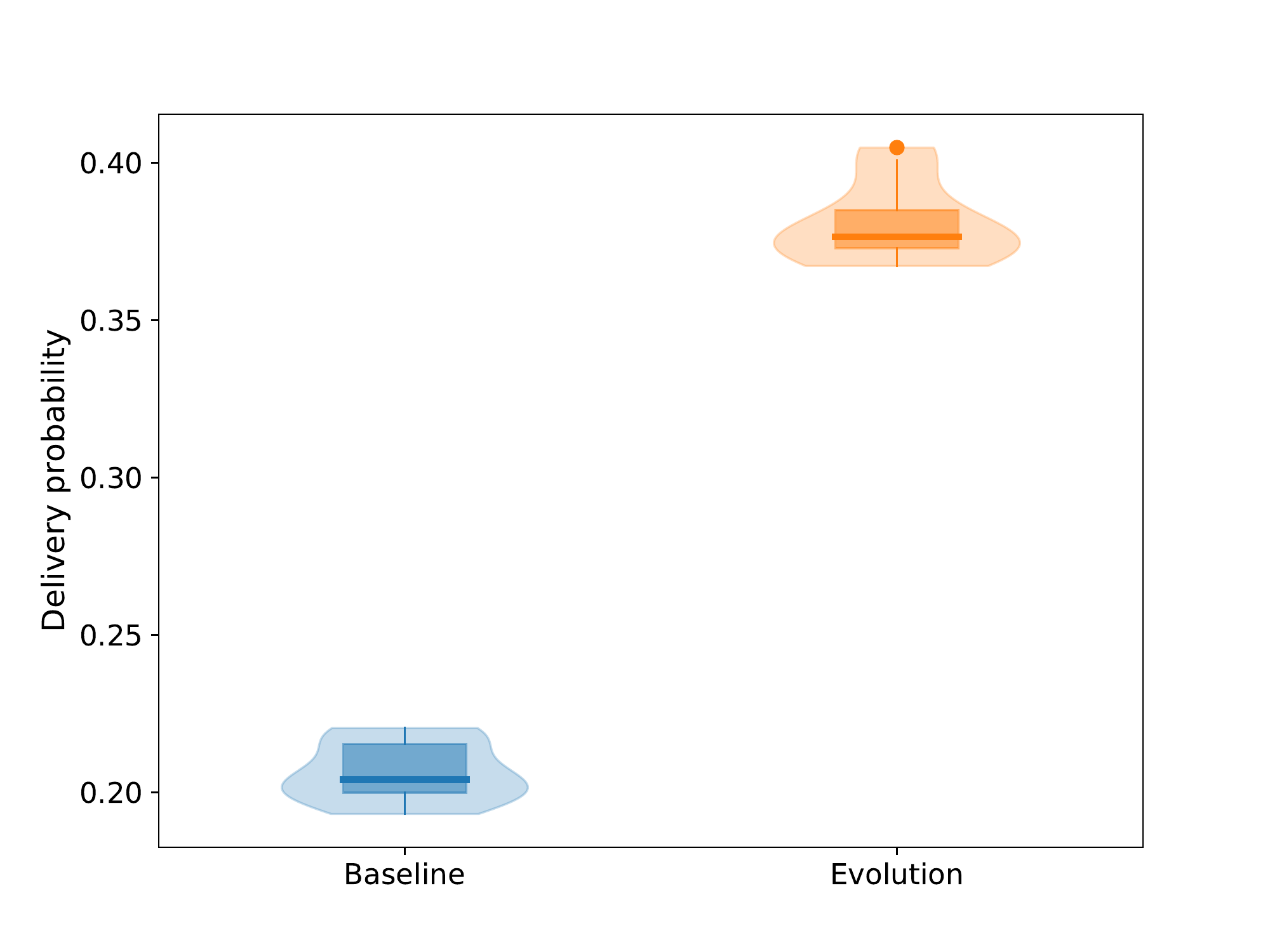}
    \caption{Default (100 hosts)}%\label{}
    \end{subfigure}
    \begin{subfigure}[b]{0.32\textwidth}
    \centering
    \includegraphics[width=\textwidth]{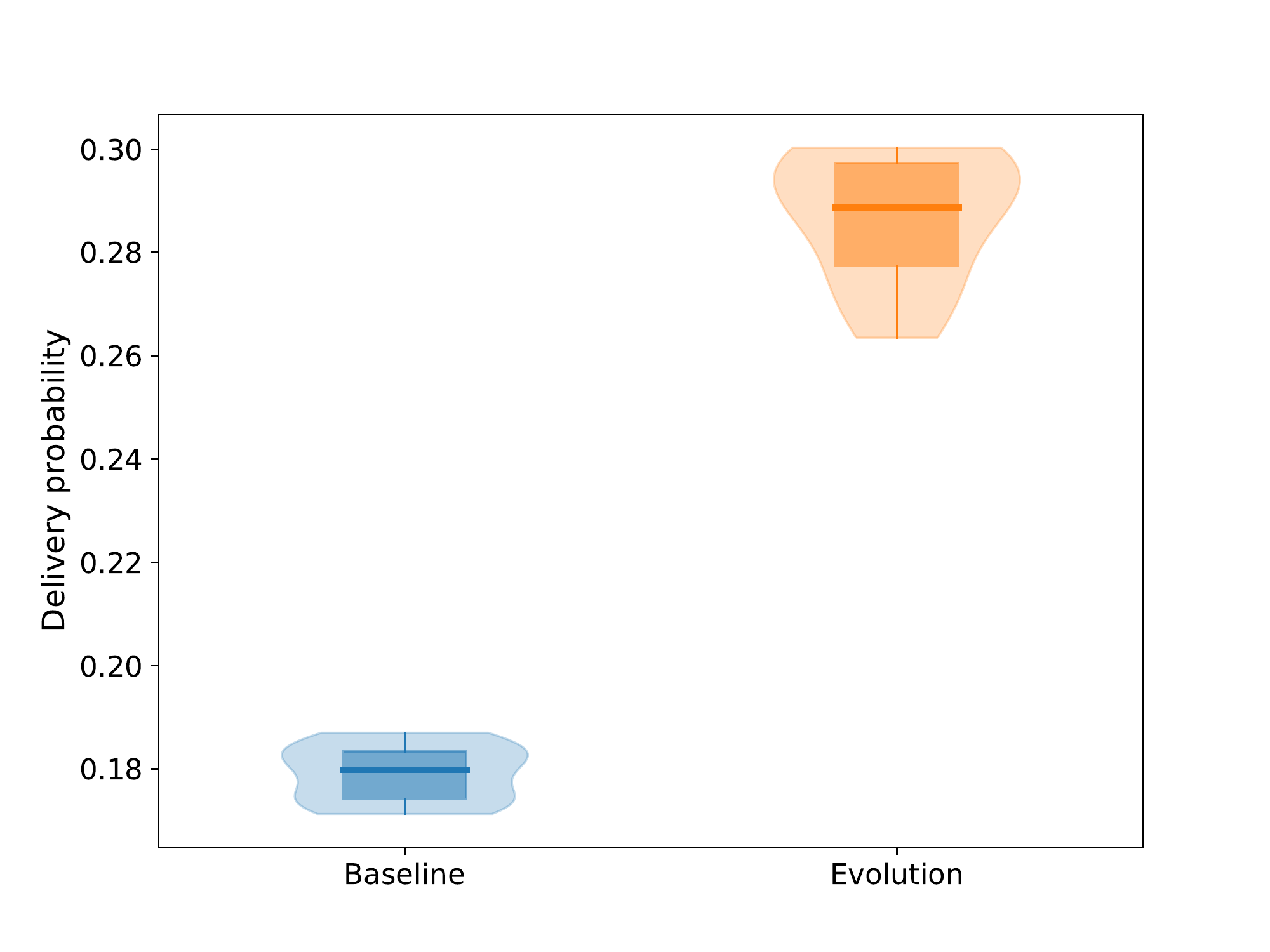}
    \caption{Helsinki (100 hosts)}%\label{}
    \end{subfigure}
    \begin{subfigure}[b]{0.32\textwidth}
    \centering
    \includegraphics[width=\textwidth]{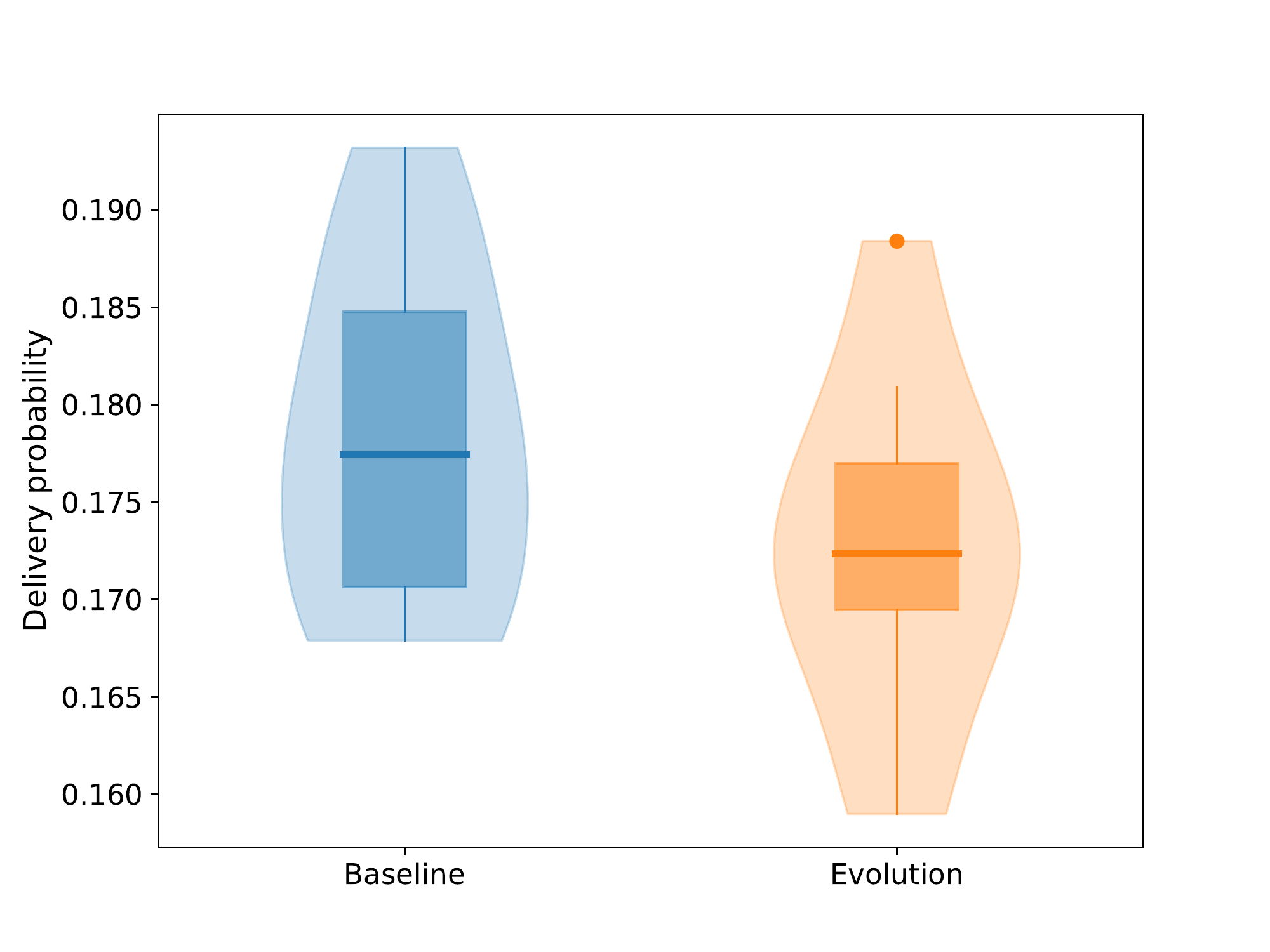}
    \caption{Manhattan (100 hosts)}%\label{}
    \end{subfigure}
    \caption{Distribution of the delivery probability obtained by the Epidemic routing protocol (\emph{Baseline}) and the best evolved protocols (\emph{Evolution}), values from \NUMRUNS~simulations.}
    \label{fig:epidemic_boxplots}
\end{figure}

\begin{figure}[ht!]
    \begin{subfigure}[b]{0.32\textwidth}
    \centering
    \includegraphics[width=\textwidth]{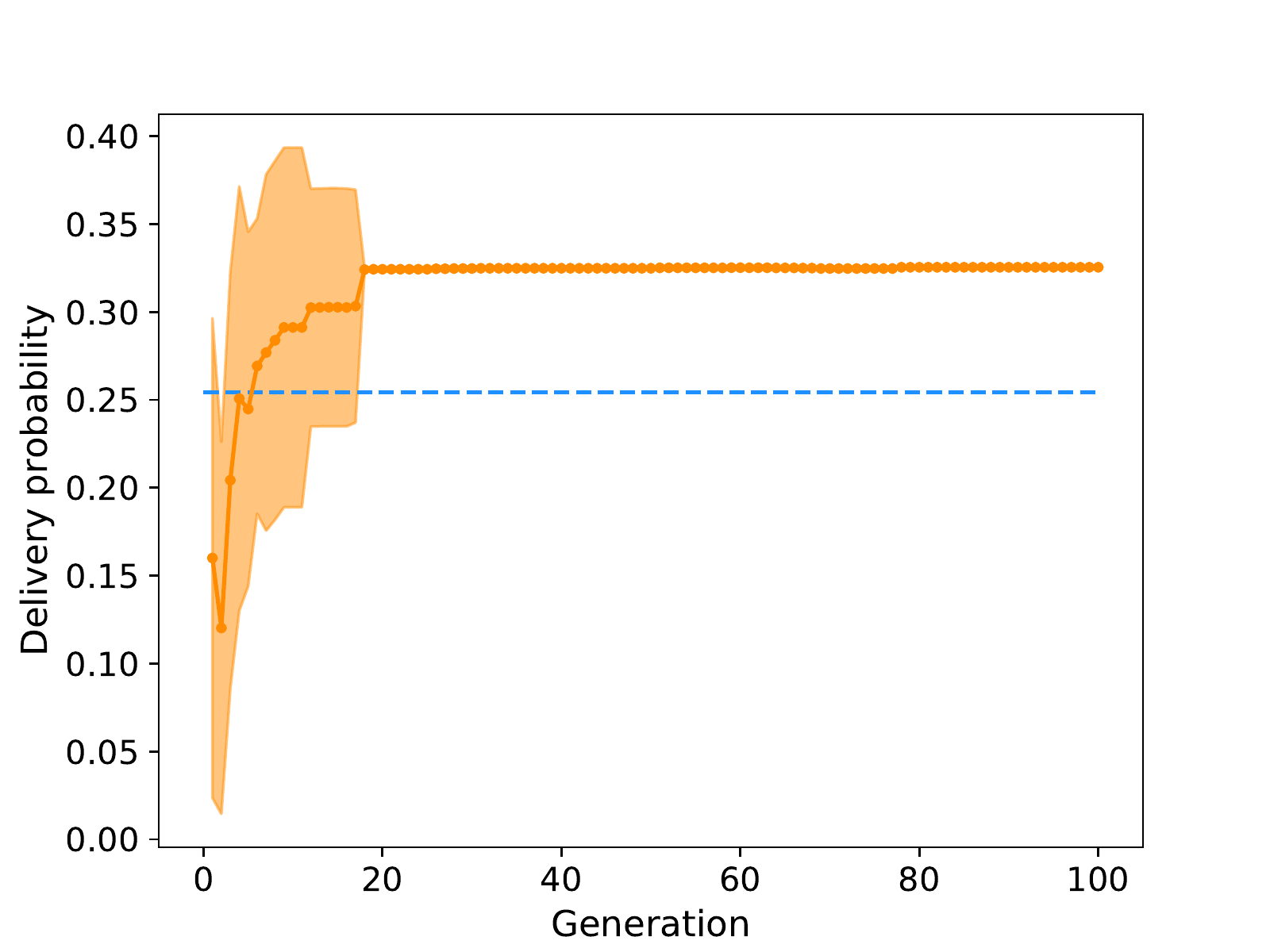}
    \caption{Default (40 hosts)}%\label{}
    \end{subfigure}
    \begin{subfigure}[b]{0.32\textwidth}
    \centering
    \includegraphics[width=\textwidth]{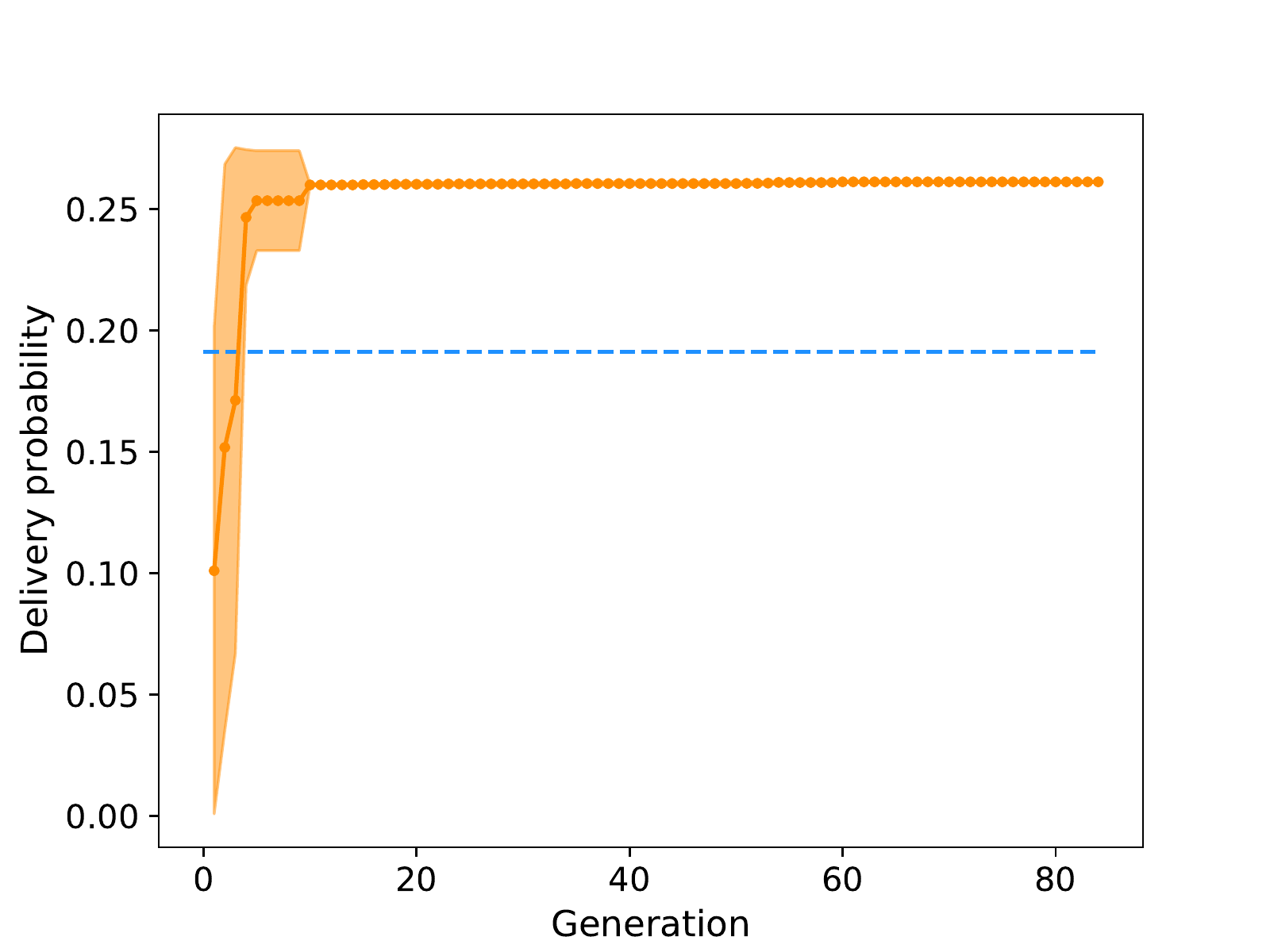}
    \caption{Helsinki (40 hosts)}%\label{}
    \end{subfigure}
    \begin{subfigure}[b]{0.32\textwidth}
    \centering
    \includegraphics[width=\textwidth]{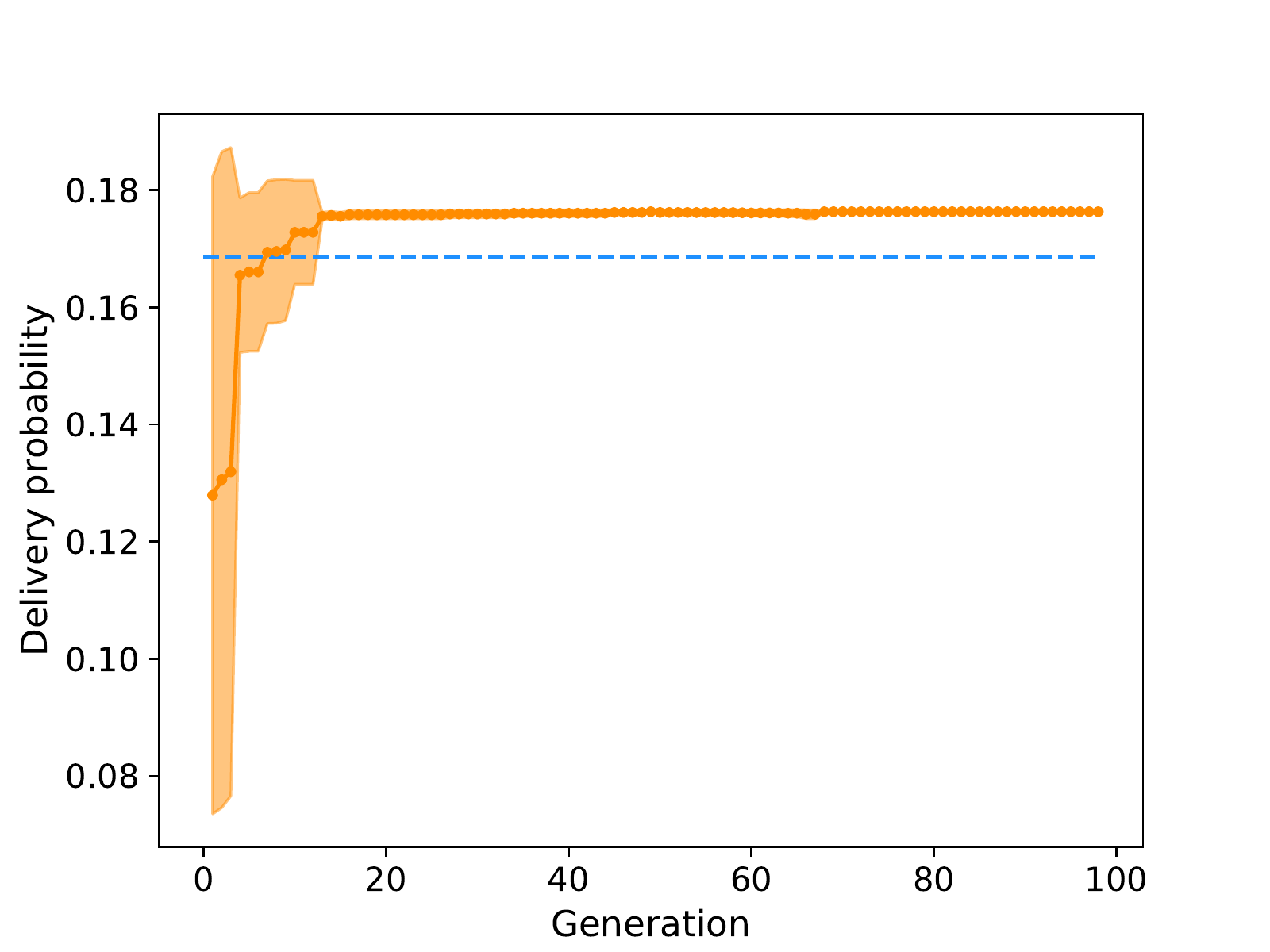}
    \caption{Manhattan (40 hosts)}%\label{}
    \end{subfigure}
    \begin{subfigure}[b]{0.32\textwidth}
    \centering
    \includegraphics[width=\textwidth]{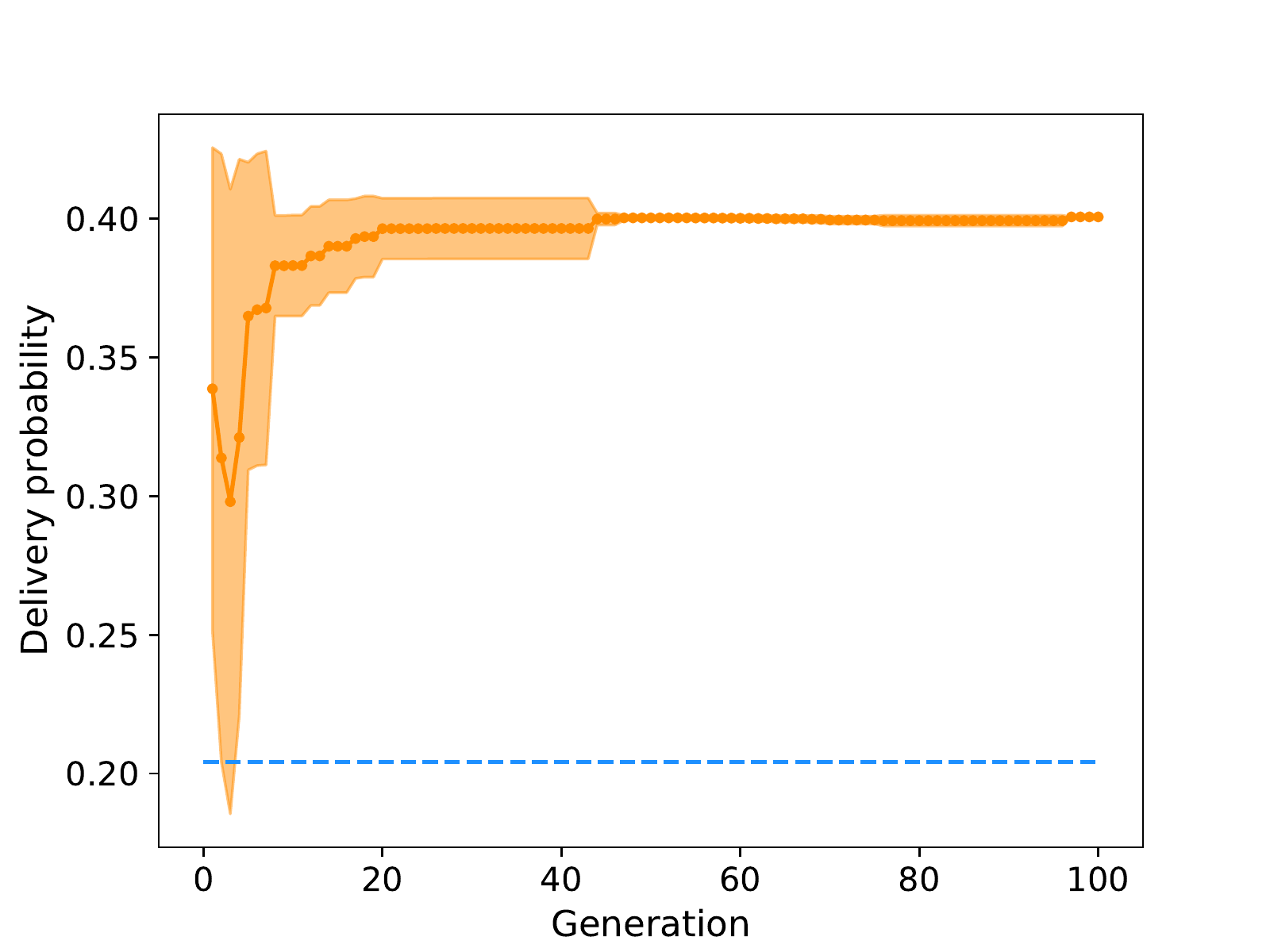}
    \caption{Default (100 hosts)}%\label{}
    \end{subfigure}
    \begin{subfigure}[b]{0.32\textwidth}
    \centering
    \includegraphics[width=\textwidth]{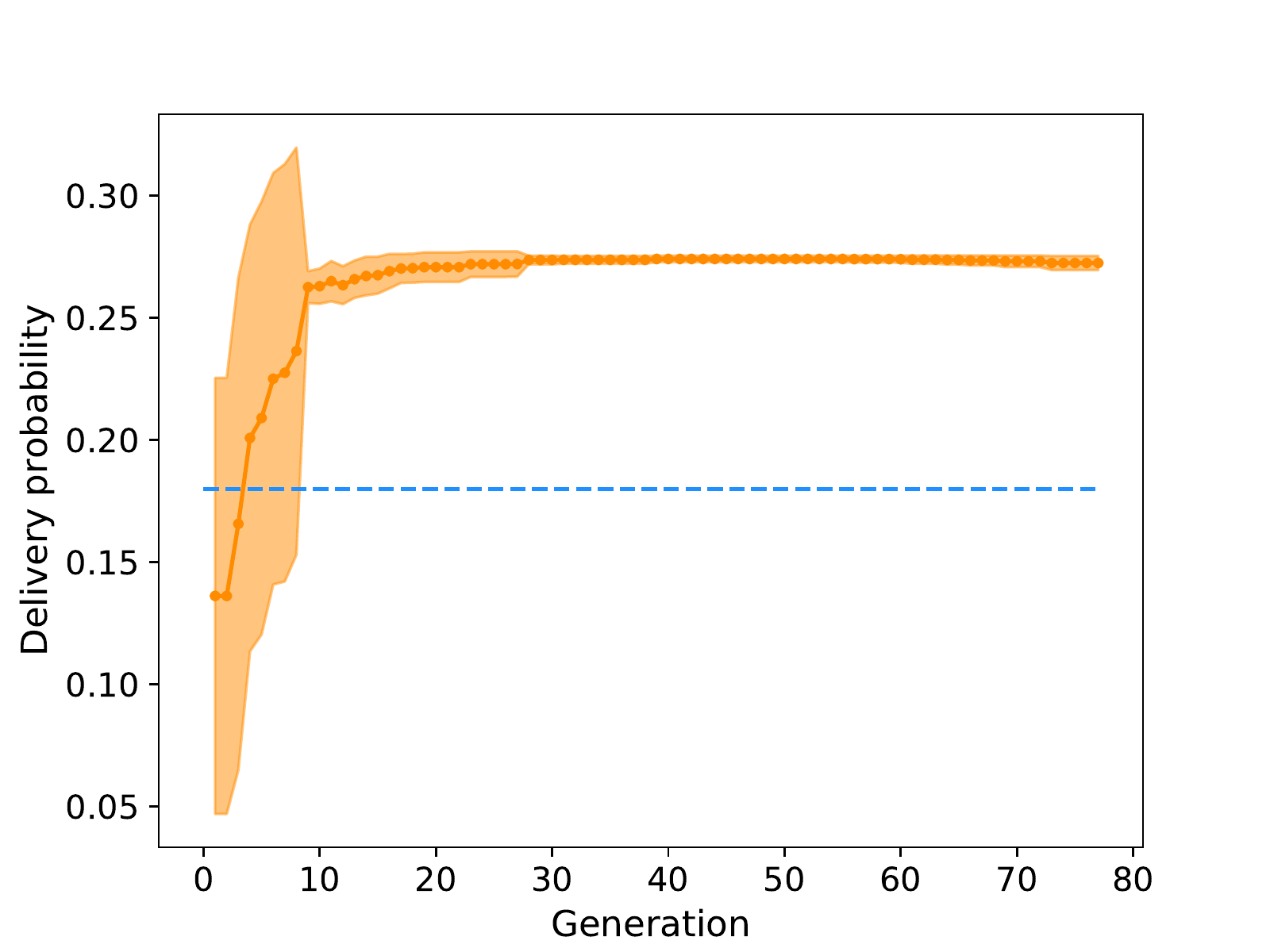}
    \caption{Helsinki (100 hosts)}%\label{}
    \end{subfigure}
    \begin{subfigure}[b]{0.32\textwidth}
    \centering
    \includegraphics[width=\textwidth]{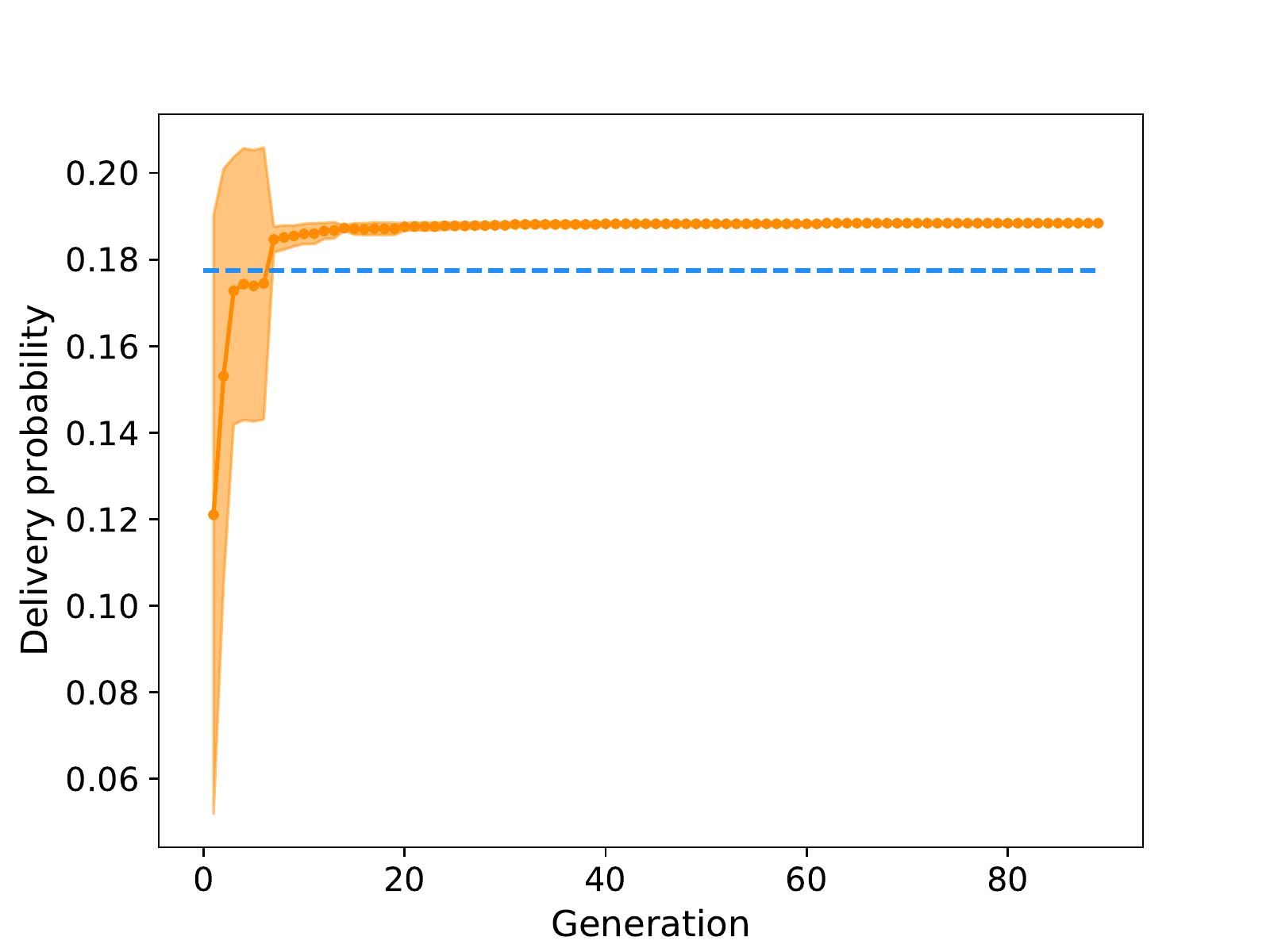}
    \caption{Manhattan (100 hosts)}%\label{}
    \end{subfigure}
    \caption{Best delivery probability at each generation (mean $\pm$ std. dev. across \NUMRUNS~runs) obtained by GP vs the Epidemic routing protocol as baseline (median across \NUMRUNS~simulations, dashed blue line). Note that the trends stop at different generations due to the steady fitness stop criterion (50 generations without improvement).}
    \label{fig:epidemic_fitness_mean}
\end{figure}

In Table \ref{tab:wilcoxon_prophet}, we report the comparative results (median across \NUMRUNS~simulations) of the delivery probability obtained by the baseline Epidemic protocol vs that obtained by the \emph{best evolved protocol} on each test case. We consider as best evolved protocol the one showing the highest delivery probability across \NUMRUNS~runs of GP, and the lowest number of nodes in case of equal delivery probability. For each pairwise comparison, we report also the p-value of the Wilcoxon rank-sum test ($N=\NUMRUNS, \alpha=0.05$). From the table, it can can observed that GP is able to obtain statistically significant improvements of delivery probability (p-value $\leq\alpha$) in the Default and Helsinki cases. On the other hand, in the Manhattan test cases the Null Hypothesis on the statistical equivalence between the delivery probability of the baseline protocol and that of the best evolved one cannot be rejected (p-value $>\alpha$). For the sake of completeness, the delivery probability distribution across \NUMRUNS~simulations of the baseline Epidemic protocol and the best evolved ones are shown, in the form of violin plots, in Figure \ref{fig:epidemic_boxplots}. Finally, the analysis of the fitness trends shown in Figure \ref{fig:epidemic_fitness_mean} (mean $\pm$ std. dev. of the best delivery probability found at each generation across \NUMRUNS~runs of GP) reveals that in 5 out of 6 test cases the initial GP population shows an average delivery probability lower than the corresponding baseline (median across \NUMRUNS~simulations, shown as a dashed blue line). In the remaining test case, i.e, the Default map with 100 hosts per group, the initial GP population is even better than the baseline. In all cases, the average delivery probability quickly increases during the evolutionary process, to stabilize after 20-40 generations on average. It can also be noted that the GP algorithm is quite robust, since the std. dev. across runs (indicated by the shaded area) decreases over time, reaching an almost-zero value towards the end of the available budget.

\begin{figure}[ht!]
    \begin{subfigure}[b]{0.32\textwidth}
    \centering
    \includegraphics[width=\textwidth]{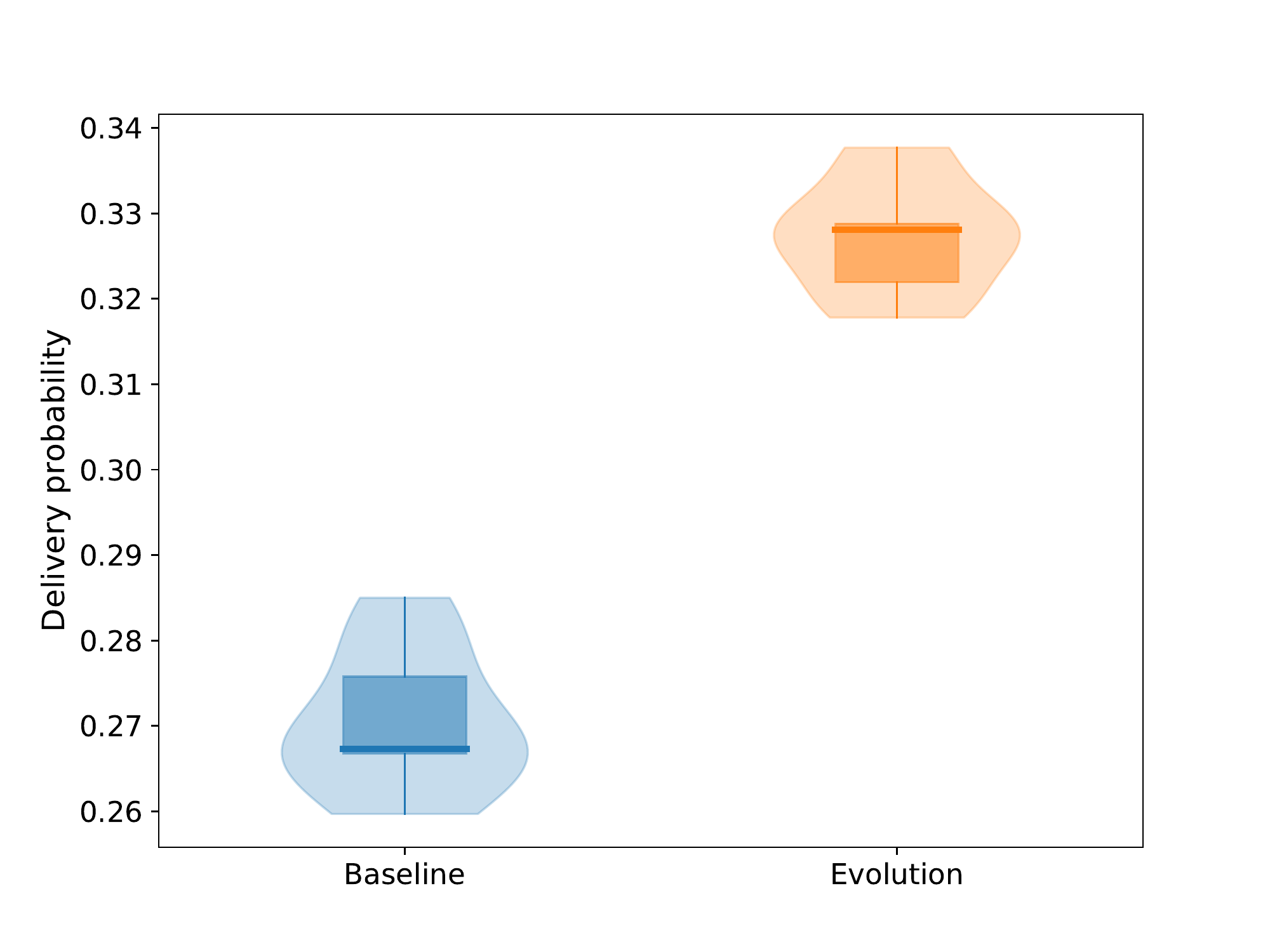}
    \caption{Default (40 hosts)}%\label{}
    \end{subfigure}
    \begin{subfigure}[b]{0.32\textwidth}
    \centering
    \includegraphics[width=\textwidth]{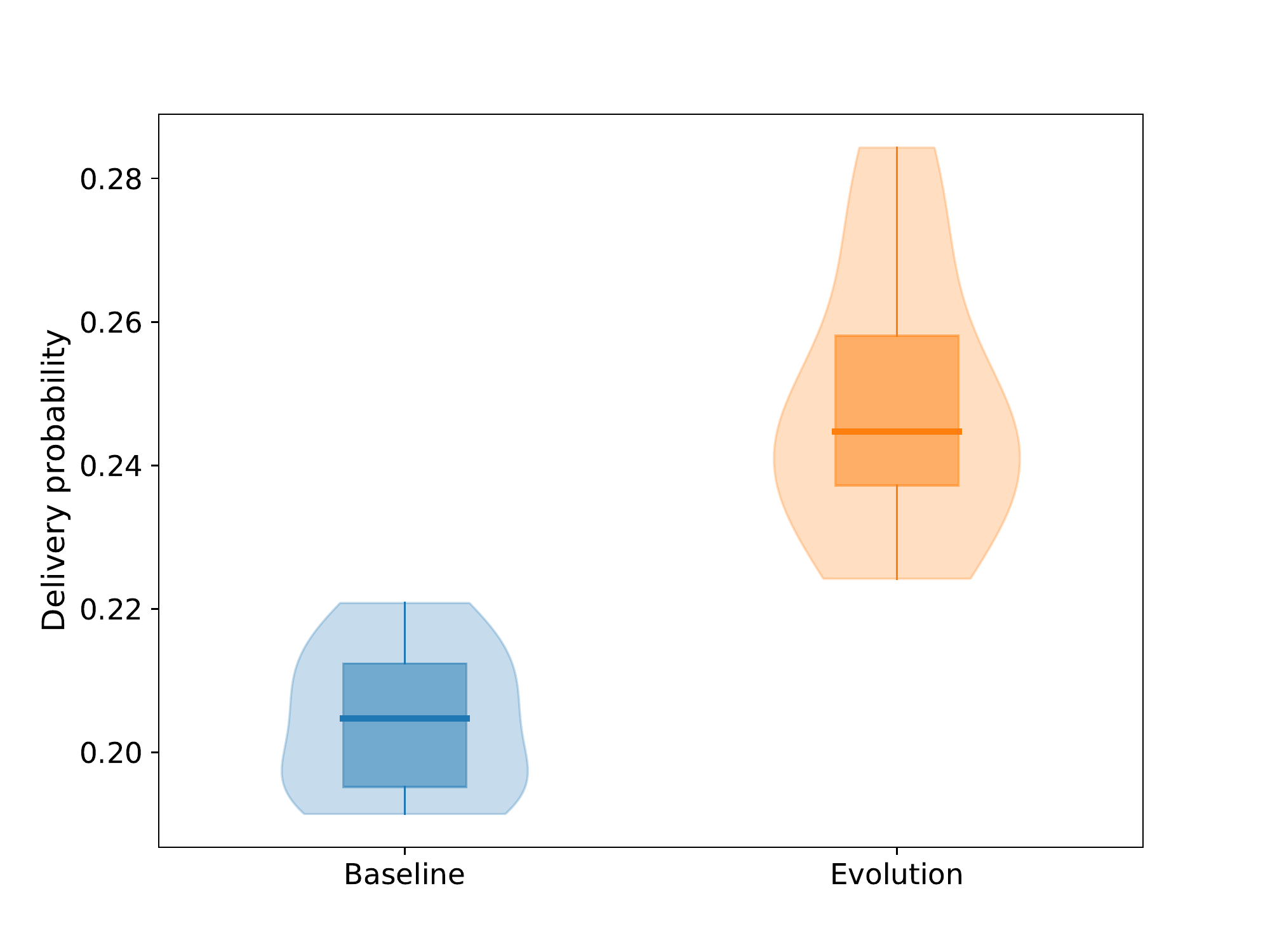}
    \caption{Helsinki (40 hosts)}%\label{}
    \end{subfigure}
    \begin{subfigure}[b]{0.32\textwidth}
    \centering
    \includegraphics[width=\textwidth]{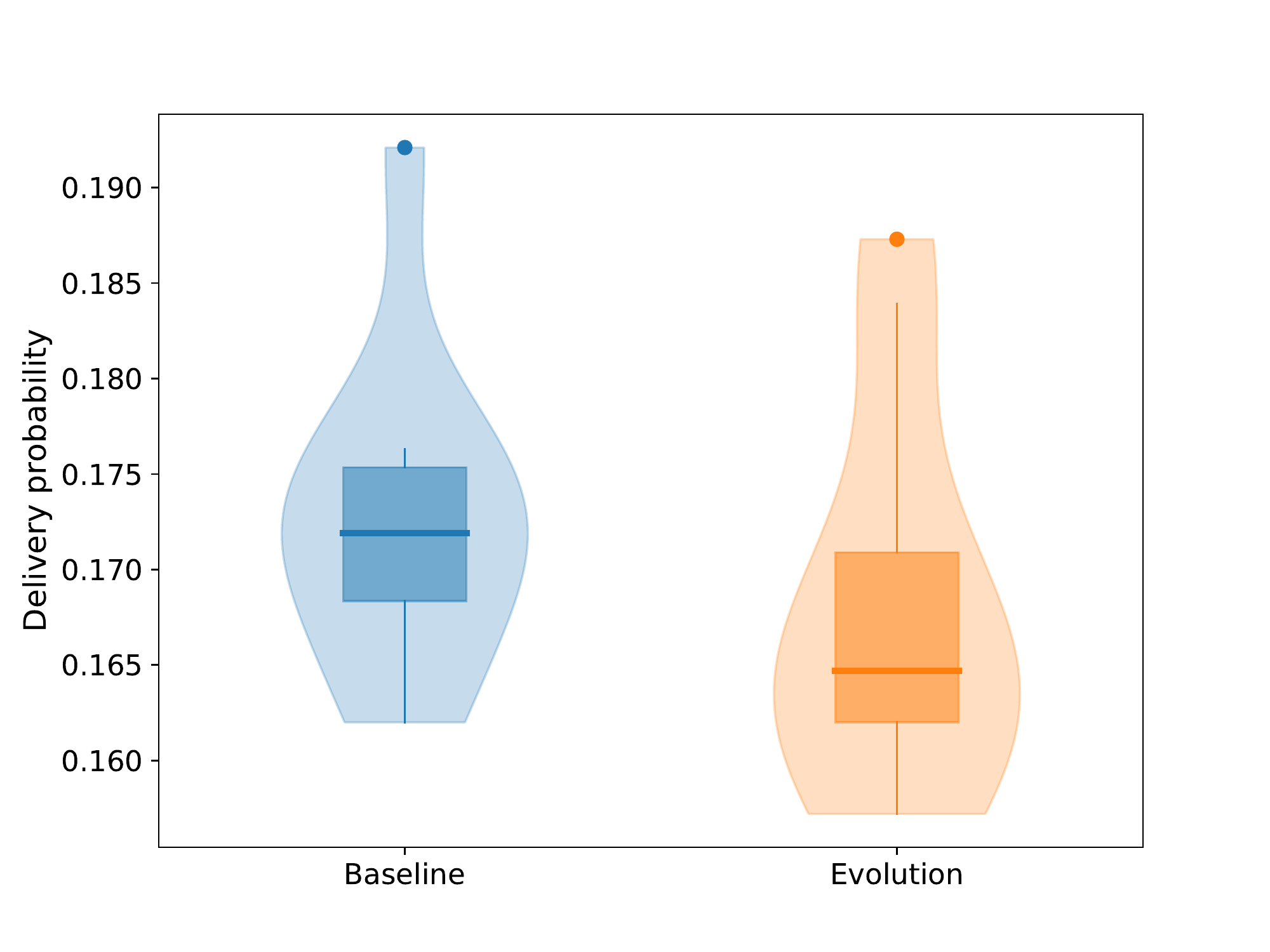}
    \caption{Manhattan (40 hosts)}%\label{}
    \end{subfigure}
    \begin{subfigure}[b]{0.32\textwidth}
    \centering
    \includegraphics[width=\textwidth]{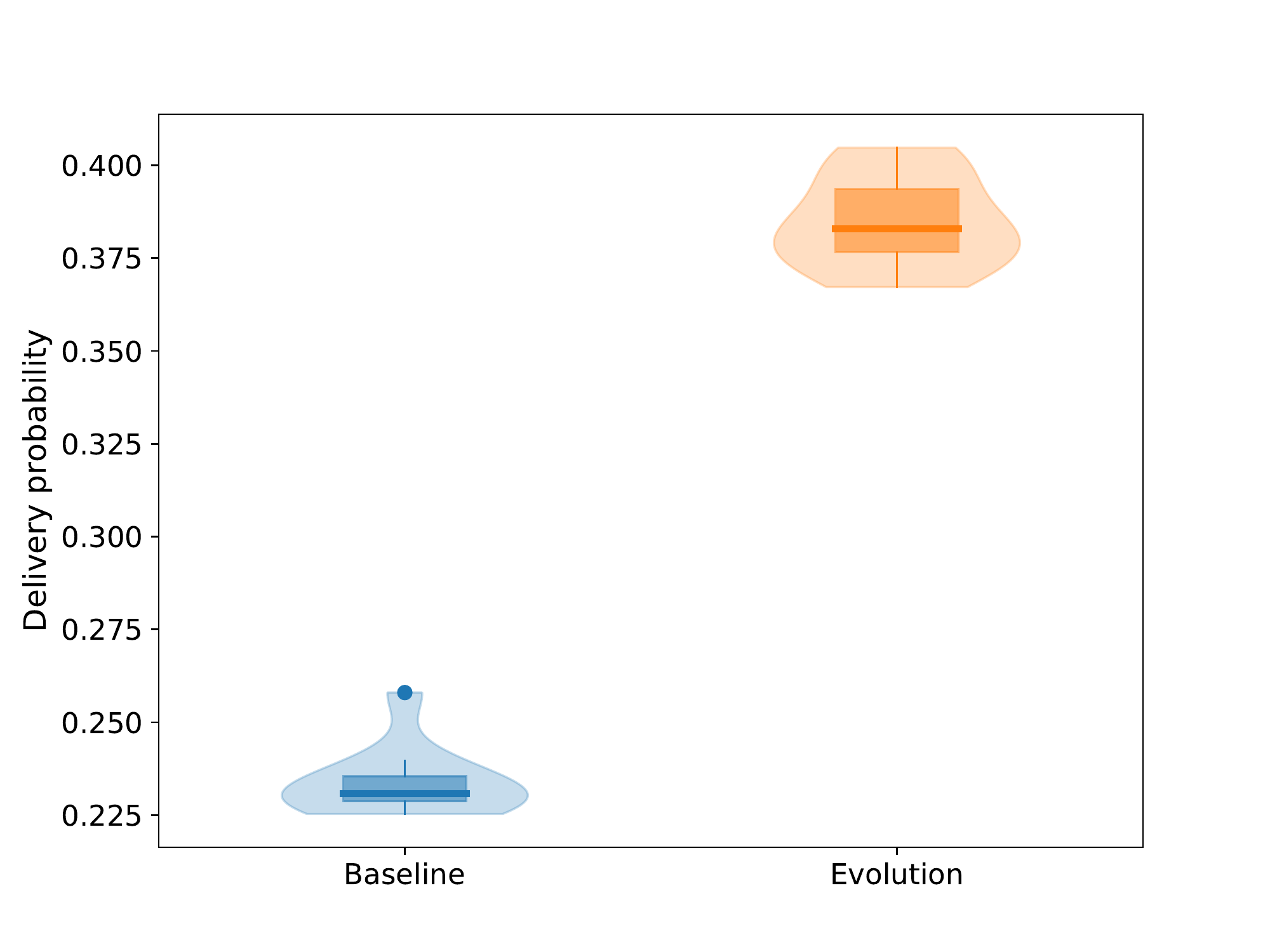}
    \caption{Default (100 hosts)}%\label{}
    \end{subfigure}
    \begin{subfigure}[b]{0.32\textwidth}
    \centering
    \includegraphics[width=\textwidth]{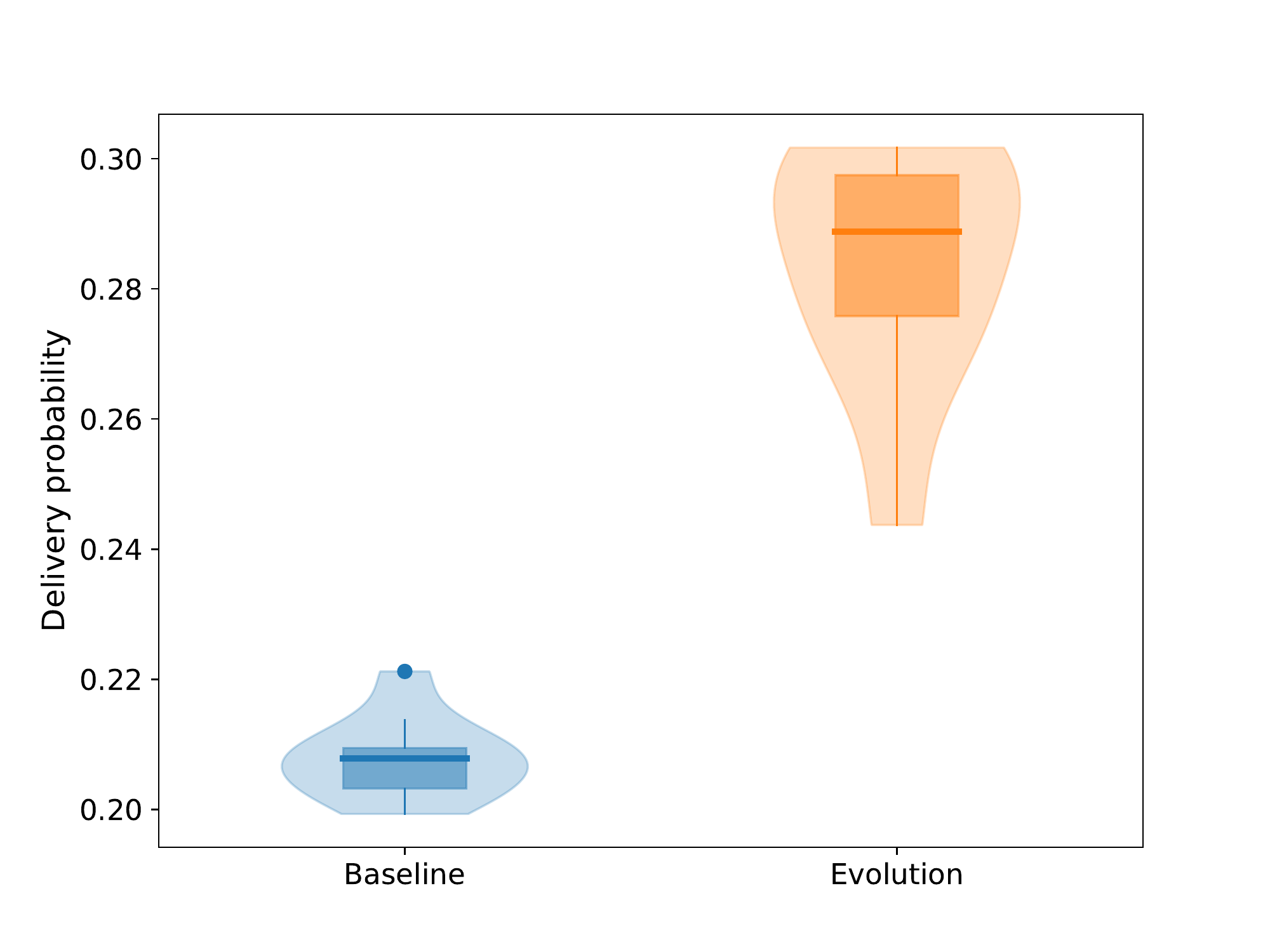}
    \caption{Helsinki (100 hosts)}%\label{}
    \end{subfigure}
    \begin{subfigure}[b]{0.32\textwidth}
    \centering
    \includegraphics[width=\textwidth]{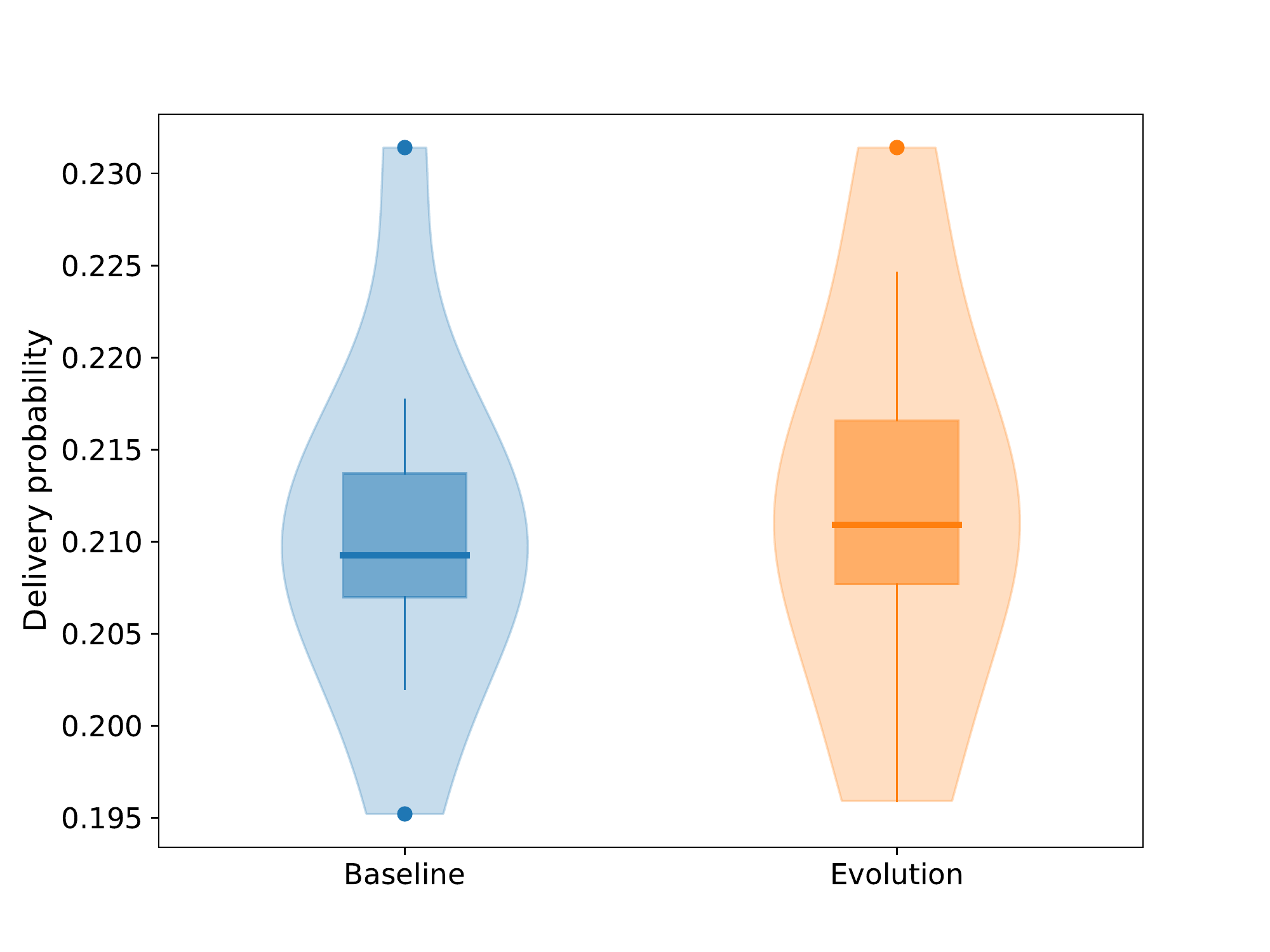}
    \caption{Manhattan (100 hosts)}%\label{}
    \end{subfigure}
    \caption{Distribution of the delivery probability obtained by the PRoPHET routing protocol (\emph{Baseline}) and the best evolved protocols (\emph{Evolution}), values from \NUMRUNS~simulations.}
    \label{fig:prophet_boxplots}
\end{figure}

\begin{figure}[ht!]
    \begin{subfigure}[b]{0.32\textwidth}
    \centering
    \includegraphics[width=\textwidth]{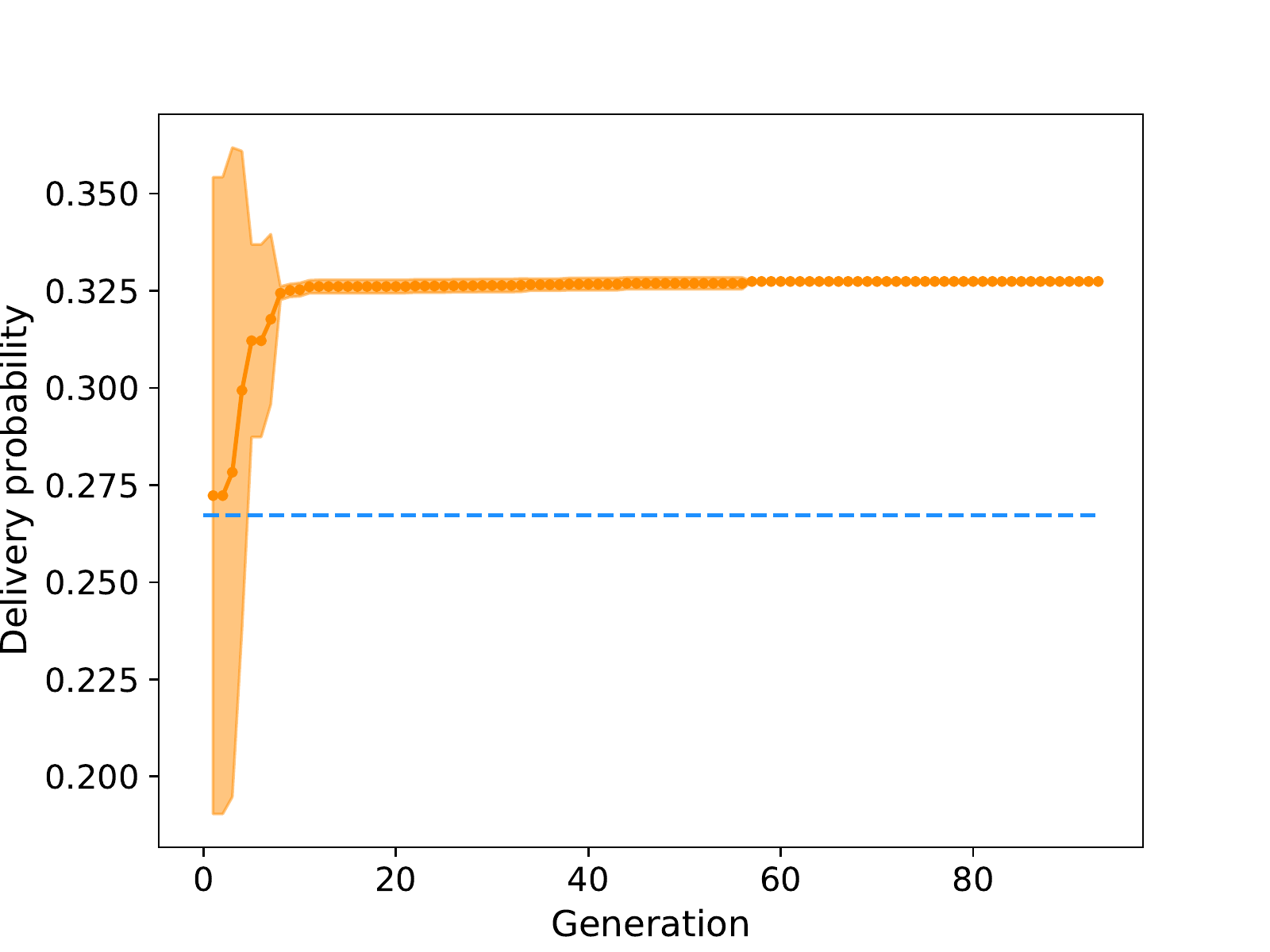}
    \caption{Default (40 hosts)}%\label{}
    \end{subfigure}
    \begin{subfigure}[b]{0.32\textwidth}
    \centering
    \includegraphics[width=\textwidth]{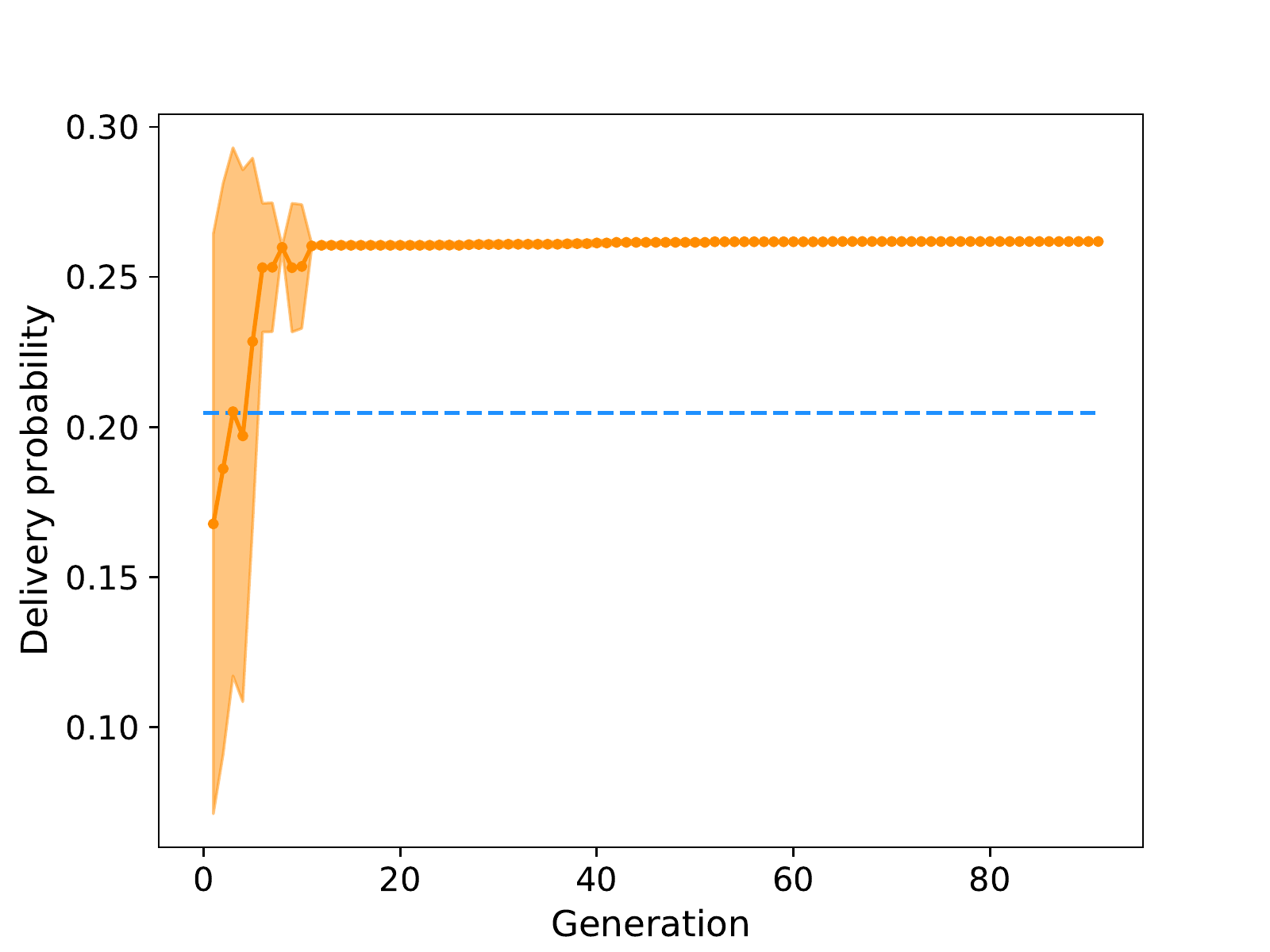}
    \caption{Helsinki (40 hosts)}%\label{}
    \end{subfigure}
    \begin{subfigure}[b]{0.32\textwidth}
    \centering
    \includegraphics[width=\textwidth]{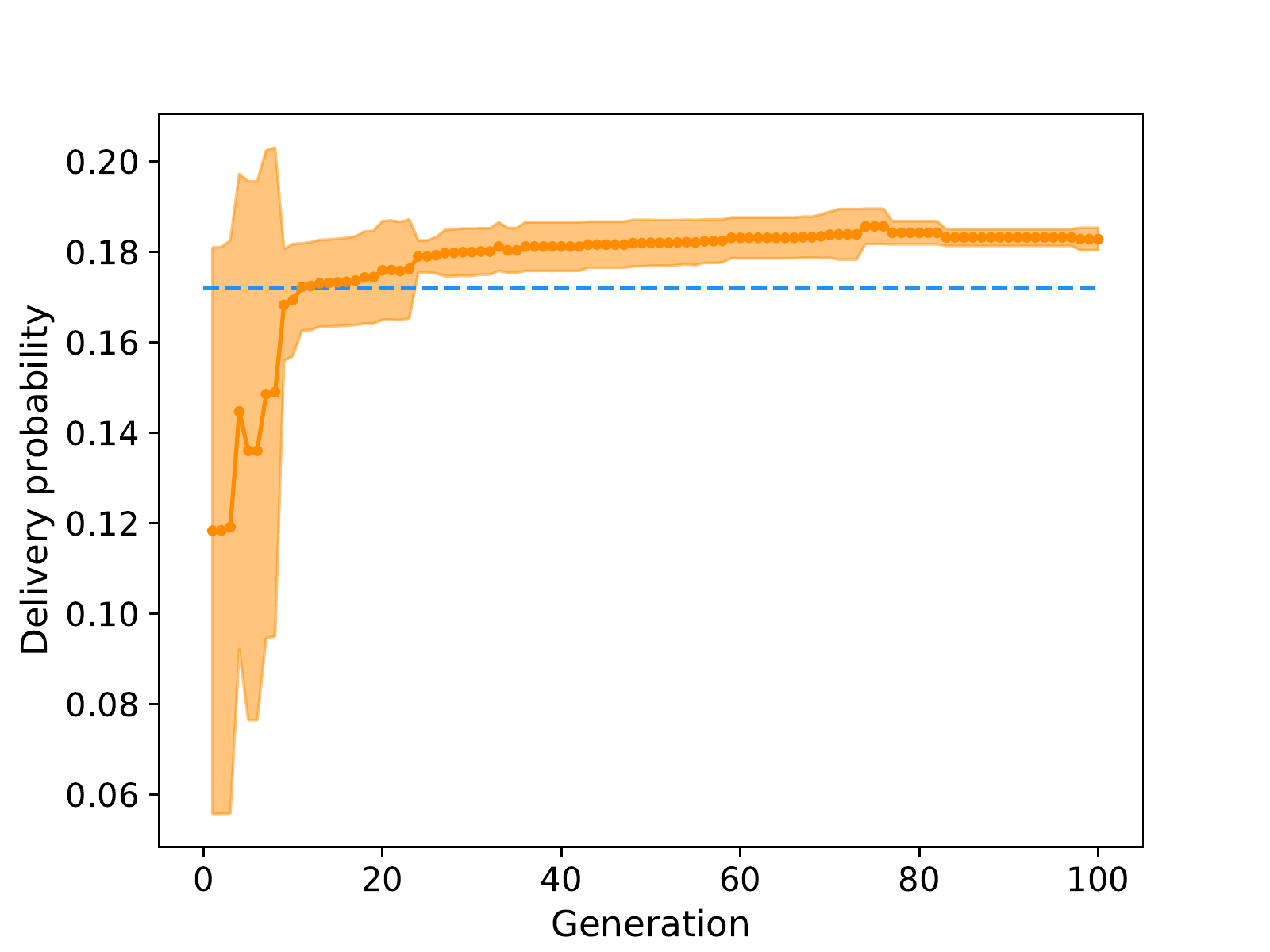}
    \caption{Manhattan (40 hosts)}%\label{}
    \end{subfigure}
    \begin{subfigure}[b]{0.32\textwidth}
    \centering
    \includegraphics[width=\textwidth]{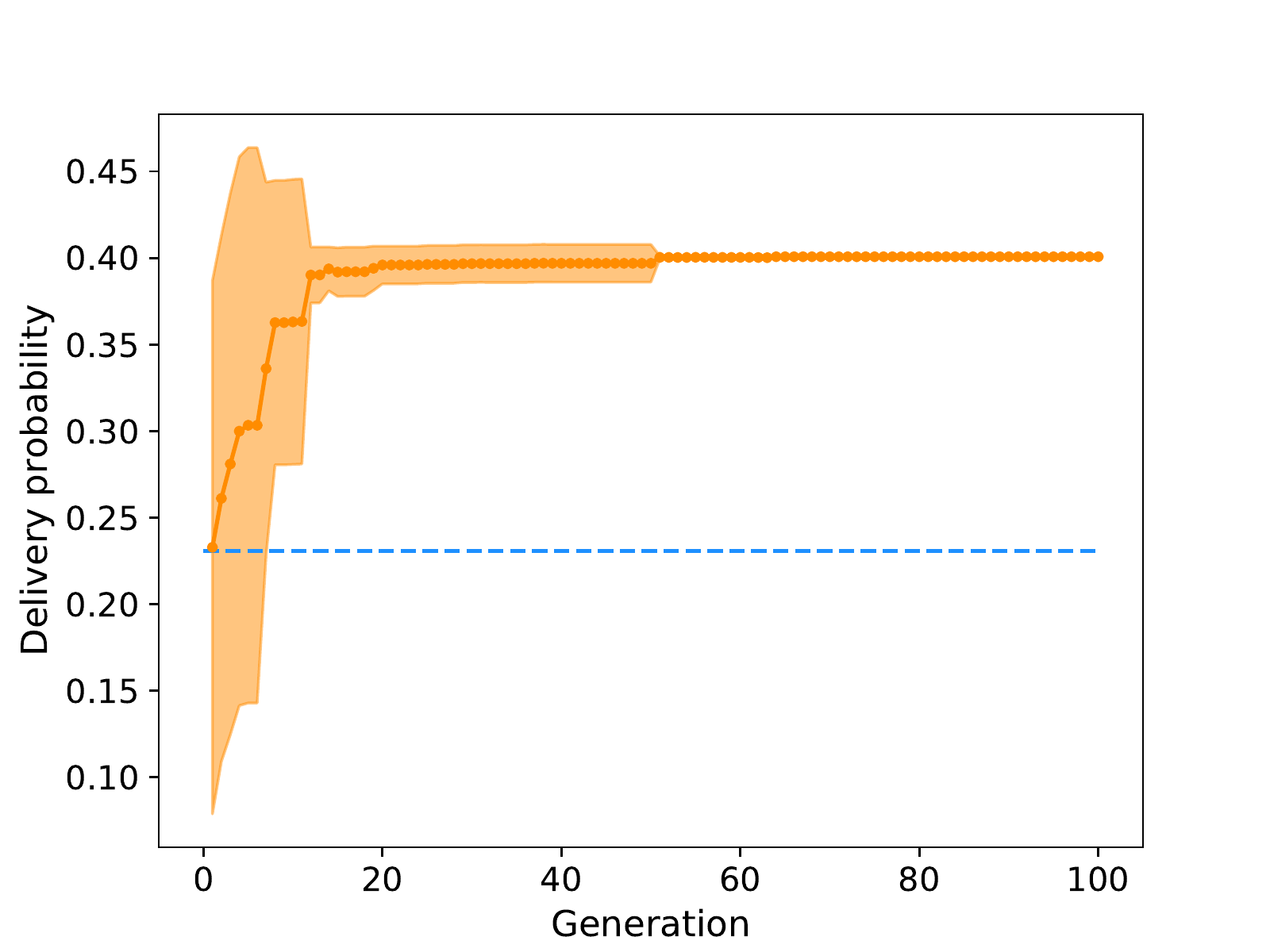}
    \caption{Default (100 hosts)}%\label{}
    \end{subfigure}
    \begin{subfigure}[b]{0.32\textwidth}
    \centering
    \includegraphics[width=\textwidth]{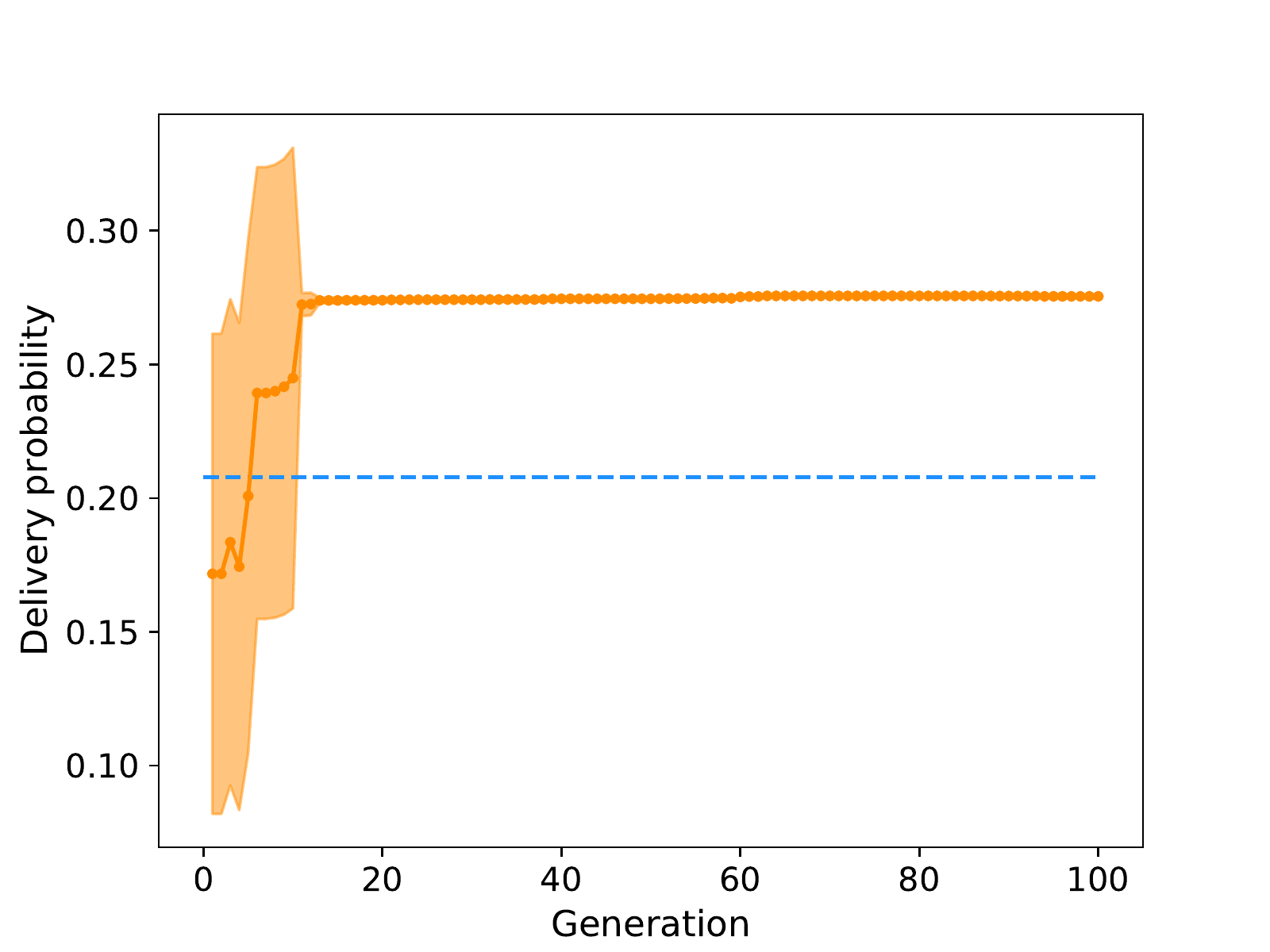}
    \caption{Helsinki (100 hosts)}%\label{}
    \end{subfigure}
    \begin{subfigure}[b]{0.32\textwidth}
    \centering
    \includegraphics[width=\textwidth]{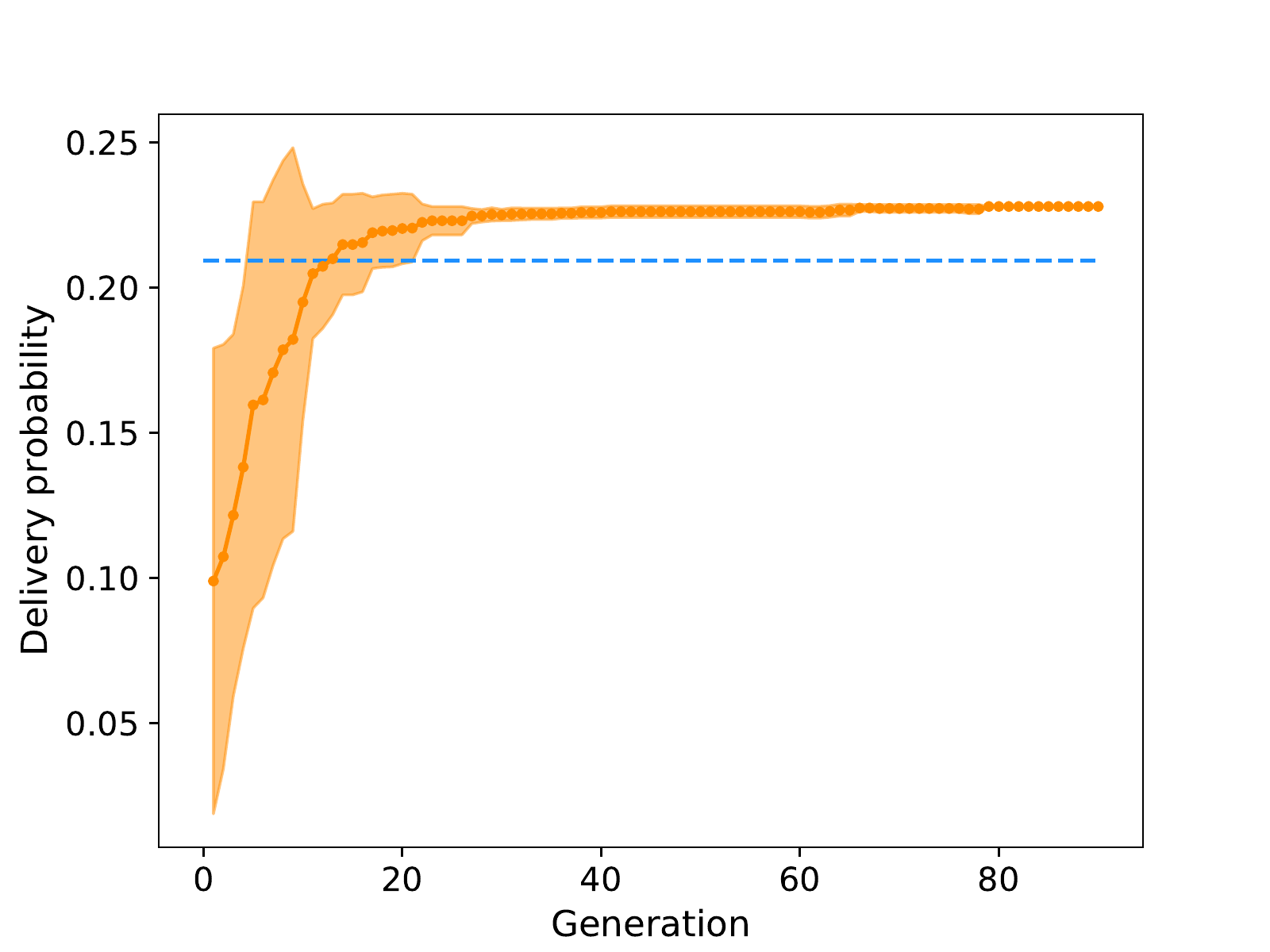}
    \caption{Manhattan (100 hosts)}%\label{}
    \end{subfigure}
    \caption{Best delivery probability at each generation (mean $\pm$ std. dev. across \NUMRUNS~runs) obtained by GP vs the PRoPHET routing protocol as baseline (median across \NUMRUNS~simulations, dashed blue line). Note that the trends stop at different generations due to the steady fitness stop criterion (50 generations without improvement).}
    \label{fig:prophet_fitness_mean}
\end{figure}

%The only difference w.r.t. Epidemic is that in this case there is an additional element in the terminal set, \texttt{tryOtherMessages}, as shown in Table \ref{tab:terminals}. This element is used in the best evolved protocols for all the PRoPHET test cases, except the Default map with 100 hosts per group, see Figures \ref{fig:gen_tree_prop_def_40}-\ref{fig:gen_tree_prop_man_100} reported in Appendix \ref{sec:appendix_best_trees}.
The same analysis has been performed comparing the PRoPHET routing protocol, as baseline, and the best evolved protocol for each test case, see Table \ref{tab:wilcoxon_prophet}. The corresponding violin plots and fitness trends are shown in Figure \ref{fig:prophet_boxplots} and \ref{fig:prophet_fitness_mean} respectively. The results reveal that also in this case GP is able to obtain statistically significant improvements of delivery probability (p-value $\leq\alpha$) in all cases except the two Manhattan test cases. As for the fitness trends, it can be noted that in the two Default test cases the average delivery probability of the initial GP population is approximately equal to that of the baseline, while in the remaining cases it is quite lower.
\vspace{-0.3cm}

\begin{table}[ht!]
\begin{minipage}{0.49\textwidth}
\caption{Delivery probability (median across \NUMRUNS~simulations) and p-value of the Wilcoxon rank-sum test ($N=\NUMRUNS, \alpha=0.05$) of the Epidemic routing protocol vs the corresponding best evolved protocol. The evolved protocols perform statistically better in the Default and Helsinki test cases.}
\centering
\label{tab:wilcoxon_epidemic}
\resizebox{\textwidth}{!}{
\begin{tabular}{lccc}
 \toprule
 \multirow{2}{*}{\textbf{Test case}} & \multicolumn{2}{c}{\textbf{Epidemic}} &  \multirow{2}{*}{\textbf{GP}}\\
 \cmidrule(l{2pt}r{2pt}){2-3}
 & \textbf{Deliv. prob.} & \textbf{p-value} & \\
 \midrule
 Default (40 hosts) & 0.2542 & 0.005 & 0.3342 \\
 Default (100 hosts) & 0.2041 & 0.005 & 0.3764 \\
 Helsinki (40 hosts) & 0.1910 & 0.005 & 0.2467 \\ 
 Helsinki (100 hosts) & 0.1798 & 0.005 & 0.2887 \\
 Manhattan (40 hosts) & 0.1685 & 0.574 & 0.1654 \\
 Manhattan (100 hosts) & 0.1774 & 0.139 & 0.1664 \\
 \bottomrule
\end{tabular}
}
\end{minipage}
\hfill
\begin{minipage}{0.49\textwidth}
\caption{Delivery probability (median across \NUMRUNS~simulations) and p-value of the Wilcoxon rank-sum test ($N=\NUMRUNS, \alpha=0.05$) of the PRoPHET routing protocol vs the corresponding best evolved protocol. The evolved protocols perform statistically better in the Default and Helsinki test cases.}
\centering
\label{tab:wilcoxon_prophet}
\resizebox{\textwidth}{!}{
\begin{tabular}{lccc}
 \toprule
 \multirow{2}{*}{\textbf{Test case}} & \multicolumn{2}{c}{\textbf{PRoPHET}} &  \multirow{2}{*}{\textbf{GP}}\\
 \cmidrule(l{2pt}r{2pt}){2-3}
 & \textbf{Deliv. prob.} & \textbf{p-value} & \\
 \midrule
 Default (40 hosts) & 0.2673 & 0.005 & 0.3281 \\
 Default (100 hosts) & 0.2307 & 0.005 & 0.3829 \\
 Helsinki (40 hosts) & 0.2047 & 0.005 & 0.2447 \\ 
 Helsinki (100 hosts) & 0.2078 & 0.005 & 0.2887 \\
 Manhattan (40 hosts) & 0.1719 & 0.333 & 0.1647 \\
 Manhattan (100 hosts) & 0.2092 & 0.646 & 0.2109 \\
 \bottomrule
\end{tabular}
}
\end{minipage}
\end{table}

\subsubsection*{Comparison vs other PRoPHET variants}

To further validate our results, we compared the best evolved protocols in the case of PRoPHET against three recent PRoPHET variants, namely:
\begin{itemize}[leftmargin=*]
    \item \textbf{PRoPHET+}, proposed in \cite{huang2010prophetplus}. It extends PRoPHET by adding a \textit{deliverability} measure, used to forward messages based on buffer size and availability, energy consumption, bandwidth, popularity and predictability.
    \item \textbf{PRoPHETv2}, proposed in \cite{prophetrtf2011}. It introduces a dependency (for the delivery predictability) based on the predictability of the encounters of any two nodes. This way, the risk of increasing the delivery predictability of a node due to repeated connections in a short interval of time is reduced.
    \item \textbf{Evict PRoPHET}, proposed in \cite{sati2020efficient}. In this variant, an efficient eviction policy is introduced to decide which message should be removed from the buffer. The policy makes use of an utility function that accounts for time-to-live (elapsed and remaining), hop count, time spent into the buffer, and re-transmissions.
\end{itemize}

The comparative analysis, reported in Table~\ref{tab:res_prophet_variants} (median across \NUMRUNS~simulations), shows that the best evolved protocols statistically outperform (p-value $\leq\alpha$) the delivery probability obtained by Evict PRoPHET and PRoPHET+ in all the Default and Helsinki test cases, while the three protocols result equivalent in the Manhattan cases. Compared to PRoPHETv2, the best evolved protocol results statistically superior in the Default and Manhattan test cases with 100 hosts, and equivalent in the others. Overall, this comparison confirms the effectiveness of the proposed technique at finding efficient protocols which are at least comparable to, or even better than, the state-of-the-art protocol implementations.

%----------------------------------------------------------------

\clearpage

%----------------------------------------------------------------

\begin{table}[ht!]
\centering
\caption{Delivery probability (median across \NUMRUNS~simulations) and p-value of the Wilcoxon rank-sum test ($N=\NUMRUNS, \alpha=0.05$) of the PRoPHET variants vs the corresponding best evolved protocol. The evolved protocols perform statistically better than Evict PRoPHET and PRoPHET+ in the Default and Helsinki test cases, and better than PRoPHETv2 in the Default and Manhattan test cases with 100 hosts.}
\label{tab:res_prophet_variants}
\resizebox{\textwidth}{!}{
\begin{tabular}{lccccccccc}
 \toprule
 \multirow{2}{*}{\textbf{Test case}} & \multicolumn{2}{c}{\textbf{PRoPHETv2}} & 
 \multicolumn{2}{c}{\textbf{Evict PRoPHET}} & 
 \multicolumn{2}{c}{\textbf{PRoPHET+}} & \multirow{2}{*}{\textbf{GP}}\\
 \cmidrule(l{2pt}r{2pt}){2-3}
 \cmidrule(l{2pt}r{2pt}){4-5}
 \cmidrule(l{2pt}r{2pt}){6-7}
 & \textbf{Deliv. prob.} & \textbf{p-value} & 
 \textbf{Deliv. prob.} & \textbf{p-value} & 
 \textbf{Deliv. prob.} & \textbf{p-value} & \\
 \midrule
 Default (40 hosts) & 0.3185 & 0.028 & 0.2710 & 0.005 & 0.2724 & 0.005 & 0.3281 \\
 Default (100 hosts) & 0.3201 & 0.005 & 0.2328 & 0.005 & 0.2348 & 0.005 & 0.3829 \\
 Helsinki (40 hosts) & 0.2320 & 0.153 & 0.2047 & 0.005 & 0.2016 & 0.005 & 0.2447 \\ 
 Helsinki (100 hosts) & 0.2546 & 0.007 & 0.2112 & 0.005 & 0.2116 & 0.005 & 0.2887 \\
 Manhattan (40 hosts) & 0.1822 & 0.059 & 0.1743 & 0.093 & 0.1709 & 0.284 & 0.1647 \\
 Manhattan (100 hosts) & 0.2419 & 0.005 & 0.2071 & 0.507 & 0.2044 & 0.414 & 0.2109 \\
 \bottomrule
\end{tabular}
}
\end{table}

%----------------------------------------------------------------

\subsection{Evolved vs baseline protocols: trade-off between data delivery probability and other network metrics}\label{sec:other_metrics}

After the positive assessment of the evolved protocols in terms of improved delivery probability, we have analyzed if there is a trade-off between the delivery probability and other network metrics of interest, in the attempt to understand if an improvement of the delivery probability entails a worsening of other network aspects. In particular, we took into account four different metrics provided in the \texttt{MessageStatsReport} log output file generated The ONE, defined below.

\begin{itemize}[leftmargin=*,label={},wide=0pt]
\item \noindent\textbf{Overhead ratio}. This metric is calculated according to the following formula:
\begin{equation*}
\text{Overhead ratio} = \frac{n_{relayed}-n_{delivered}}{n_{delivered}} \quad\quad \text{if}~~n_{delivered}>0
\end{equation*}
where $n_{relayed}$ indicates how many messages are transmitted over the network (including duplicates), while $n_{delivered}$ indicates how many messages reach their destination.
%the lower the better

\item\noindent\textbf{Latency (avg)}. This metric is calculated according to the following formula:
\begin{equation*}
\text{Latency (avg)} = \frac{1}{n_{delivered}}\sum_{i=1}^{n_{delivered}}\left({T_{delivery}^i-T_{creation}^i}\right)
\end{equation*}
where $T_{delivery}^i$ and $T_{creation}^i$ indicate respectively the timestep at which each delivered message reaches its destination and the timestep at which that message was created, $T_{delivery}^i > T_{creation}^i$.
%(which is randomized during the simulation)
%the lower the better

\item\noindent\textbf{Hop count (avg)}. This metric is calculated according to the following formula:
\begin{equation*}
\text{Hop count (avg)} = \frac{1}{n_{delivered}}\sum_{i=1}^{n_{delivered}}\left({HC^i}\right)
\end{equation*}
where $HC^i$ is the hop count of the i-th message, i.e., how many nodes each delivered message passed through before reaching its destination.
%the lower the better

\item\noindent\textbf{Buffer time (avg)}. Every node in the network keeps a limited-size buffer to store the received messages waiting to be forwarded to other nodes. The different routing protocols implemented in The ONE have different ways to handle when a message is deleted from the buffer, implemented in the \texttt{deleteMessage()} of the classes contained in the \texttt{routing} package. The \texttt{EpidemicRouter} and \texttt{ProphetRouter} classes, inheriting the behavior of \texttt{ActiveRouter}, delete a message from the buffer either when a copy of that message has already reached its destination, in which case the deleted message is marked as \emph{removed} (however, this condition never occurs in the Epidemic and PRoPHET nor in the evolved protocols, see Figures \ref{fig:epidemic_messages}-\ref{fig:prophet_messages} in Appendix \ref{sec:logics}) or when the node needs to make room in its buffer for a new incoming message (deleting oldest messages first), in which case the deleted message is marked as \emph{dropped}. A message can also be deleted when it reaches its time-to-live. The TTL is defined in the simulation settings file (in our experiments we set a TTL of 5 hours, out of a total simulation duration of 12). When the TTL is exceeded, the message is also marked as \emph{dropped} (see below for a detailed analysis of the message status). Regardless the deletion reason, The ONE keeps track of the average waiting time of the messages deleted from the buffers:
\begin{equation*}
\text{Buffer time (avg)} = \frac{1}{n_{deleted}}\sum_{i=1}^{n_{deleted}}\left({T_{deletion}^i-T_{receive}^i}\right)
\end{equation*}
where $T_{deletion}^i$ and $T_{receive}^i$ indicate respectively the timestep at which a message was deleted from one node's buffer and the timestep at which it was received at that node, $T_{deletion}^i > T_{receive}^i$.
%the lower the better
\end{itemize}

%(functional requirements)
%(non-functional requirements)
From an application point of view, it is desirable to increase the delivery probability (how many messages reach their destination) and decrease the latency (how long they take to do so). On the other hand, for a better network resource management, i.e., to optimize the amount of data transmitted over the network and the amount of memory in use in each node, it would be desirable to reduce the overhead and, possibly, the buffer time. This is especially crucial in DTNs, and MANETs in general, where the nodes can be battery-powered (and sending a message is the most energy-consuming event) and memory-limited, see e.g. \cite{iacca2013distributed} for a discussion of these two aspects in the context of Wireless Sensor Networks. Therefore, it is important to assess if improving the delivery probability produces a worsening of these other metrics.

\begin{figure}[ht!]
    \centering
    \includegraphics[width=.9\textwidth]{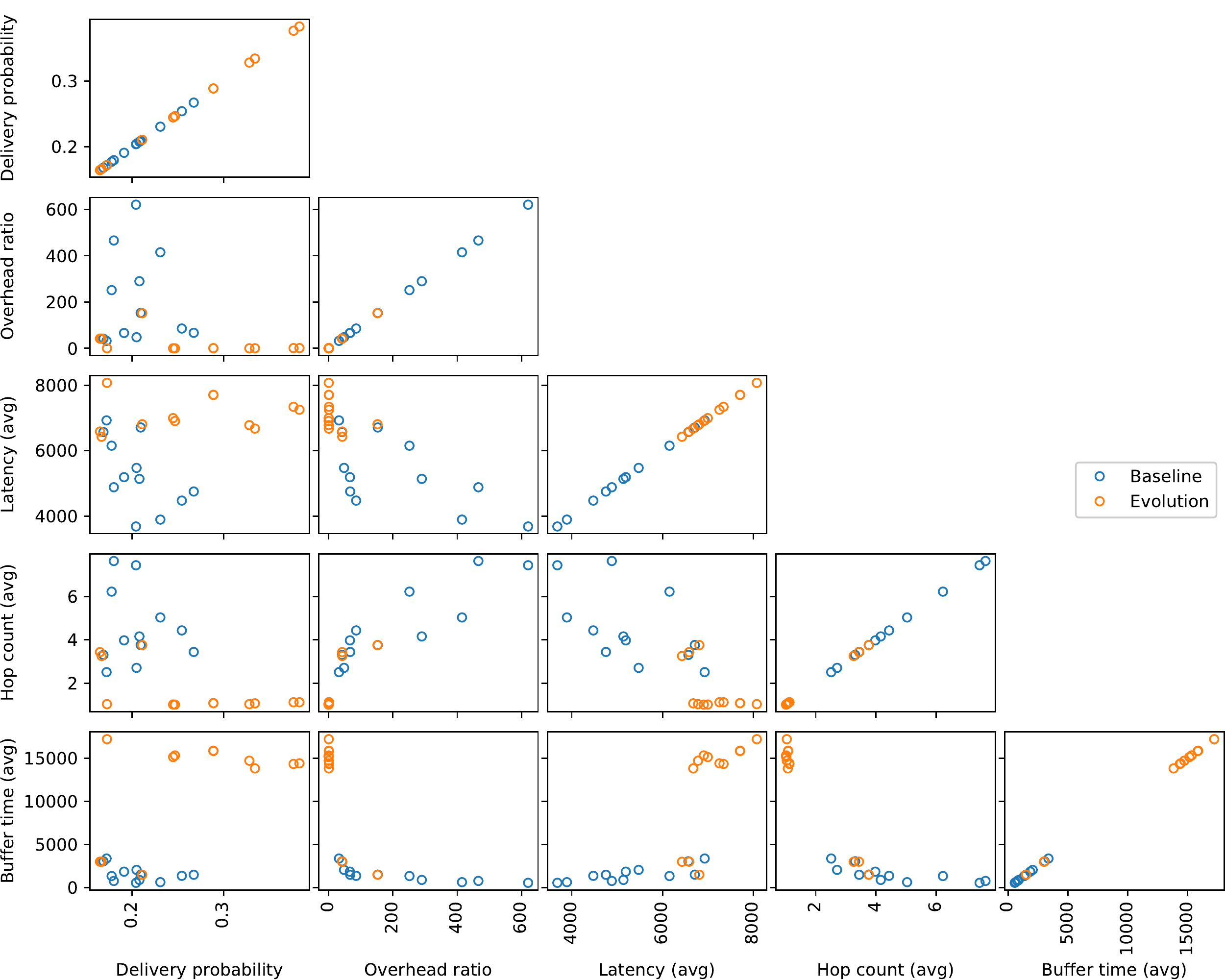}
    \caption{Trade-off between the delivery probability and the different metrics reported for the different test cases in Table \ref{tab:metrics_epidemic} and Table \ref{tab:metrics_prophet}.}
    \label{fig:metrics}
\vspace{-0.4cm}
\end{figure}

We report the results of the aforementioned metrics in Tables \ref{tab:metrics_epidemic} and \ref{tab:metrics_prophet}, respectively for the Epidemic and PRoPHET protocols. A graphical representation of these values, also w.r.t. the corresponding data delivery probability, is reported in Figure \ref{fig:metrics} in the form of a matrix of 2D scatter plots. Our analysis reveals some interesting, and in some cases counterintuitive, findings. On the one hand, there are two positive ``byproducts'' of the evolution: 1) the overhead of the evolved protocols is dramatically lower than that of the baseline protocols (in some cases, even 3 orders of magnitude smaller), which is different from what we would have expected to observe (i.e., a larger overhead corresponding to an increased delivery probability: the more duplicates, the higher the chance to reach a destination); 2) apart from the Manhattan test cases, the hop count of the evolved protocols is in most cases very close to 1, compared to values between 2 and 8 observed with the baseline protocols. On the other hand, it can be noted that the evolved protocols are characterized by a higher latency w.r.t. the baseline protocols (in the worst case, i.e., the Default test case with both Epidemic and PRoPHET, it is twice as big), as well as a higher buffer time that in the worst case (again, the Default test cases with both baseline protocols) is even 20 times bigger.

Overall, these results suggest that the evolved protocols are very efficient in terms of delivery probability as well as hop count and overhead, but there exists a trade-off between these three metrics and latency and buffer time. This trade-off is particularly evident in Figure \ref{fig:metrics}, see e.g. the $\langle$Buffer time (avg)$\rangle$ vs $\langle$Overhead ratio$\rangle$ and the $\langle$Buffer time (avg)$\rangle$ vs $\langle$Hop count (avg)$\rangle$ subplots, where a Pareto front can be clearly identified: in both cases, the evolved protocols appear on one corner of the Pareto front (characterized by low overheads and hop counts but high buffer times), while the baseline protocols seem to be designed to have a low buffer time but a high overhead and hop count. Concerning the correlation between the delivery probability, i.e., the goal of the optimization performed by GP, and the other metrics, a somehow less ``sharp'' (but, still evident) Pareto front can be identified in the $\langle$Overhead ratio$\rangle$ vs $\langle$Delivery probability$\rangle$ and $\langle$Hop count (avg)$\rangle$ vs $\langle$Delivery probability$\rangle$ subplots. The $\langle$Latency$\rangle$ vs $\langle$Delivery probability$\rangle$ and the $\langle$Buffer time (avg)$\rangle$ vs $\langle$Delivery probability$\rangle$ reveal instead two well-defined clusters separating the results of the evolved protocols from those of the baseline protocols. Finally, considering the other combinations of metrics, it appears that the latency and hop count correlate negatively and positively, respectively, with the overhead. Furthermore, the hop count and buffer time correlate negatively and positively, respectively, with the latency. As for this last aspect, while the negative correlation between hop count and latency might be counterintuitive, it should be considered that latency does not depend only on the number of hops, but also on the time spent by each message in the local buffer of each node it goes through.

All in all, this analysis indicates that a specific characteristic of the evolved protocols is that they tend to generate less duplicates than their baseline counterparts, thus avoiding filling the buffers too frequently. This makes it possible to keep messages longer in the buffers (higher buffer times), and eventually transmit them before they are deleted. Furthermore, apart for the Manhattan cases, the messages reach their destination with an average hop count close to 1, i.e., they are delivered as soon as they are within reach of the source node. This, in turn, contributes to having less duplicates, thus a lower overhead, as shown by the positive correlation between hop count and overhead. The resulting effect of this behavior is then an increase of the delivery probability.

\begin{table}[ht!]
\centering
\caption{Network metrics of the evolved protocols vs the Epidemic routing protocol (median across \NUMRUNS~simulations). The evolved protocols show lower overhead and hop count but higher latency and buffer time.}
\label{tab:metrics_epidemic}
\resizebox{\textwidth}{!}{
%\begin{tabular}{lllllllll}
\begin{tabular}{lcccccccc}
 \toprule
 \multirow{2}{*}{\textbf{Test case}}
 & \multicolumn{2}{c}{\textbf{Overhead ratio}} & \multicolumn{2}{c}{\textbf{Latency (avg)}} & \multicolumn{2}{c}{\textbf{Hop count (avg)}} & \multicolumn{2}{c}{\textbf{Buffer time (avg)}} \\
 %\cline{2-9}
 \cmidrule(l{2pt}r{2pt}){2-3}
 \cmidrule(l{2pt}r{2pt}){4-5}
 \cmidrule(l{2pt}r{2pt}){6-7}
 \cmidrule(l{2pt}r{2pt}){8-9}
 & \textbf{Epidemic} & \textbf{GP} & \textbf{Epidemic} & \textbf{GP} & \textbf{Epidemic} & \textbf{GP} & \textbf{Epidemic} & \textbf{GP} \\
 \midrule
 Default (40 hosts) & 85 & 0.4 & 4473 & 6676 & 4.4 & 1.1 & 1364 & 13836 \\
 Default (100 hosts) & 621 & 1 & 3681 & 7343 & 7.4 & 1.1 & 553 & 14354 \\
 Helsinki (40 hosts) & 66 & 0.1 & 5189 & 6908 & 4.0 & 1.0 & 1847 & 15341 \\
 Helsinki (100 hosts) & 466 & 0.7 & 4881 & 7706 & 7.6 & 1.0 & 770 & 15851 \\
 Manhattan (40 hosts) & 42 & 42 & 6569 & 6427 & 3.3 & 3.2 & 3051 & 2998 \\
 Manhattan (100 hosts) & 251 & 0.3 & 6152 & 8076 & 6.2 & 1.0 & 1342 & 17231 \\
 \bottomrule
\end{tabular}
}
\end{table}
\vspace{-0.3cm}
\begin{table}[ht!]
\centering
\caption{Network metrics of the evolved protocols vs the PRoPHET routing protocol (median across \NUMRUNS~simulations). The evolved protocols show lower overhead and hop count but higher latency and buffer time.}
\label{tab:metrics_prophet}
\resizebox{\textwidth}{!}{
%\begin{tabular}{lllllllll}
\begin{tabular}{lcccccccc}
 \toprule
 \multirow{2}{*}{\textbf{Test case}}
 & \multicolumn{2}{c}{\textbf{Overhead ratio}} & \multicolumn{2}{c}{\textbf{Latency (avg)}} & \multicolumn{2}{c}{\textbf{Hop count (avg)}} & \multicolumn{2}{c}{\textbf{Buffer time (avg)}} \\
  %\cline{2-9}
 \cmidrule(l{2pt}r{2pt}){2-3}
 \cmidrule(l{2pt}r{2pt}){4-5}
 \cmidrule(l{2pt}r{2pt}){6-7}
 \cmidrule(l{2pt}r{2pt}){8-9}
 & \textbf{PRoPHET} & \textbf{GP} & \textbf{PRoPHET} & \textbf{GP} & \textbf{PRoPHET} & \textbf{GP} & \textbf{PRoPHET} & \textbf{GP} \\
 \midrule
 Default (40 hosts) & 67 & 0.2 & 4751 & 6781 & 3.4 & 1.0 & 1486 & 14731 \\
 Default (100 hosts) & 415 & 1.0 & 3891 & 7254 & 5.0 & 1.1 & 629 & 14426 \\
 Helsinki (40 hosts) & 48 & 0.1 & 5472 & 6996 & 2.7 & 1.0 & 2054 & 15154 \\
 Helsinki (100 hosts) & 290 & 0.7 & 5137 & 7708 & 4.2 & 1.1 & 888 & 15891 \\
 Manhattan (40 hosts) & 32 & 42 & 6927 & 6585 & 2.5 & 3.4 & 3373 & 3006 \\
 Manhattan (100 hosts) & 153 & 152 & 6710 & 6809 & 3.8 & 3.8 & 1503 & 1493 \\
 \bottomrule
\end{tabular}
}
\end{table}

% -------------------------------------------------------------------------
% -------------------------------------------------------------------------
% -------------------------------------------------------------------------

\section{Conclusions}
\label{sec:conclusions}

\subsubsection*{Main findings} We applied Genetic Programming to \emph{genetically improve} two replication-based routing protocols widely adopted in intermittently connected networks, namely Epidemic and PRoPHET. In four out of six test cases, GP was able to find protocol implementations that produced significantly better delivery probabilities w.r.t. the two baseline protocols. In the two Manhattan-like test cases, on the other hand, no significant improvement was obtained. A similar difference in performance was also observed when comparing our improved PRoPHET protocols against three variants of PRoPHET proposed in the recent literature. We also observed that the evolved protocols are in general characterized by a reduced overhead yet larger latencies w.r.t. the baseline protocols. Furthermore, we found that apart from the Manhattan cases the evolved protocols could generalize across test cases (results reported in Appendix \ref{sec:generalization}). Finally, it is worth noticing that the genetically improved protocols are not expensive to implement (as they use the same components of the baseline protocols) and they can cope with the typical hardware/computational constraints of the devices for which the existing protocols are usually intended.

%, as a form of emergent property in networked systems
\subsubsection*{Limitations} This work represents a first attempt to establish a more general methodology to evolve protocols, and stacks thereof. As such, the present methodology has some clear limitations that it would be worth to overcome: for instance, it would be useful to implement a mechanism to automatically extract the fundamental components of the existing protocols (currently this is done manually), or to evolve also other parts of the routing protocol (i.e., not only the \texttt{update()} method). %, and reduce the complexity of the evolved logics
Another improvement might be the introduction of an anti-bloat mechanism as well as a history of the evaluated solutions based on syntactic or semantic similarity, in order to avoid fitness re-evaluation and improve diversity during the search.

%This work paves the road to a number of exciting research directions beyond the improvements above. 
\subsubsection*{Future works} A straightforward extension of this work would be to test the proposed methodology to other DTN protocols, such as RAPID \cite{balasubramanian2007dtn}, MaxProp \cite{burgess2006maxprop}, or Spray \& Wait \cite{spyropoulos2005spray}. It would be possible to extend it also to routing protocols used in other kinds of MANETs (i.e., non-DTN), such as table-driven protocols, e.g. Optimized Link State Routing Protocol (OLSR) \cite{clausen2003optimized} and Destination Sequence Distance Vector (DSDV) \cite{perkins1994highly}, or on-demand (reactive) protocols, e.g. Ad hoc On-demand Distance Vector (AODV) \cite{perkins1999ad} and Dynamic Source Routing (DSR) \cite{johnson1996dynamic}. However, since these protocols are in general more complex than Epidemic and PRoPHET, the search space projected by their components is potentially much larger. Thus, more computational resources and/or specific search operators might be needed in order to find improved protocols. Going beyond routing, it would be possible to apply the proposed methodology to other network layers, for instance to improve existing MAC or congestion protocols. As we have discussed in Section \ref{sec:related}, albeit some previous research has applied GP to other network layers, the existing works either evolve \emph{from scratch} a protocol, starting from its specifications, or optimize specific elements of the protocols, such as e.g. the formulas used to vary the congestion window, rather than performing an actual Genetic Improvement in terms of software code. Another possibility would be to use not only existing high-level components (obtained, as we have seen, from the decomposition of the existing protocols), but also components that are evolved \emph{ex novo}. While the search space would be tremendously larger, the additional degrees of freedom offered might provide important opportunities to improve the existing protocols. An intriguing possibility would also be to implement forms of transfer learning across different networks, and online adaptation in response to network changes.
A further point of attention is the trade-off between different network metrics: rather than improving a protocol w.r.t. only one metric of interest, it would possible be to apply multi-objective GP and look explicitly for the different trade-offs existing between two or more metrics of interest, so to eventually define \emph{a posteriori} the right protocol to use. Finally, it would be valuable to assess in hardware the performance of the evolved protocols.

% -------------------------------------------------------------------------
% -------------------------------------------------------------------------
% -------------------------------------------------------------------------

\bibliographystyle{ACM-Reference-Format}
\bibliography{paper}

\clearpage

% -------------------------------------------------------------------------
% -------------------------------------------------------------------------
% -------------------------------------------------------------------------

\appendix

%----------------------------------------------------------------

\section{Generalization of the evolved protocols}\label{sec:generalization}

%For this analysis, we limited our experiments to six the Default and Helsinki test cases, with 40 and 100 hosts, excluding the Manhattan cases due to their computationally heavier simulations.
We performed additional experiments aimed at assessing the generalizability of the evolved protocols between different test cases, i.e., testing how a protocol evolved on a specific test case performs on another one. Due to the large number of combinations of test cases, we limited this analysis to six selected cases that we considered representative of the generalization capabilities of the evolved protocols, i.e., covering at least one case for each protocol, map, and number of hosts per group. Also in this case, each test case was simulated \NUMRUNS~times.

Table \ref{tab:res_generalization} shows the results of the test cases considered in this part of the experimentation. From the table, it can be seen that the evolved protocols are able to generalize in the first four cases, where the performance (\emph{Baseline}) of the protocol evolved on the \emph{Target test case} is statistically equivalent (p-value $>\alpha$) to the performance (\emph{Tested}) of the protocol evolved on the \emph{Source test case}. In the remaining two cases, the \emph{Tested} performance is statistically lower (p-value $\leq\alpha$) than the \emph{Baseline} one, meaning that the protocol specifically evolved on the \emph{Target test case} shows a higher delivery probability than that of the protocol evolved on the \emph{Source test case}. In particular, it appears that the Manhattan cases are harder to generalize (both from and to), see the last two rows in the table. This is likely due to the lower node density that characterizes the Manhattan cases (on the effect of density on the delivery probability in Manhattan-like test cases, see e.g. \cite{thong2004performance}), that is comparably lower than in the two other maps: observing Table \ref{tab:one_settings} and Figure \ref{fig:maps} in the main text, it is worth highlighting the fact that the world size (i.e, the size of the map) in the Manhattan case is much larger than in the Default map, while it is the same as in the Helsinki map, although the latter presents way less roads and thus nodes therein have a much more confined mobility. This is likely the same reason that makes it harder to improve the performance of Epidemic and PRoPHET protocols on the Manhattan map.

Finally, it should be noted that apart from the Manhattan cases, in all the other cases the \emph{Tested} performances are still much higher than those of the baseline Epidemic and PRoPHET protocols. In fact, comparing the delivery probability shown in the \emph{Tested} column in the first four rows in Table \ref{tab:res_generalization} with the corresponding results shown in the main text (Tables \ref{tab:wilcoxon_epidemic}-\ref{tab:wilcoxon_prophet}), it results: 0.2874 vs 0.1798, 0.3260 vs 0.2542, 0.2519 vs 0.2047, 0.3641 vs 0.2307. In the Manhattan cases, the comparison results instead in 0.2075 vs 0.2041 and 0.1493 vs 0.1685, respectively for the last two rows of Table \ref{tab:res_generalization}.

\begin{table}[ht!]
\centering
\caption{Results of the generalizability experiments (median across \NUMRUNS~simulations). The \emph{Baseline} column indicates the delivery probability obtained by the best routing protocol evolved on the \emph{Target test case}, while the \emph{Tested} column indicates the delivery probability obtained by the best routing protocol evolved on the \emph{Source test case} when this is applied to the \emph{Target test case}. For each experiment, we report the p-value of the Wilcoxon rank-sum test ($N=\NUMRUNS, \alpha=0.05$). The evolved protocols are able to generalize in the first four cases (no statistical difference between \emph{Baseline} and \emph{Tested}), while they are not in the remaining ones.}
\label{tab:res_generalization}
\resizebox{\textwidth}{!}{
\begin{tabular}{llccc}
 \toprule
 \textbf{Source test case} & \textbf{Target test case} & \textbf{Baseline} & \textbf{Tested} & \textbf{p-value} \\
 \midrule
 Epidemic, Default (40 hosts) & Epidemic, Helsinki (100 hosts) & 0.2887 & 0.2874 & 0.726 \\
 Epidemic, Helsinki (100 hosts) & Epidemic, Default (40 hosts) & 0.3342 & 0.3260 & 0.093 \\
 PRoPHET, Default (100 hosts) & PRoPHET, Helsinki (40 hosts) & 0.2447 & 0.2519 & 0.878 \\ 
 PRoPHET, Helsinki (40 hosts) & PRoPHET, Default (100 hosts) & 0.3829 & 0.3641 & 0.022 \\
 Epidemic, Manhattan (40 hosts) & Epidemic, Default (100 hosts) & 0.3764 & 0.2075 & 0.005 \\
 Epidemic, Default (100 hosts) & Epidemic, Manhattan (40 hosts) & 0.1654 & 0.1493 & 0.005 \\
 \bottomrule
\end{tabular}
}
\end{table}

%----------------------------------------------------------------

\section{Analysis of the evolved protocols}\label{sec:logics}

Comparing the logics (in the form of GP tree) of the baseline protocols, shown in Figures \ref{fig:tree_baseline_epidemic}-\ref{fig:tree_baseline_prophet} in the main text, with that of the best evolved protocols, shown in Figures in \ref{fig:gen_tree_ep_def_40}-\ref{fig:gen_tree_prop_man_100} in Appendix \ref{sec:appendix_best_trees}, it can be noted that the evolved protocols are quite different from the baselines. In particular, apart from the case of Epidemic on the Default map with 40 hosts per group (Figure \ref{fig:gen_tree_ep_def_40}), the evolved protocols appear more complex, in terms of number of nodes and depth, than their corresponding baseline. The main difference in the routing logic consists in the condition in which the evolved protocols can send messages. In the baseline protocols, \texttt{isTransferring()} is a condition that does not allow the transmission of messages. Instead, in most of the evolved protocols, when \texttt{isTransferring()} is true, the protocol still tries to transmit its messages (i.e., puts them in the sending buffer). A second important difference is the number of attempts of transmissions that the evolved protocols perform at each update. In fact, the baseline protocols perform at most two attempts per update, while the evolved protocols perform up to 4 attempts.
% see isTransferring() code here: https://github.com/akeranen/the-one/blob/v1.6.0/routing/ActiveRouter.java#L526
% TODO: this part is actually a bit critical

\begin{figure}[ht!]
    \begin{subfigure}[b]{0.32\textwidth}
    \centering
    \includegraphics[width=\textwidth, trim=0 0.6cm 0 0, clip]{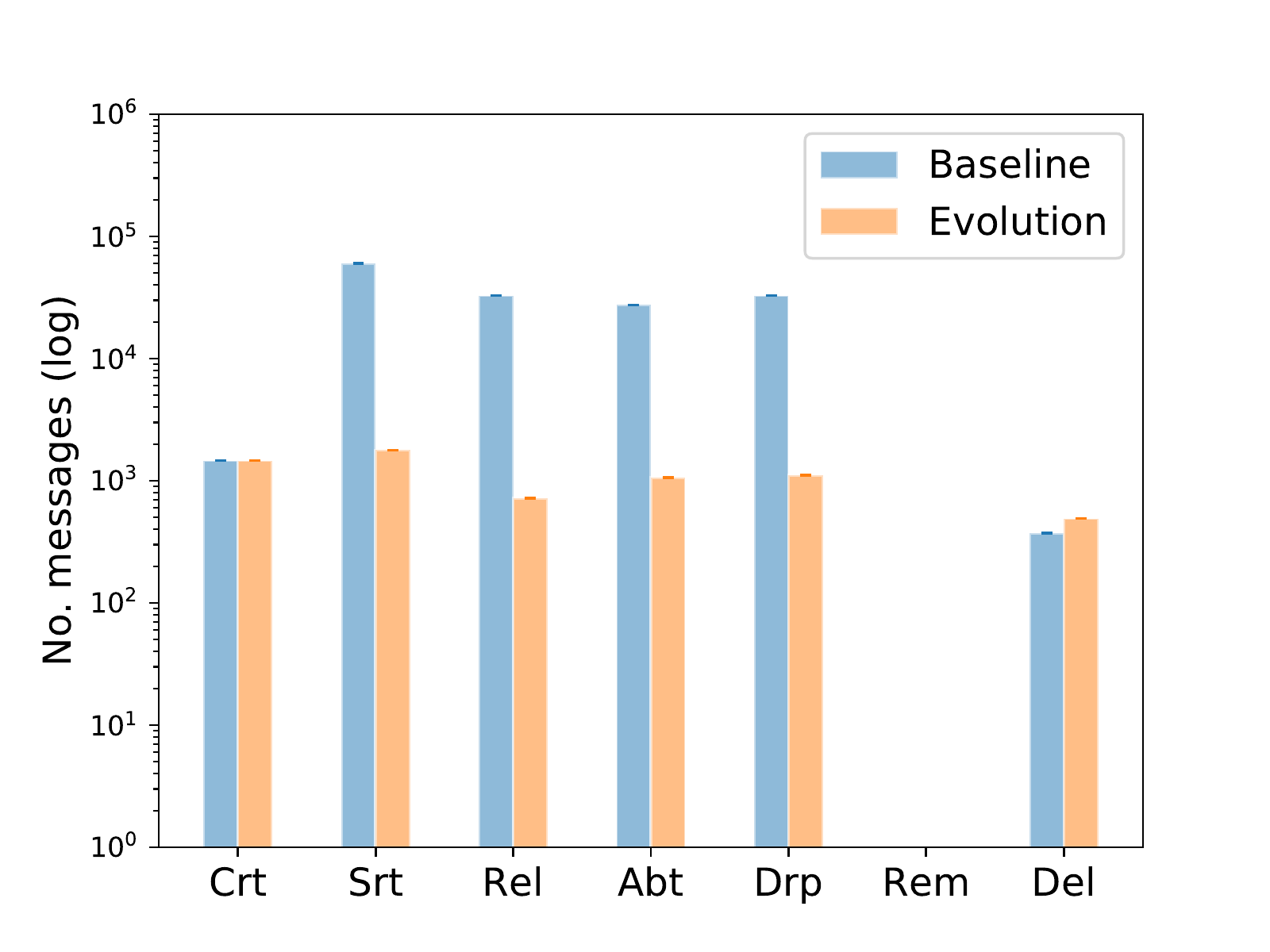}
    \caption{Default (40 hosts)}%\label{}
    \end{subfigure}
    \begin{subfigure}[b]{0.32\textwidth}
    \centering
    \includegraphics[width=\textwidth, trim=0 0.6cm 0 0, clip]{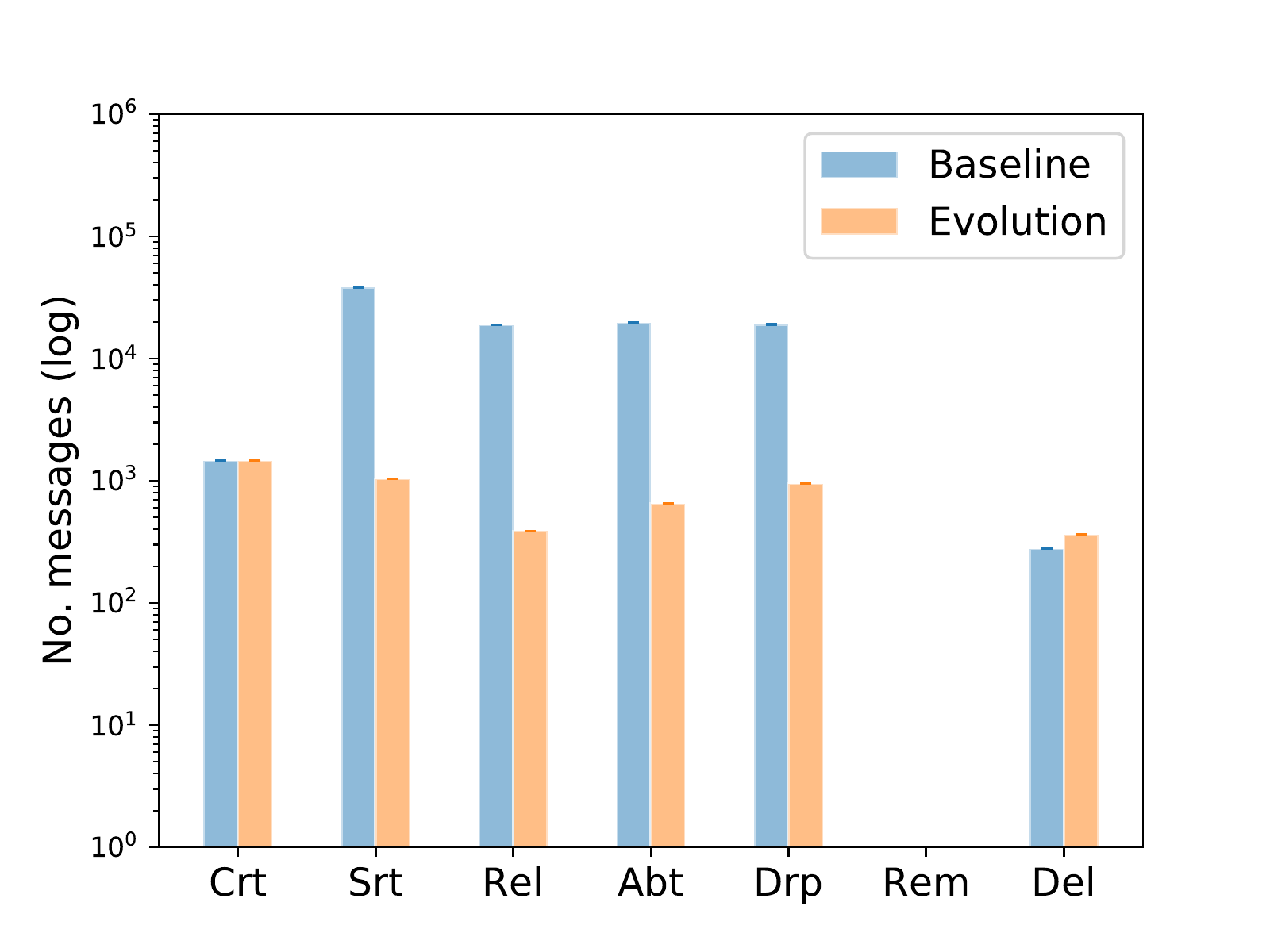}
    \caption{Helsinki (40 hosts)}%\label{}
    \end{subfigure}
    \begin{subfigure}[b]{0.32\textwidth}
    \centering
    \includegraphics[width=\textwidth, trim=0 0.6cm 0 0, clip]{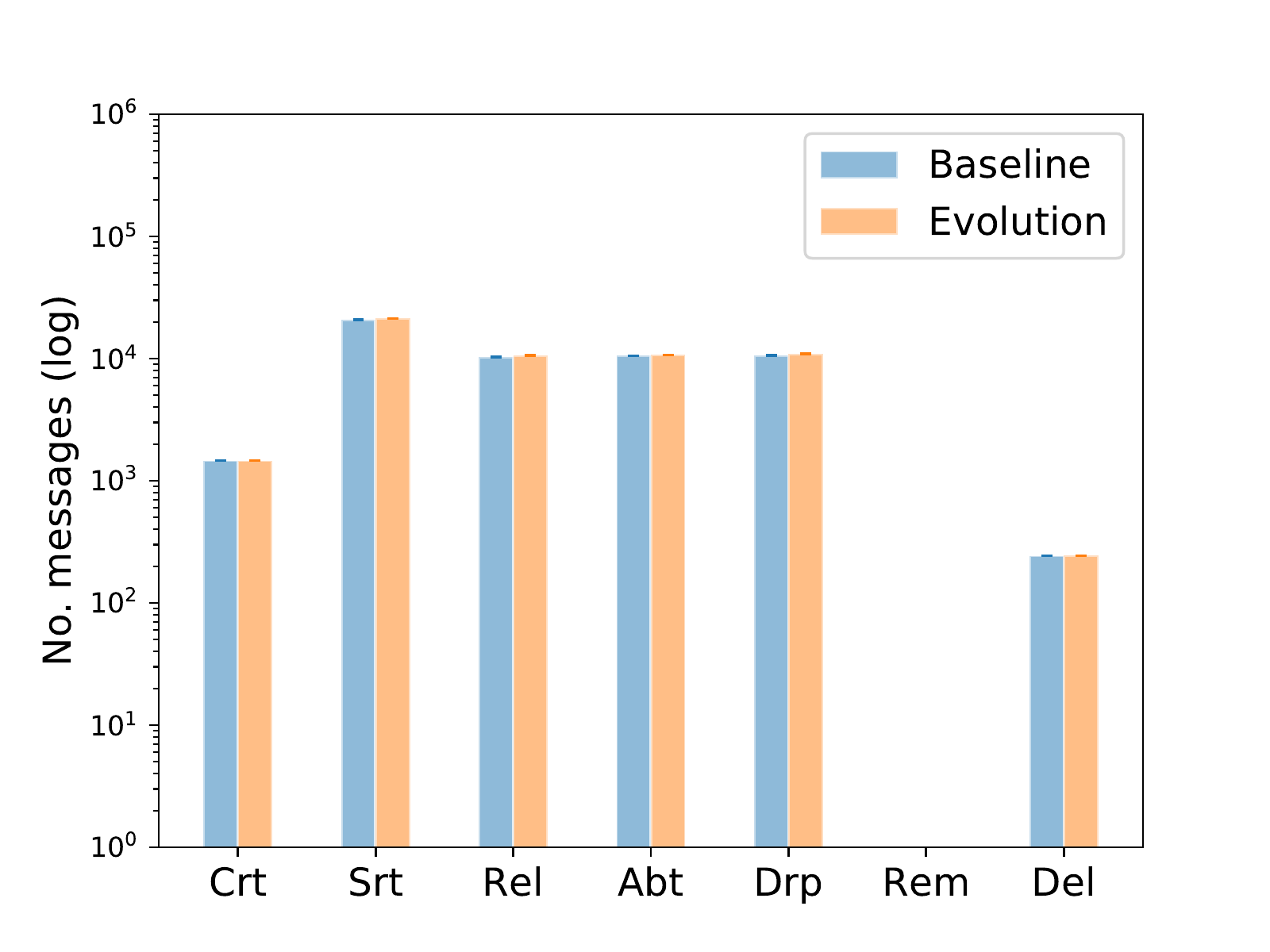}
    \caption{Manhattan (40 hosts)}%\label{}
    \end{subfigure}
    \begin{subfigure}[b]{0.32\textwidth}
    \centering
    \includegraphics[width=\textwidth, trim=0 0.6cm 0 0, clip]{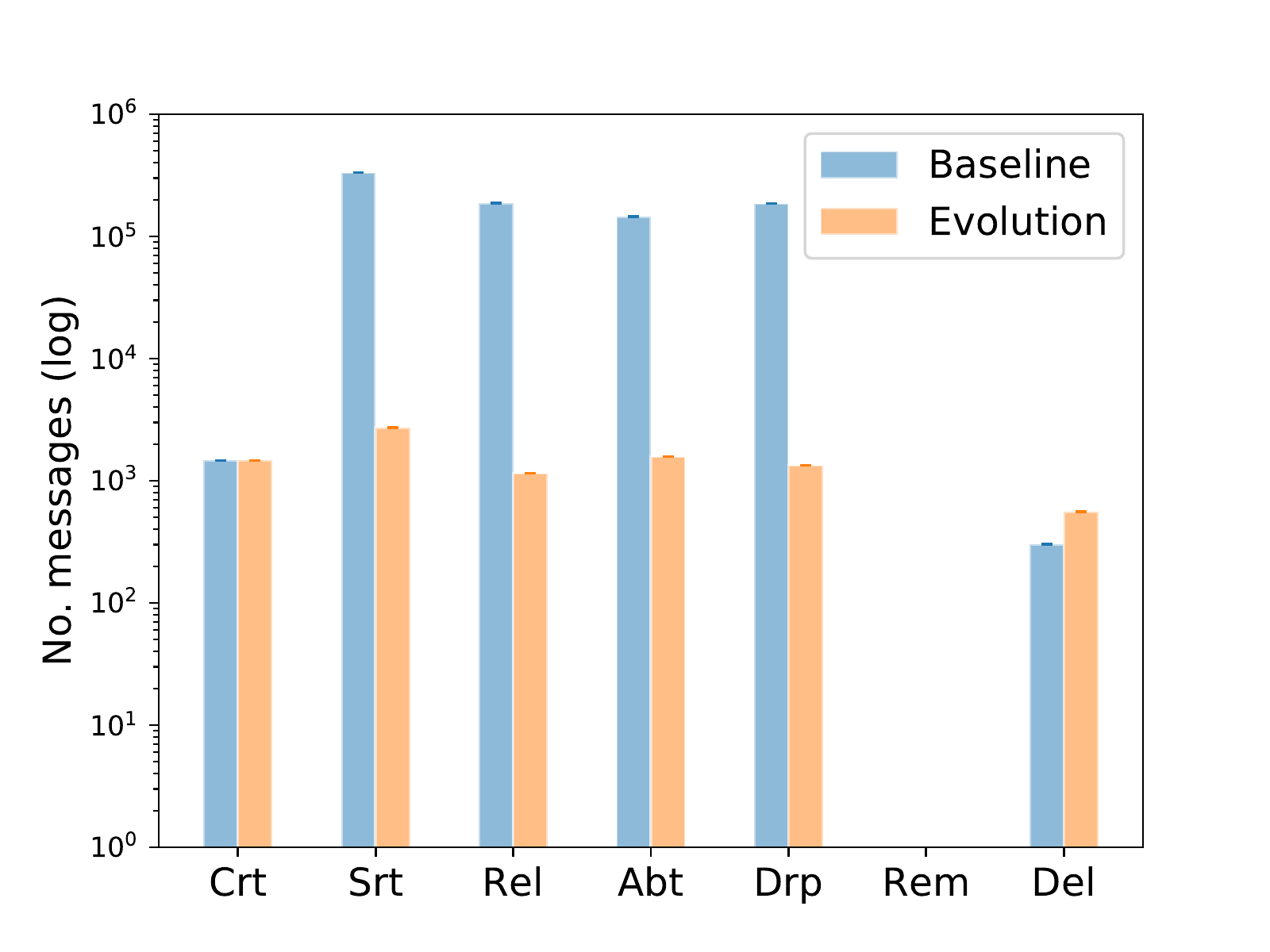}
    \caption{Default (100 hosts)}%\label{}
    \end{subfigure}
    \begin{subfigure}[b]{0.32\textwidth}
    \centering
    \includegraphics[width=\textwidth, trim=0 0.6cm 0 0, clip]{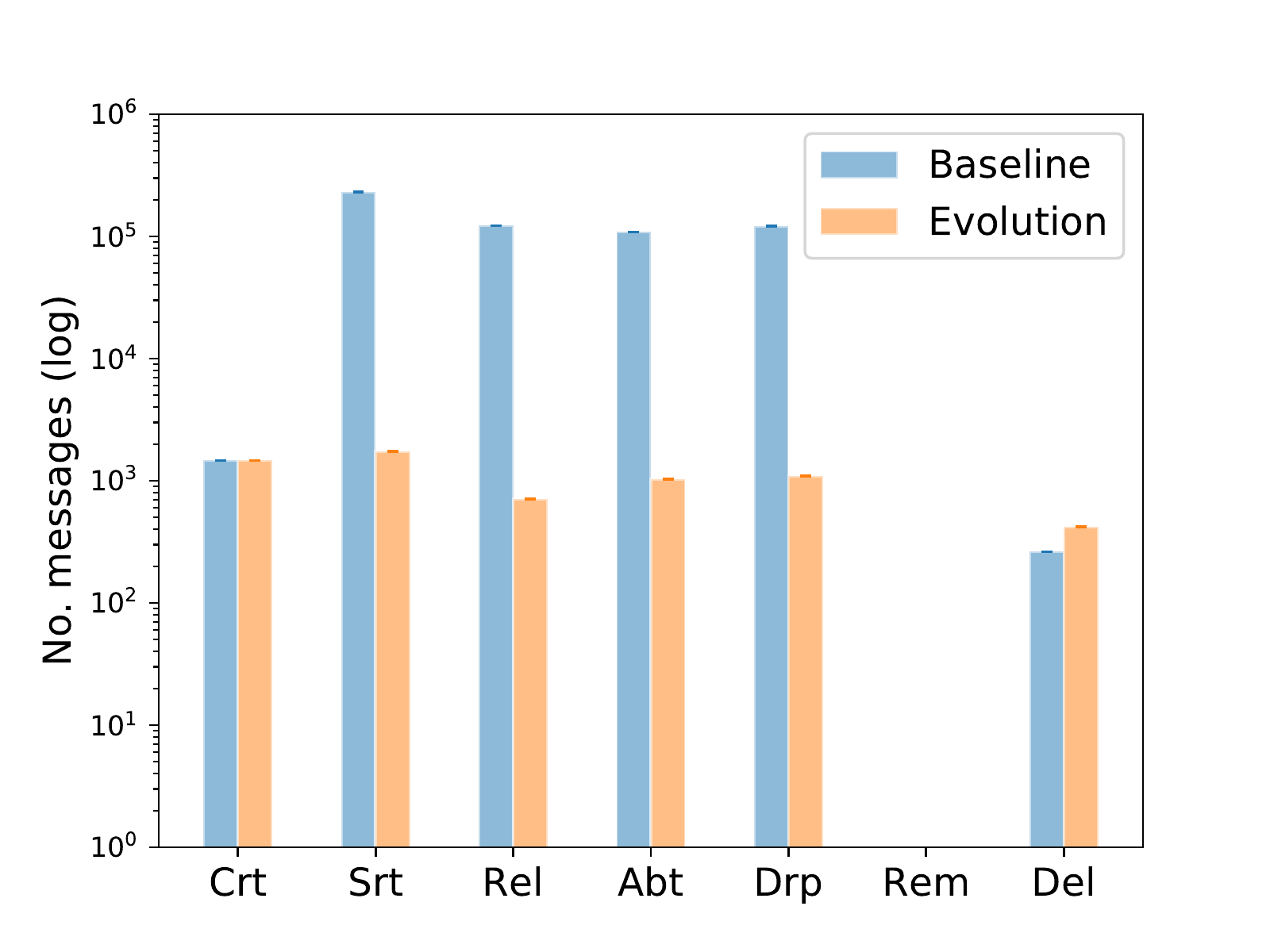}
    \caption{Helsinki (100 hosts)}%\label{}
    \end{subfigure}
    \begin{subfigure}[b]{0.32\textwidth}
    \centering
    \includegraphics[width=\textwidth, trim=0 0.6cm 0 0, clip]{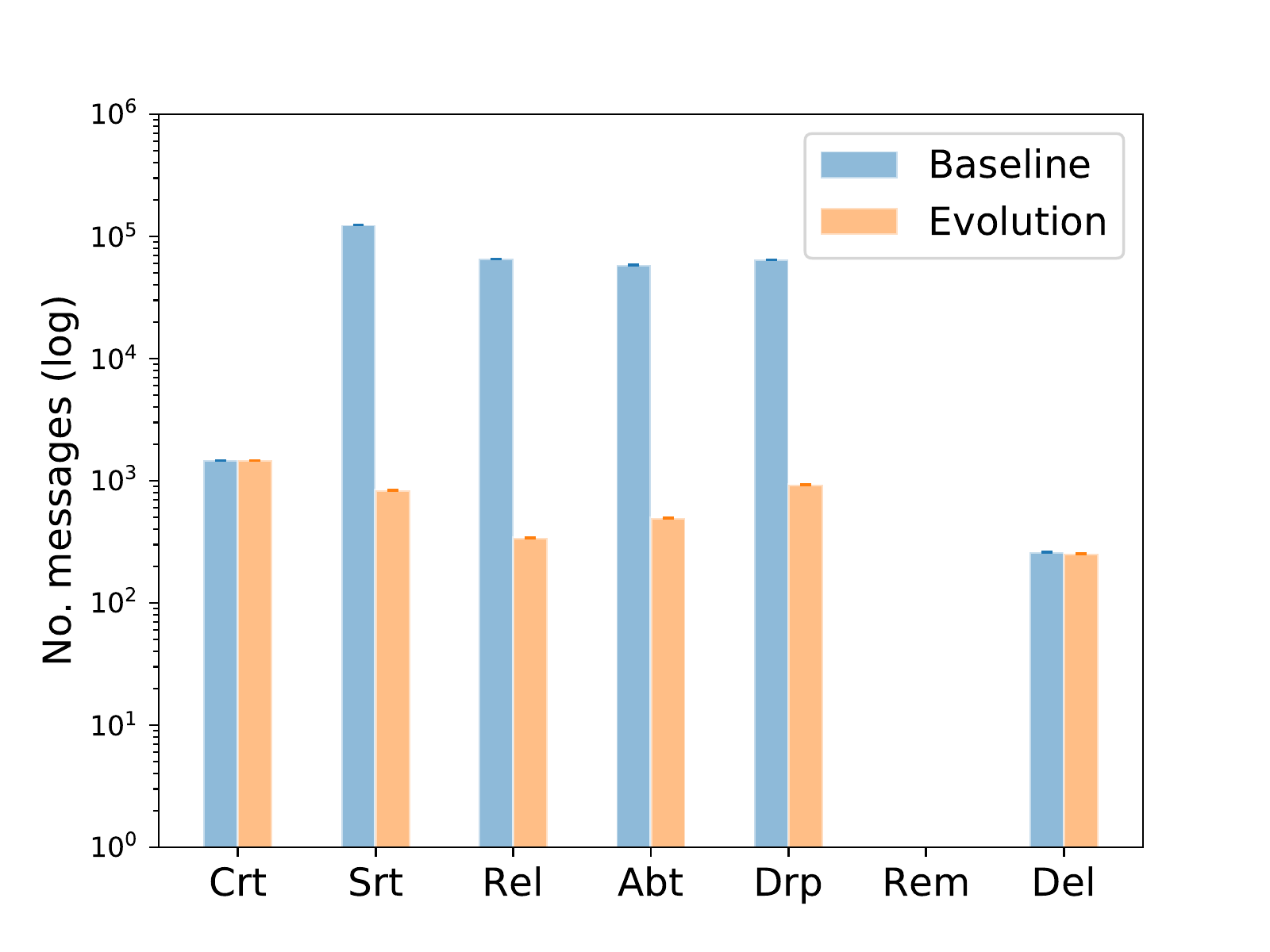}
    \caption{Manhattan (100 hosts)}%\label{}
    \end{subfigure}
    \vspace{-0.2cm}
    \caption{Number of created (Crt), started (Srt), relayed (Rel), aborted (Abt), dropped (Drp), removed (Rem) and delivered (Del) messages with the Epidemic routing protocol (\emph{Baseline}) and the best evolved protocols (\emph{Evolution}), mean $\pm$ std. dev. (log scale) across \NUMRUNS~simulations.}
    \label{fig:epidemic_messages}
\end{figure}

\begin{figure}[ht!]
    \begin{subfigure}[b]{0.32\textwidth}
    \centering
    \includegraphics[width=\textwidth, trim=0 0.6cm 0 0, clip]{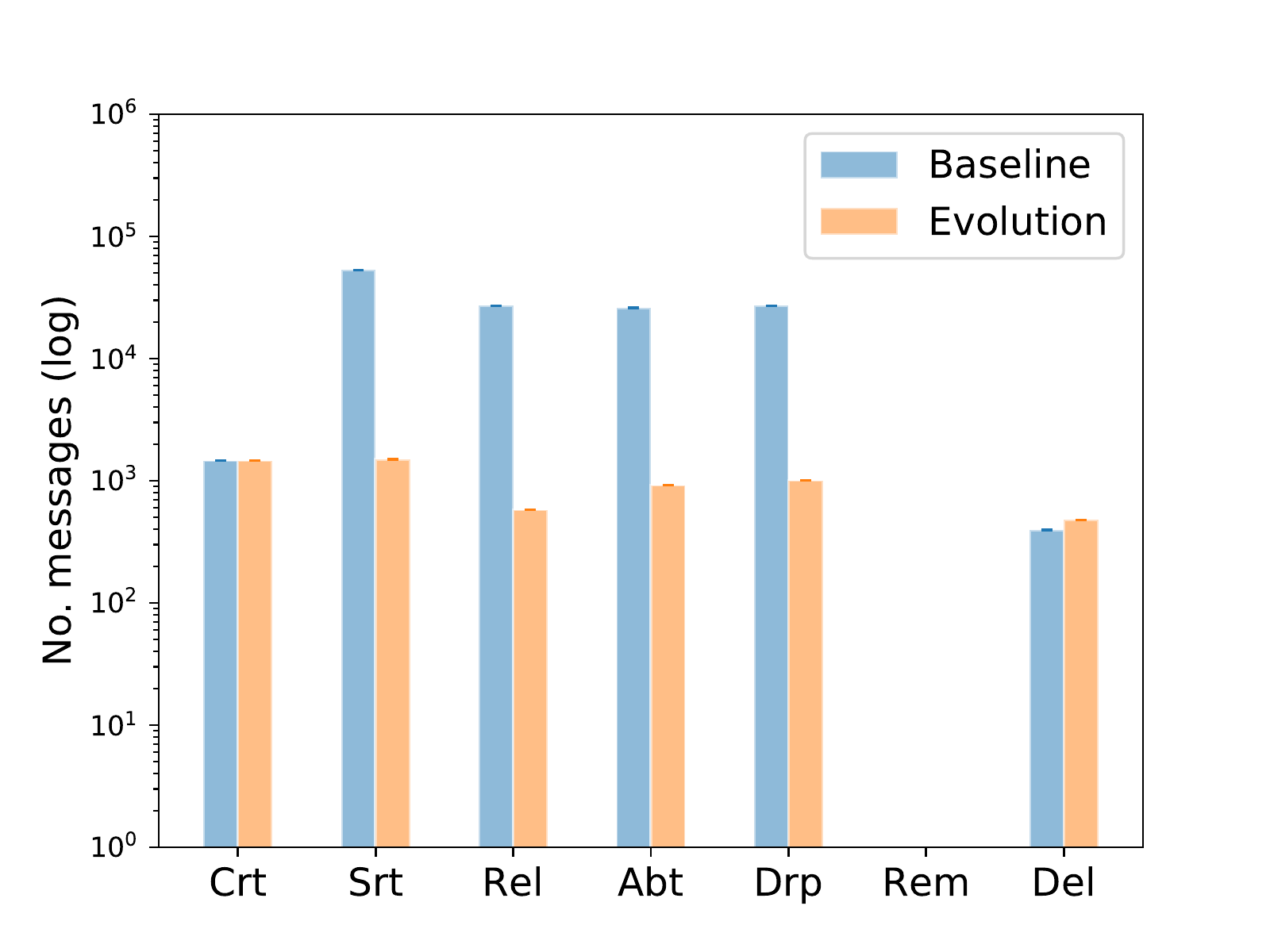}
    \caption{Default (40 hosts)}%\label{}
    \end{subfigure}
    \begin{subfigure}[b]{0.32\textwidth}
    \centering
    \includegraphics[width=\textwidth, trim=0 0.6cm 0 0, clip]{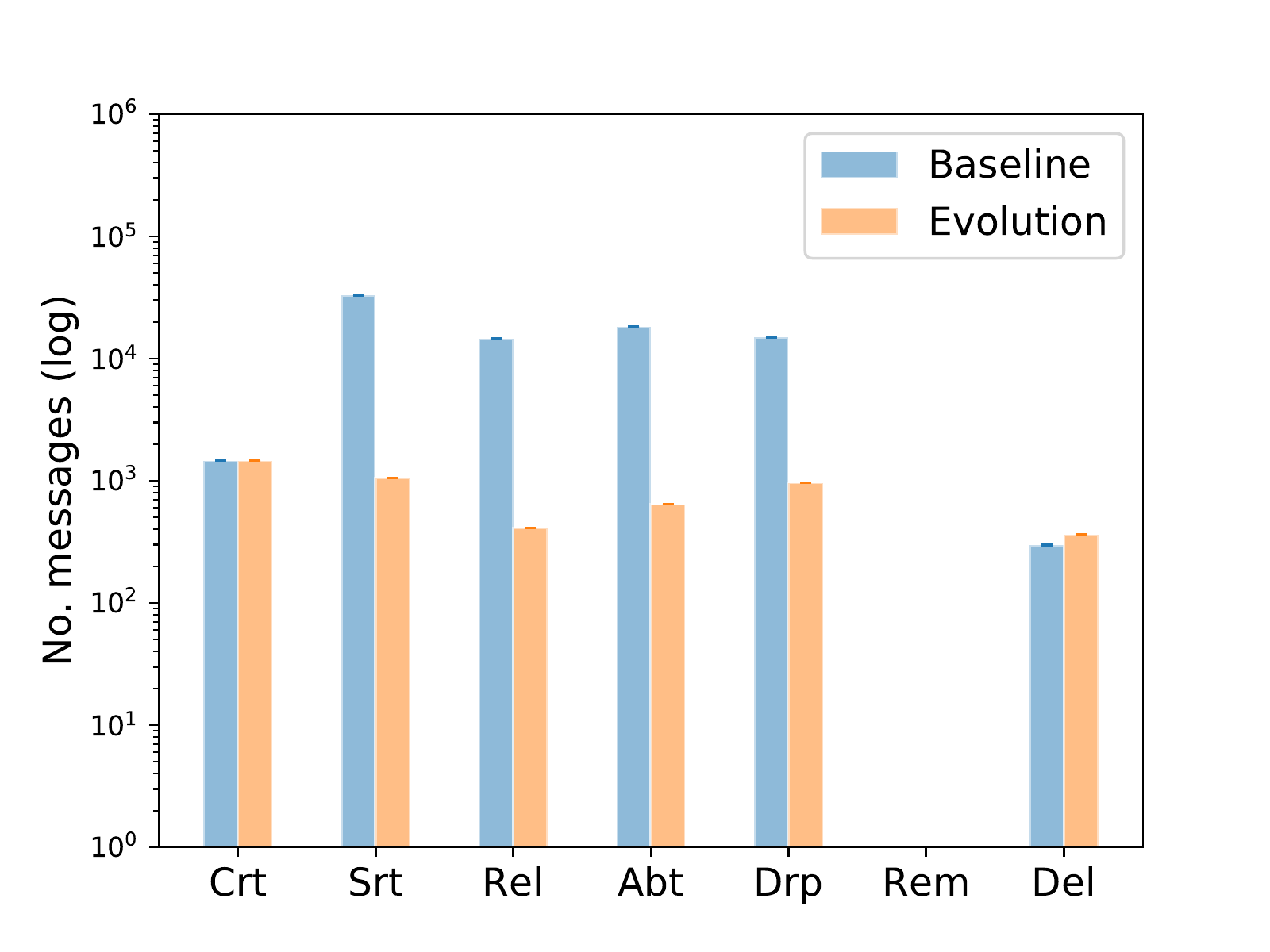}
    \caption{Helsinki (40 hosts)}%\label{}
    \end{subfigure}
    \begin{subfigure}[b]{0.32\textwidth}
    \centering
    \includegraphics[width=\textwidth, trim=0 0.6cm 0 0, clip]{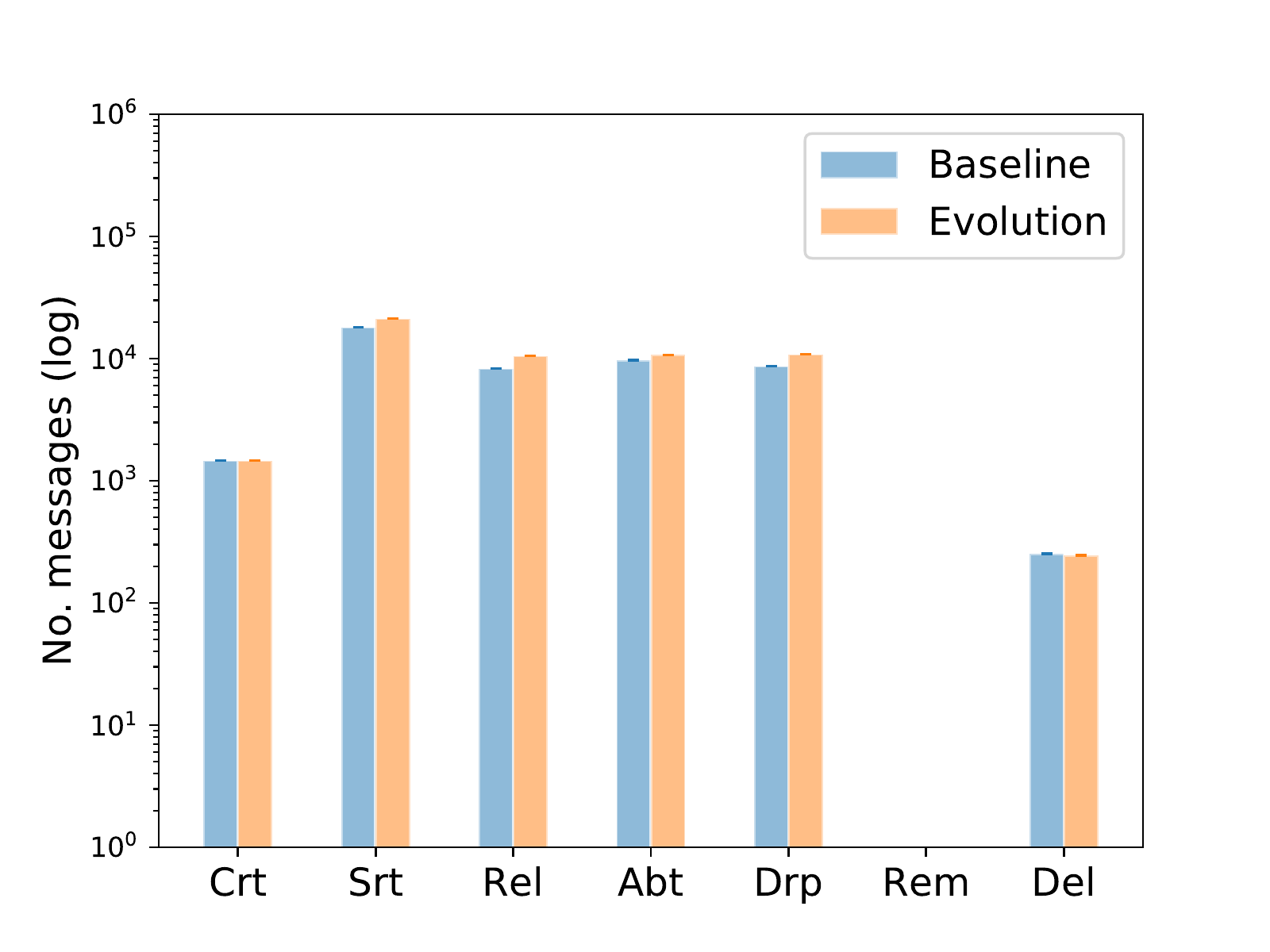}
    \caption{Manhattan (40 hosts)}%\label{}
    \end{subfigure}
    \begin{subfigure}[b]{0.32\textwidth}
    \centering
    \includegraphics[width=\textwidth, trim=0 0.6cm 0 0, clip]{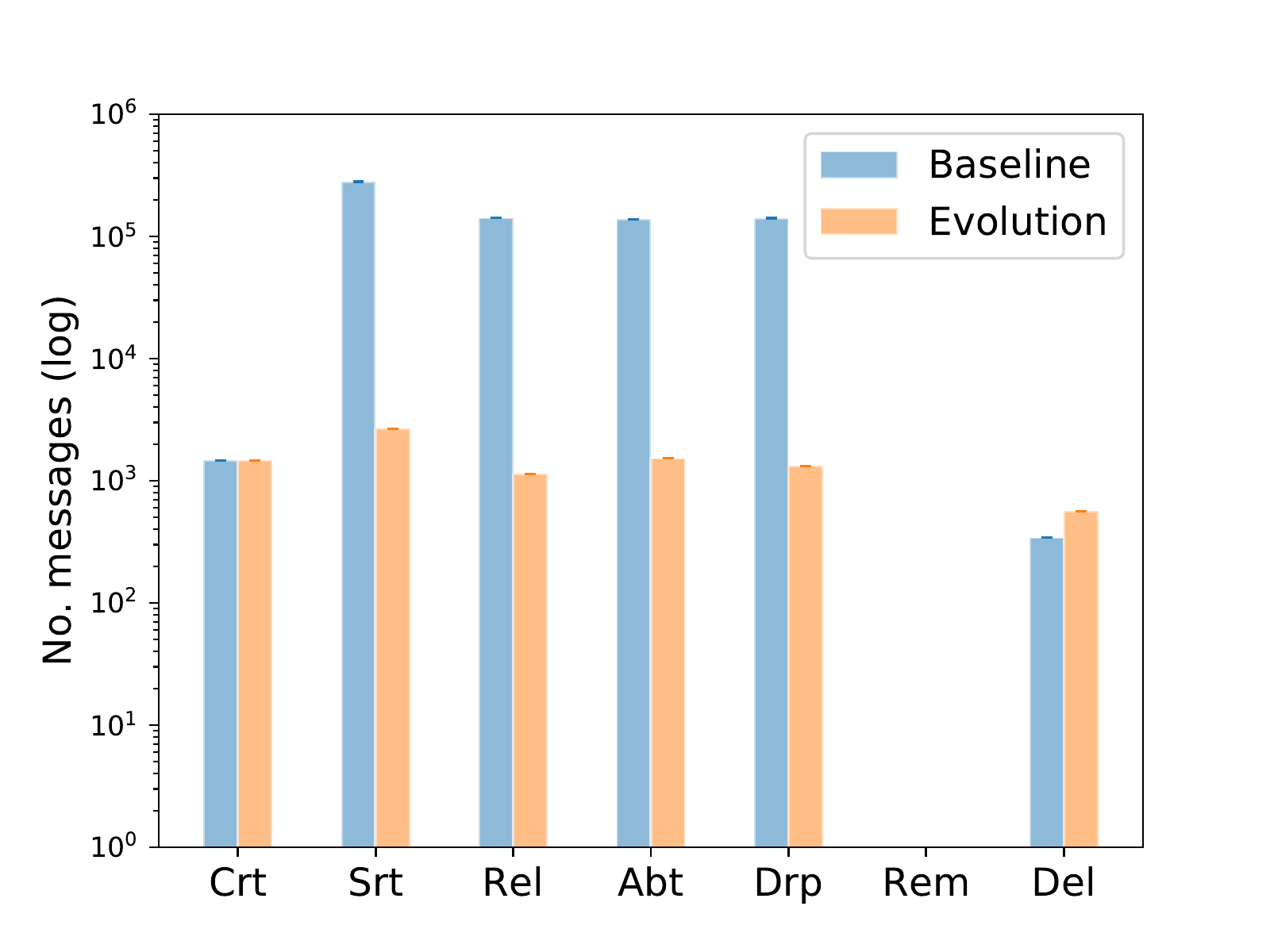}
    \caption{Default (100 hosts)}%\label{}
    \end{subfigure}
    \begin{subfigure}[b]{0.32\textwidth}
    \centering
    \includegraphics[width=\textwidth, trim=0 0.6cm 0 0, clip]{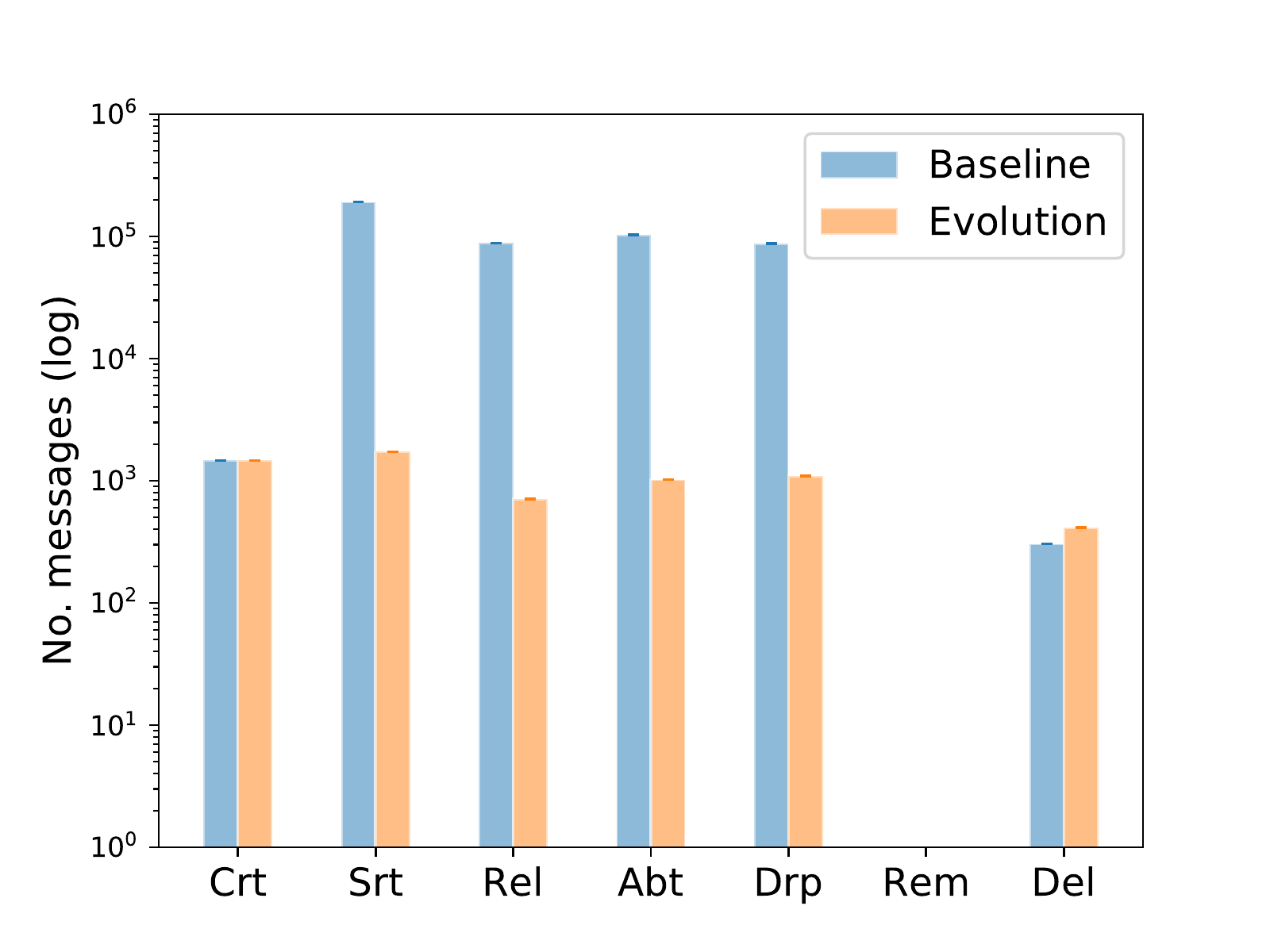}
    \caption{Helsinki (100 hosts)}%\label{}
    \end{subfigure}
    \begin{subfigure}[b]{0.32\textwidth}
    \centering
    \includegraphics[width=\textwidth, trim=0 0.6cm 0 0, clip]{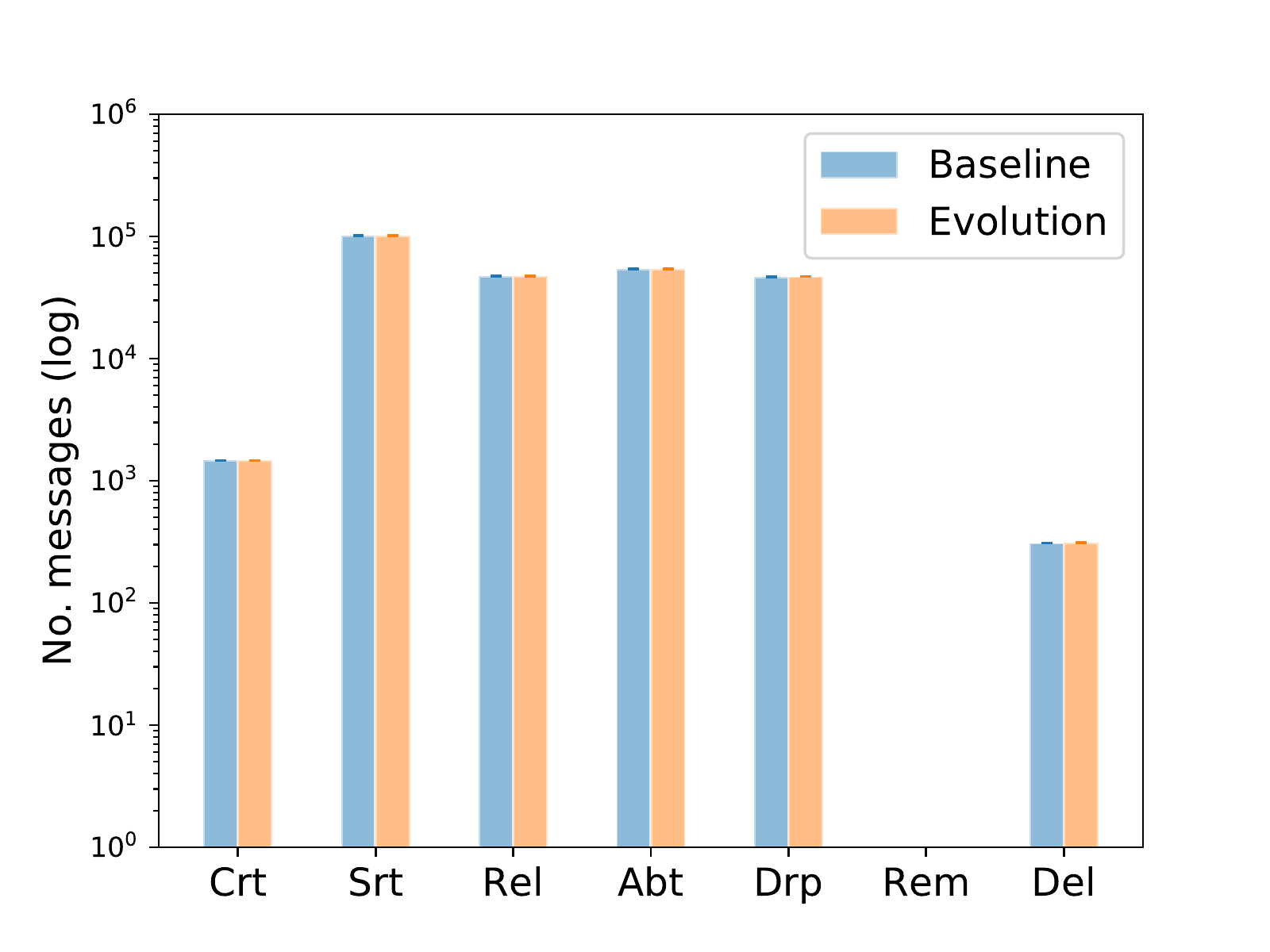}
    \caption{Manhattan (100 hosts)}%\label{}
    \end{subfigure}
    \vspace{-0.2cm}
    \caption{Number of created (Crt), started (Srt), relayed (Rel), aborted (Abt), dropped (Drp), removed (Rem) and delivered (Del) messages with the PRoPHET routing protocol (\emph{Baseline}) and the best evolved protocols (\emph{Evolution}), mean $\pm$ std. dev. (log scale) across \NUMRUNS~simulations.}
    \label{fig:prophet_messages}
\end{figure}

To gain further insight into the comparison between the behavior of the baseline and the best evolved protocols, we also analyzed the number of messages and their status during the simulations. Also this information is provided as an output of The ONE simulations\footnote{In particular, the simulator defines seven possible message status: \emph{created} (i.e., when new message is created at its source node), \emph{started} (i.e., when a message transmission starts, either from the source or from an intermediate hop), \emph{relayed} (i.e., when a message transmission to any receiving node, either the destination or an intermediate node, is completed; note that this check is done on a receiving node), \emph{aborted} (i.e., when a message transmission fails, i.e., not all the bytes are received correctly; also this check is done on a receiving node), \emph{dropped} (i.e., when a message, oldest first, is deleted from a node's buffer to make room for a new incoming message), \emph{removed} (i.e., when a message is deleted from a node's buffer since it has already reached its destination; note however that this information is not available in the Epidemic and PRoPHET protocols), and \emph{delivered} (i.e., the message is relayed and reaches its destination; the ratio \emph{delivered}/\emph{created} is the delivery probability considered throughout this study).}. Figures \ref{fig:epidemic_messages}-\ref{fig:prophet_messages} display the mean $\pm$ std. dev. (across \NUMRUNS~simulations of the baseline vs best evolved protocol, log scale), of the number of messages marked as created, started, relayed, aborted, dropped, removed and delivered, respectively for Epidemic and PRoPHET. It can be seen that the number of removed messages is zero, by construction of the baseline protocols. As for the other values, while the number of created messages is obviously the same in all simulations, the most interesting thing to note is that, apart from the Manhattan simulations, the number of started/relayed/aborted/dropped messages is consistently (up to two orders of magnitude) lower for the evolved protocols than for the baseline ones, which is consistent with the dramatically reduced overhead and average hop count observed in the Default and Helsinki test cases. Furthermore, the reduced number of dropped messages associated with the higher buffer time observed before, seems to suggest that in practice the evolved protocols do drop less messages, but these are kept much longer in the buffers.

Combining these observations with the trade-off analysis discussed in the main text (Section \ref{sec:other_metrics}), we can try to identify the reason for the improved delivery probability of the evolved protocols, at least in the Default and Helsinki test cases. On the one hand, the higher number of transmission attempts per update intuitively leads to a higher delivery probability: the more retransmissions a node tries, the higher the chance the messages in its local buffer will be eventually delivered. Furthermore, allowing the protocol to transmit also when \texttt{isTransferring()} is true makes it even more likely to transmit messages to nodes that are one hop away from their destination, which is consistent with the close to 1 average hop count measured with the evolved protocols. In practice, since nodes attempt more frequent transmissions than the baseline protocols (and this happens \emph{while the nodes are moving}), messages are more likely to be delivered with just one hop. On the other hand, contrarily to what we would have expected, this increased number of transmissions per update does not reflect in a higher number of started/relayed messages (and thus an increased overhead), or even less so aborted ones, but rather in a lower number of messages that are kept in the buffer for a longer time (as shown by the increased buffer time), before they get dropped due to new incoming messages. This longer buffer time is the cause of the observed increased latency.
\vspace{-0.3cm}
\vspace{-0.6cm}

%----------------------------------------------------------------

\section{Best evolved trees}
\label{sec:appendix_best_trees}

We report below the best performing trees evolved by Genetic Programming in all the test cases considered in our experimentation, namely all the combinations of $\langle$protocol, map, number of hosts per group$\rangle$ in: \{Epidemic, PRoPHET\} $\times$ \{Default, Helsinki, Manhattan\} $\times$ \{40 hosts, 100 hosts\}.

\begin{figure}[ht!]
    \centering
    \includegraphics[width=0.35\textwidth]{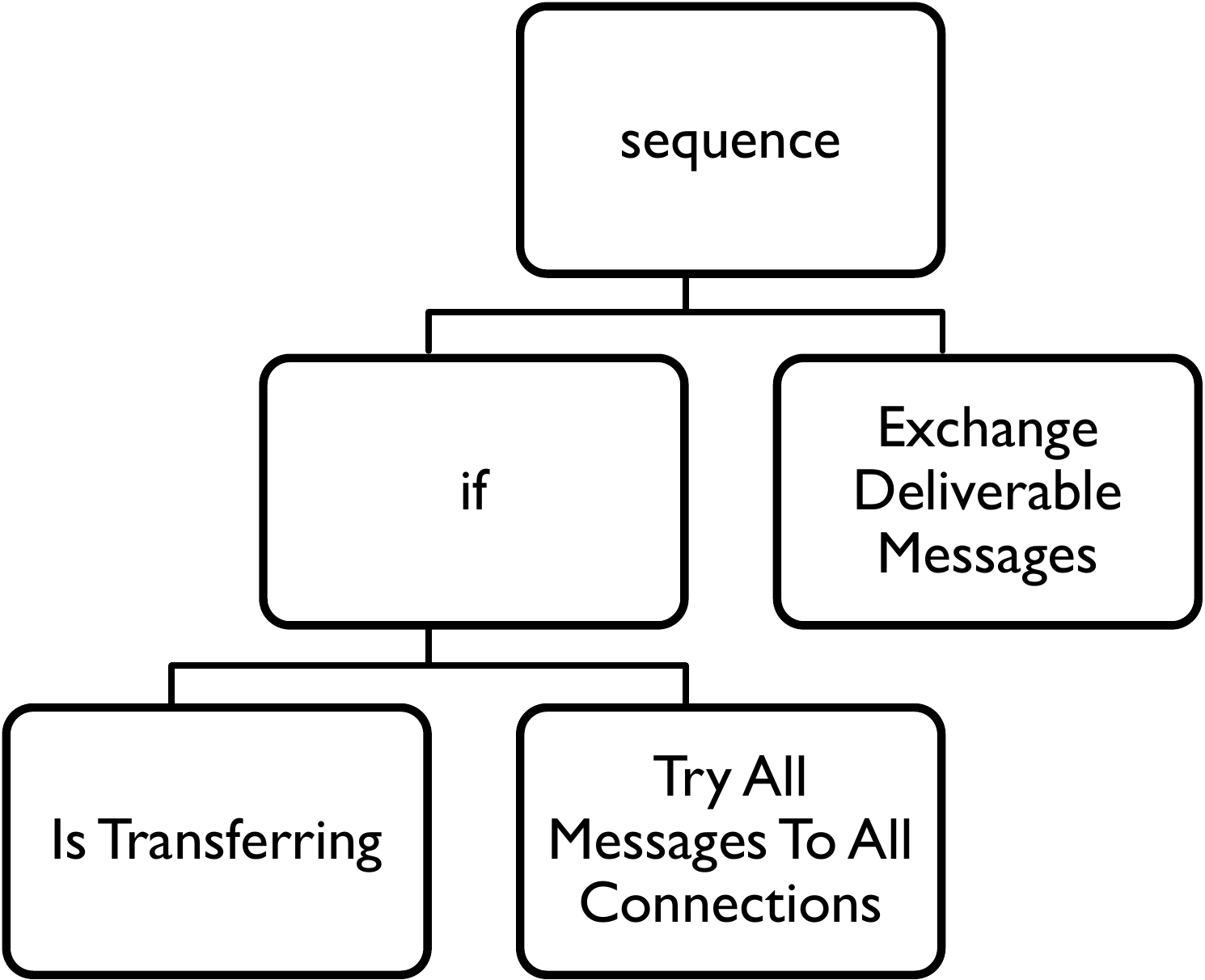}
    \caption{Best evolved tree in the test case: Epidemic, Default map and 40 hosts per group.}
    \label{fig:gen_tree_ep_def_40}
\end{figure}

\begin{figure}[ht!]
    \centering
    \includegraphics[width=1\textwidth]{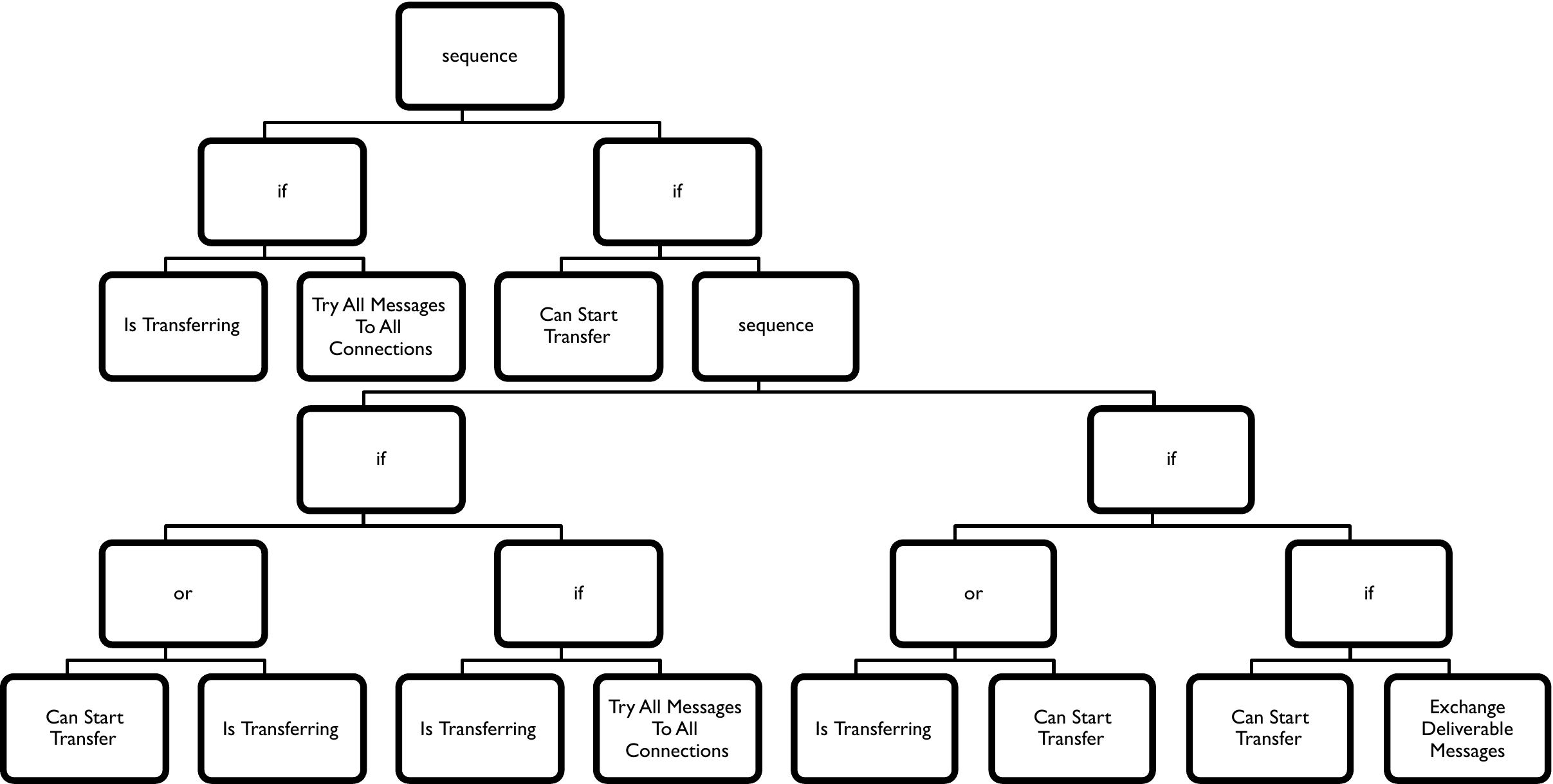}
    \caption{Best evolved tree in the test case: Epidemic, Default map and 100 hosts per group.}
    \label{fig:gen_tree_ep_def_100}
\end{figure}

\clearpage

\begin{figure}[ht!]
    \centering
    \includegraphics[width=1\textwidth]{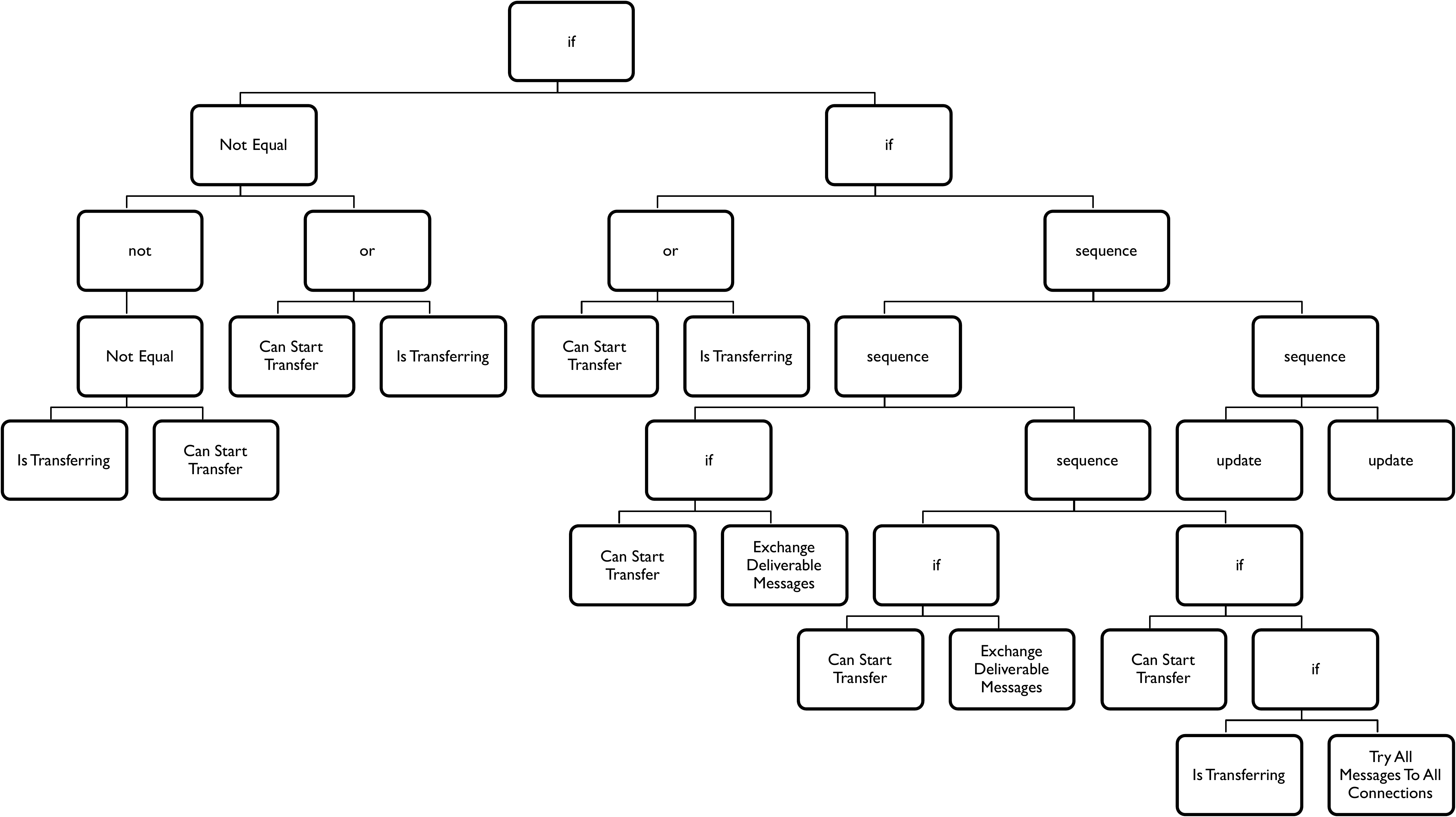}
    \caption{Best evolved tree in the test case: Epidemic, Helsinki map and 40 hosts per group.}
    \label{fig:gen_tree_ep_hel_40}
\end{figure}

\begin{figure}[ht!]
    \centering
    \includegraphics[width=0.6\textwidth]{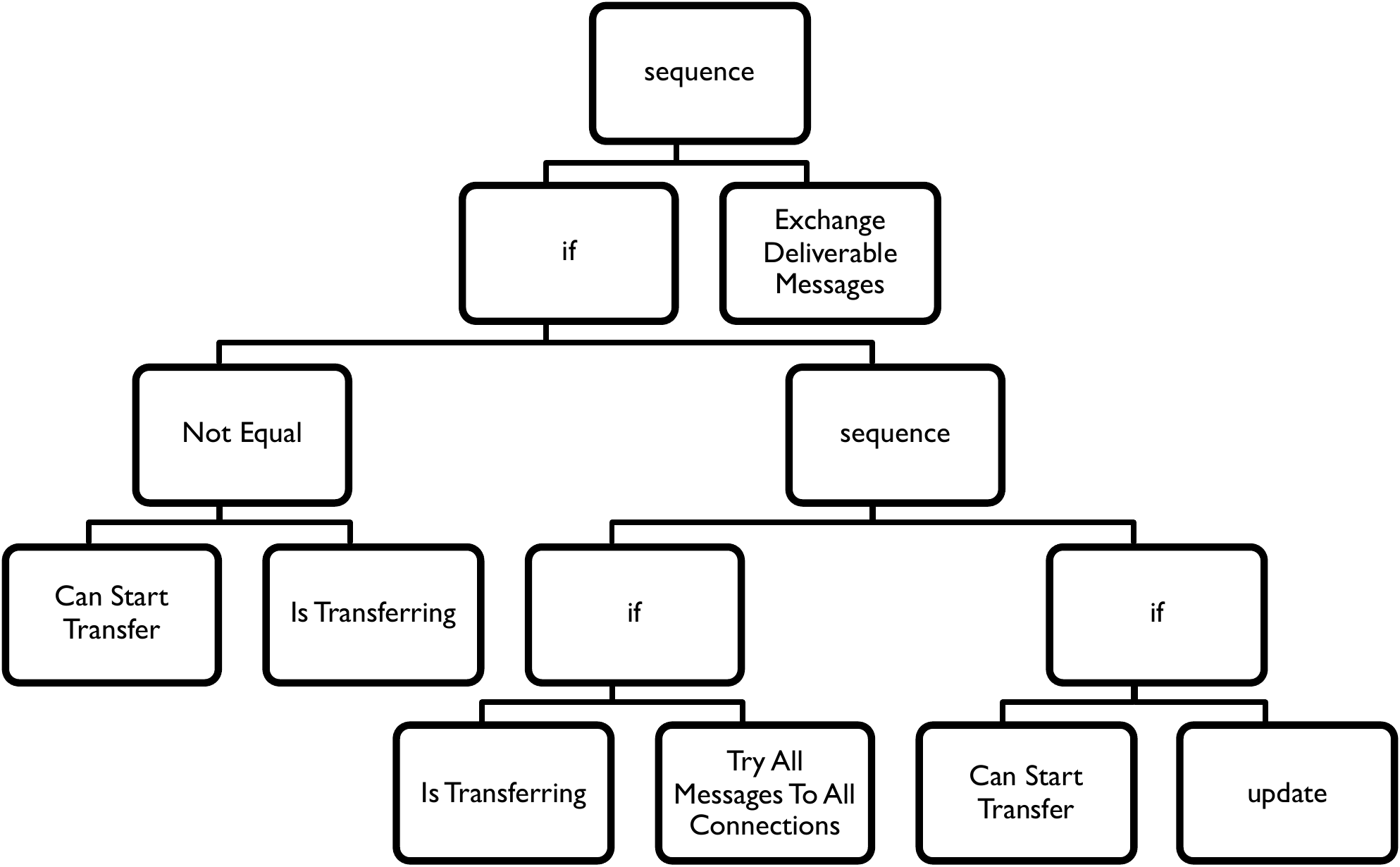}
    \caption{Best evolved tree in the test case: Epidemic, Helsinki map and 100 hosts per group.}
    \label{fig:gen_tree_ep_hel_100}
\end{figure}

\clearpage

\begin{figure}[ht!]
    \centering
    \includegraphics[width=1\textwidth]{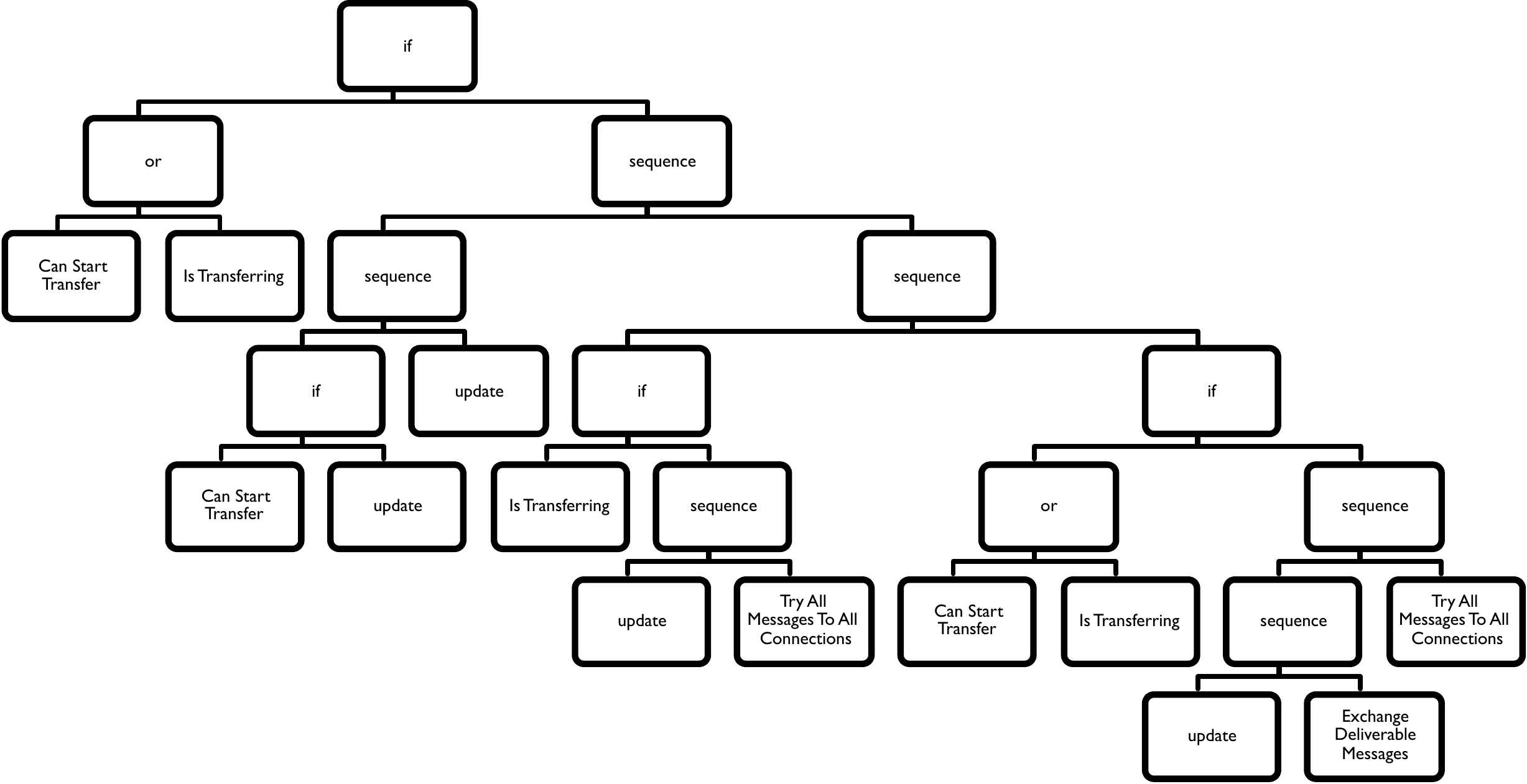}
    \caption{Best evolved tree in the test case: Epidemic, Manhattan map and 40 hosts per group.}
    \label{fig:gen_tree_ep_man_40}
\end{figure}

\begin{figure}[ht!]
    \centering
    \includegraphics[width=0.7\textwidth]{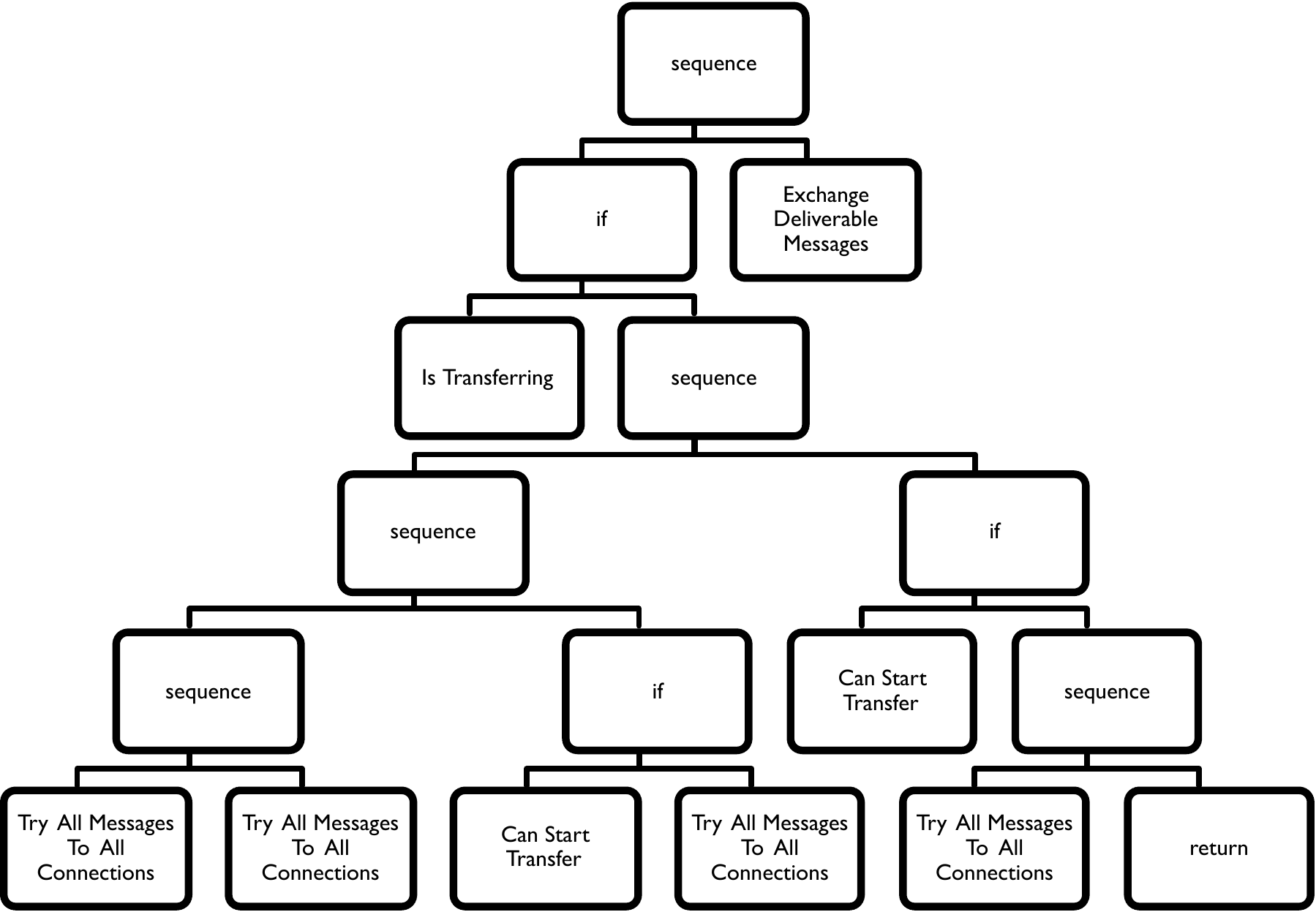}
    \caption{Best evolved tree in the test case: Epidemic, Manhattan map and 100 hosts per group.}
    \label{fig:gen_tree_ep_man_100}
\end{figure}

\clearpage

\begin{figure}[ht!]
    \centering
    \includegraphics[width=0.8\textwidth]{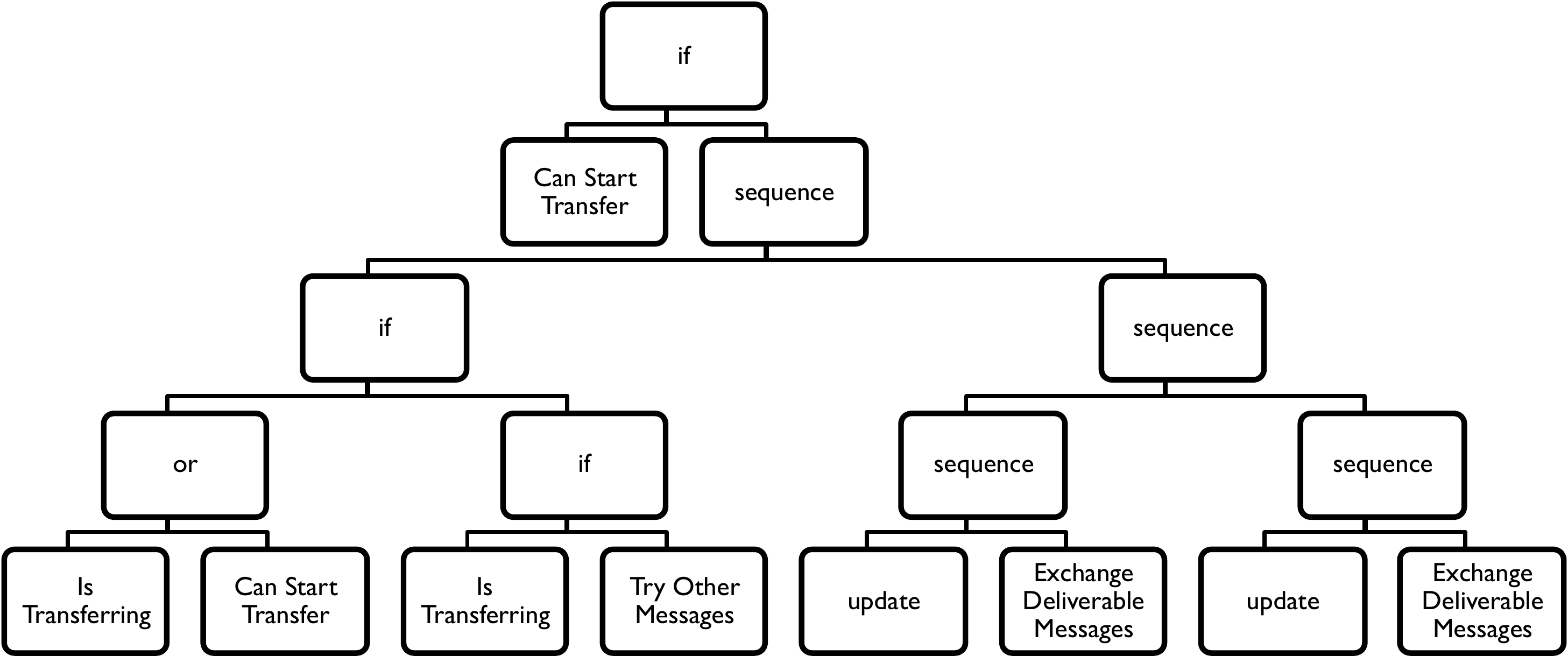}
    \caption{Best evolved tree in the test case: PRoPHET, Default map and 40 hosts per group.}
    \label{fig:gen_tree_prop_def_40}
\end{figure}

\begin{figure}[ht!]
    \centering
    \includegraphics[width=1\textwidth]{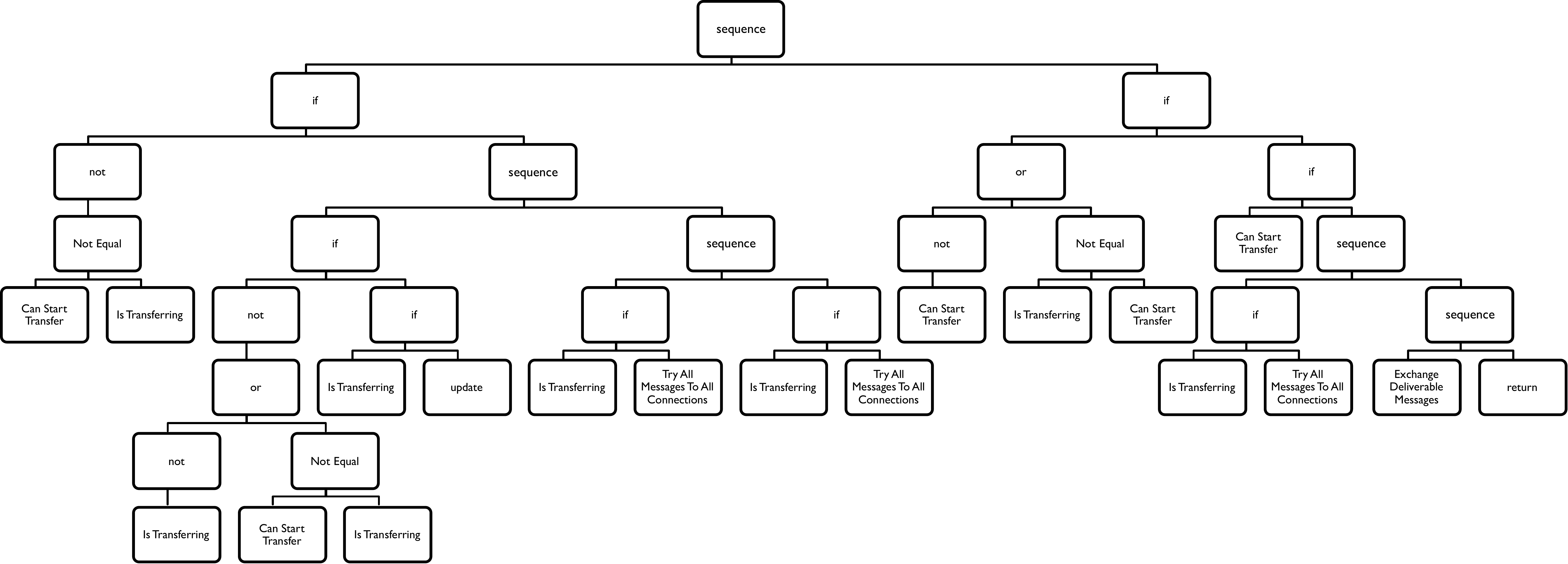}
    \caption{Best evolved tree in the test case: PRoPHET, Default map and 100 hosts per group.}
    \label{fig:gen_tree_prop_def_100}
\end{figure}

\clearpage

\begin{figure}[ht!]
    \centering
    \includegraphics[width=1\textwidth]{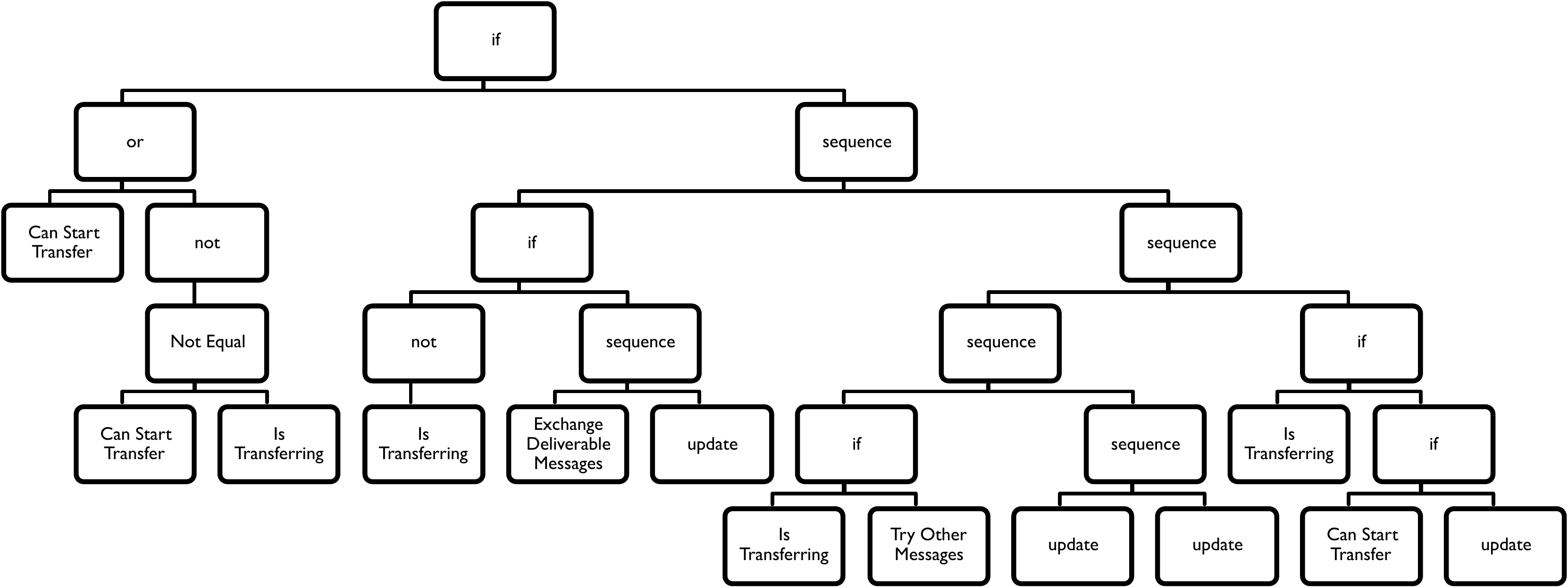}
    \caption{Best evolved tree in the test case: PRoPHET, Helsinki map and 40 hosts per group.}
    \label{fig:gen_tree_prop_hel_40}
\end{figure}

\begin{figure}[ht!]
    \centering
    \includegraphics[width=1\textwidth]{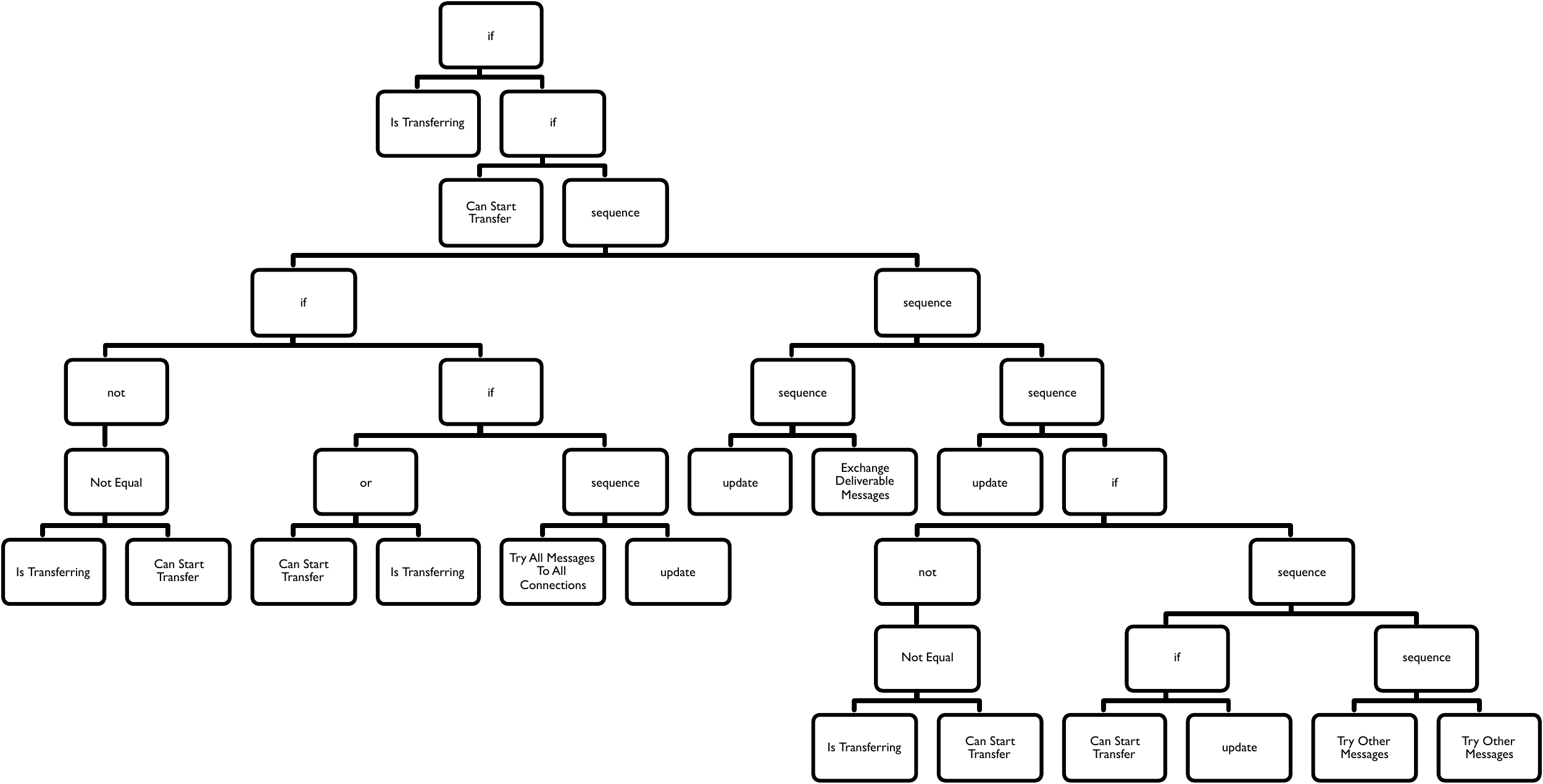}
    \caption{Best evolved tree in the test case: PRoPHET, Helsinki map and 100 hosts per group.}
    \label{fig:gen_tree_prop_hel_100}
\end{figure}

\clearpage

\begin{figure}[ht!]
    \centering
    \includegraphics[width=1\textwidth]{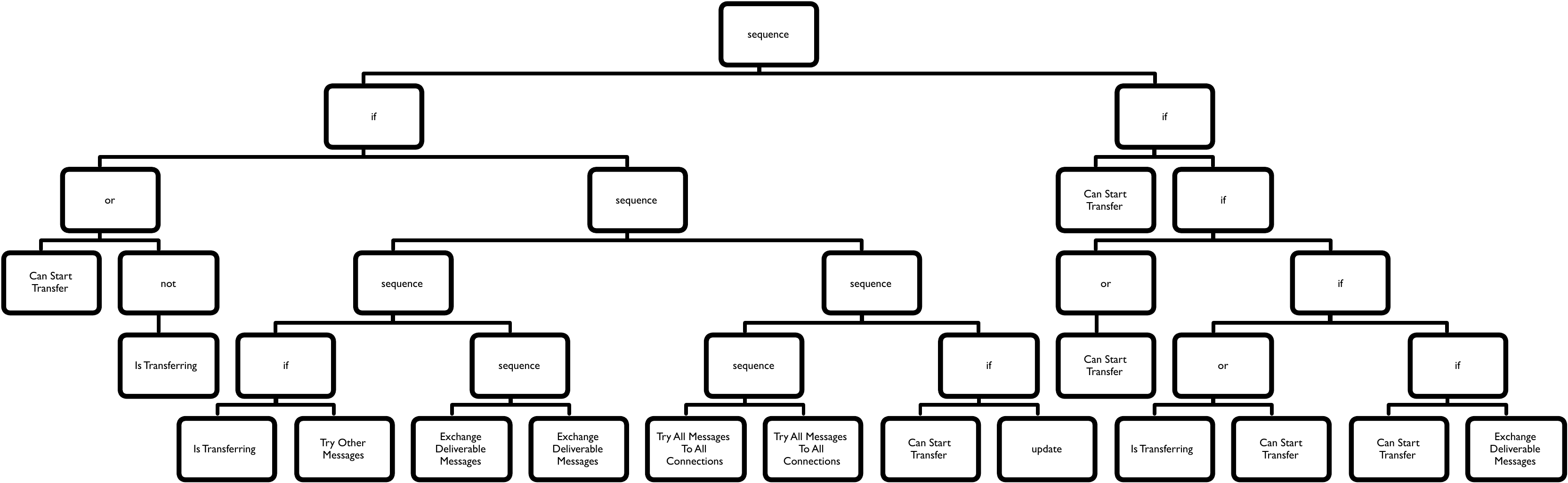}
    \caption{Best evolved tree in the test case: PRoPHET, Manhattan map and 40 hosts per group.}
    \label{fig:gen_tree_prop_man_40}
\end{figure}

\begin{figure}[ht!]
    \centering
    \includegraphics[width=0.7\textwidth]{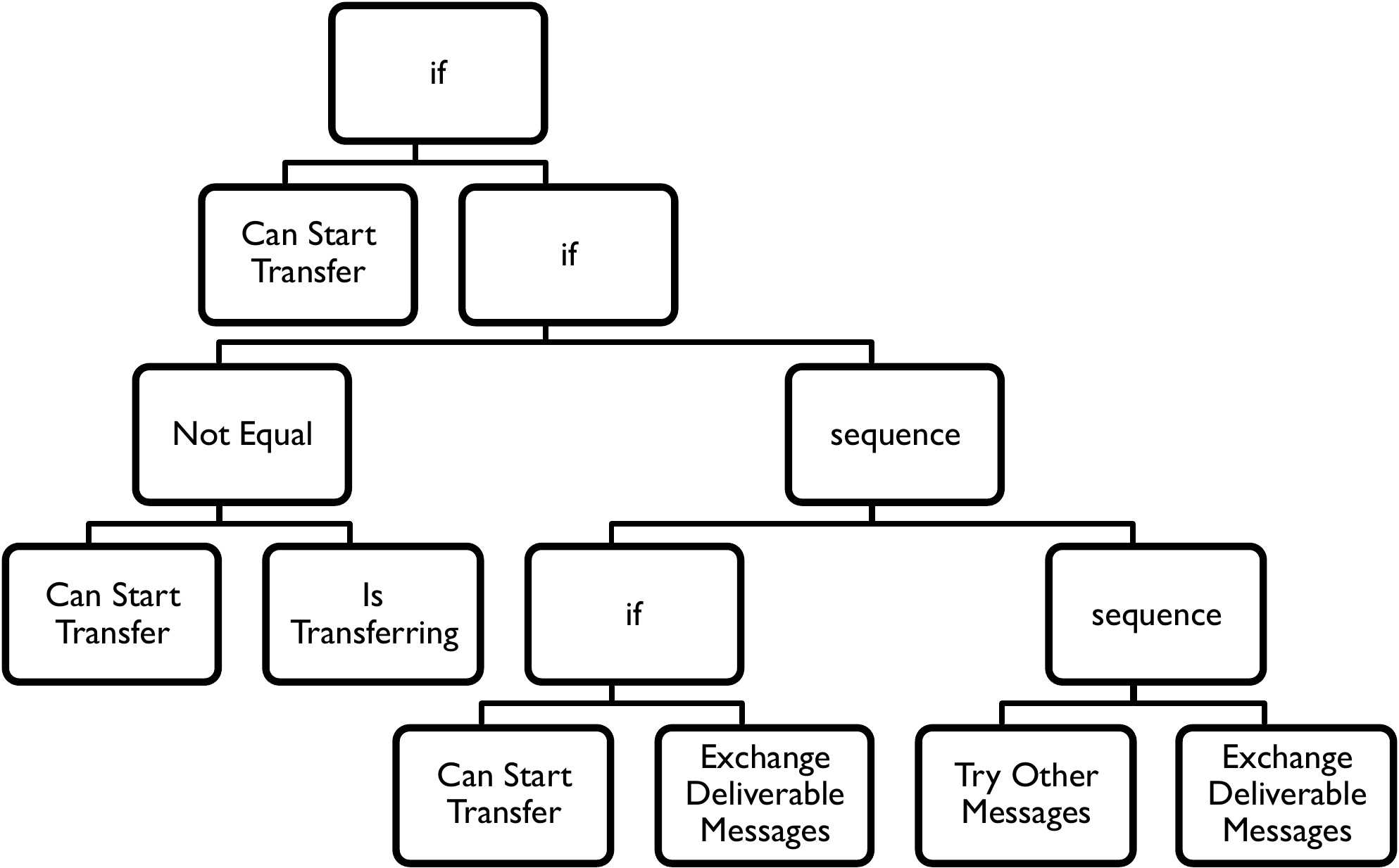}
    \caption{Best evolved tree in the test case: PRoPHET, Manhattan map and 100 hosts per group.}
    \label{fig:gen_tree_prop_man_100}
\end{figure}

\clearpage

%------------------------------------------------------------

\section{Routing protocol templates}
\label{sec:appendix_templates}

We report below the (simplified) class templates used for the Epidemic and PRoPHET routing protocols implemented in The ONE. We omit for brevity most comments and function implementations\footnote{The complete code implementation is available at \url{https://github.com/akeranen/the-one/tree/master/src/routing}.}. The \texttt{update()} method, highlighted in green, is the fragment of code improved by means of Genetic Programming, see Section \ref{sec:methods} in the main text for details.

% \begin{lstlisting}[
%   numberstyle=\color{gray}\tiny\highlight{32}{47}{green!15},
%   caption={EpidemicRouter class template},captionpos=b,
%   label={lst:epidemic}
% ]
% /*
%  * Copyright 2010 Aalto University, ComNet
%  * Released under GPLv3. See LICENSE.txt for details.
%  */
% package routing;

% import core.Settings;

% /**
%  * Epidemic message router with drop-oldest buffer and only single transferring
%  * connections at a time.
%  */
% public class EpidemicRouter extends ActiveRouter {

% 	/**
% 	 * Constructor. Creates a new message router based on the settings in
% 	 * the given Settings object.
% 	 * @params The settings object
% 	 */
% 	public EpidemicRouter(Settings s) {
% 		super(s);
% 	}

% 	/**
% 	 * Copy constructor.
% 	 * @param r The router prototype where setting values are copied from
% 	 */
% 	protected EpidemicRouter(EpidemicRouter r) {
% 		super(r);
% 	}

% 	$$@Override
% 	public void update() {
% 		super.update();
% 		if (isTransferring() || !canStartTransfer()) {
% 			return; // transferring, don't try other connections yet
% 		}

% 		// Try first the messages that can be delivered to final recipient
% 		if (exchangeDeliverableMessages() != null) {
% 			return; // started a transfer, don't try others (yet)
% 		}

% 		// then try any/all message to any/all connection
% 		this.tryAllMessagesToAllConnections();
% 	}

% 	$$@Override
% 	public EpidemicRouter replicate() {
% 		return new EpidemicRouter(this);
% 	}

% }
% \end{lstlisting}

\begin{lstlisting}[
  numberstyle=\color{gray}\tiny\highlight{22}{35}{green!15},
  caption={EpidemicRouter class template},captionpos=b,
  label={lst:epidemic}
]
/*
 * Copyright 2010 Aalto University, ComNet
 * Released under GPLv3. See LICENSE.txt for details.
 */
package routing;

import core.Settings;

/**
 * Epidemic message router with drop-oldest buffer and only single transferring
 * connections at a time.
 */
public class EpidemicRouter extends ActiveRouter {
    public EpidemicRouter(Settings s) {
		super(s);
	}
	
	protected EpidemicRouter(EpidemicRouter r) {
		super(r);
	}

	$$@Override
	public void update() {
		super.update();
		if (isTransferring() || !canStartTransfer()) {
			return; // transferring, don't try other connections yet
		}
		// Try first the messages that can be delivered to final recipient
		if (exchangeDeliverableMessages() != null) {
			return; // started a transfer, don't try others (yet)
		}
		// then try any/all message to any/all connection
		this.tryAllMessagesToAllConnections();
	}

	$$@Override
	public EpidemicRouter replicate() {
		return new EpidemicRouter(this);
	}
}
\end{lstlisting}

\clearpage

\begin{lstlisting}[
  numberstyle=\color{gray}\tiny\highlight{44}{56}{green!15},
  caption={ProphetRouter class template},captionpos=b,
  label={lst:prophet}
]
/*
 * Copyright 2010 Aalto University, ComNet
 * Released under GPLv3. See LICENSE.txt for details.
 */
package routing;

import java.util.ArrayList;
// other imports
// ...

/**
 * Implementation of PRoPHET router as described in <I>Probabilistic routing 
 * in intermittently connected networks</I> by Anders Lindgren et al.
 */
public class ProphetRouter extends ActiveRouter {
	// public and private fields
    // ...
	
	public ProphetRouter(Settings s) {
		super(s);
		//...
	}
	
	protected ProphetRouter(ProphetRouter r) {
		super(r);
		//...
	}
	
	private void initPreds() { /* ... */ }

	$$@Override
	public void changedConnection(Connection con) { /* ... */ }
	
	private void updateDeliveryPredFor(DTNHost host) { /* ... */ }
	
	public double getPredFor(DTNHost host) { /* ... */ }
	
	private void updateTransitivePreds(DTNHost host) { /* ... */ }
	
	private void ageDeliveryPreds() { /* ... */ }
	
	private Map<DTNHost, Double> getDeliveryPreds() { /* ... */ }

	$$@Override
	public void update() {
		super.update();
		if (!canStartTransfer() ||isTransferring()) {
			return; // nothing to transfer or is currently transferring
		}
		// try messages that could be delivered to final recipient
		if (exchangeDeliverableMessages() != null) {
			return;
		}
		tryOtherMessages();
	}
	
	private Tuple<Message, Connection> tryOtherMessages() { /* ... */ }

	private class TupleComparator implements Comparator <Tuple<Message, Connection>> { /* ... */ }

	$$@Override
	public RoutingInfo getRoutingInfo() { /* ... */ }

	$$@Override
	public MessageRouter replicate() { /* ... */ }
}
\end{lstlisting}

%------------------------------------------------------------

\section{The ONE configuration file}
\label{sec:theone_config}

We report below the configuration files used in The ONE simulations. We highlight in green the settings that are modified (depending on the specific test case) during the evolutionary runs. We omit the parameters related to The ONE GUI, which are not relevant for our experimentation.

\begin{lstlisting}[
  numberstyle=\color{gray}\tiny\highlightlist{1,17,25,38,50,59,69,72,73,76,78,79,80,81,83,86,89}{green!15},
  caption={Default The ONE settings file},captionpos=b,
  label={lst:theonesetting}
]
Scenario.name = default_scenario
Scenario.simulateConnections = true
Scenario.updateInterval = 0.1
Scenario.endTime = 43200

btInterface.type = SimpleBroadcastInterface
btInterface.transmitSpeed = 250k
btInterface.transmitRange = 10

highspeedInterface.type = SimpleBroadcastInterface
highspeedInterface.transmitSpeed = 10M
highspeedInterface.transmitRange = 1000

Scenario.nrofHostGroups = 6

Group.movementModel = ShortestPathMapBasedMovement
Group.router = EpidemicRouter
Group.bufferSize = 5M
Group.waitTime = 0, 120
Group.nrofInterfaces = 1
Group.interface1 = btInterface
Group.speed = 0.5, 1.5
Group.msgTtl = 300

Group.nrofHosts = 40

Group1.groupID = p

Group2.groupID = c
Group2.okMaps = 1
Group2.speed = 2.7, 13.9

Group3.groupID = w

Group4.groupID = t
Group4.bufferSize = 50M
Group4.movementModel = MapRouteMovement
Group4.routeFile = data/tram3.wkt
Group4.routeType = 1
Group4.waitTime = 10, 30
Group4.speed = 7, 10
Group4.nrofHosts = 2
Group4.nrofInterfaces = 2
Group4.interface1 = btInterface
Group4.interface2 = highspeedInterface

Group5.groupID = t
Group5.bufferSize = 50M
Group5.movementModel = MapRouteMovement
Group5.routeFile = data/tram4.wkt
Group5.routeType = 2
Group5.waitTime = 10, 30
Group5.speed = 7, 10
Group5.nrofHosts = 2

Group6.groupID = t
Group6.bufferSize = 50M
Group6.movementModel = MapRouteMovement
Group6.routeFile = data/tram10.wkt
Group6.routeType = 2
Group6.waitTime = 10, 30
Group6.speed = 7, 10
Group6.nrofHosts = 2

Events.nrof = 1
Events1.class = MessageEventGenerator
Events1.interval = 25,35
Events1.size = 500k,1M
Events1.hosts = 0,126
Events1.prefix = M

MovementModel.rngSeed = 1
MovementModel.worldSize = 4500, 3400
MovementModel.warmup = 1000

MapBasedMovement.nrofMapFiles = 4

MapBasedMovement.mapFile1 = data/roads.wkt
MapBasedMovement.mapFile2 = data/main_roads.wkt
MapBasedMovement.mapFile3 = data/pedestrian_paths.wkt
MapBasedMovement.mapFile4 = data/shops.wkt

Report.nrofReports = 2
Report.warmup = 0
Report.reportDir = reports/
Report.report1 = MessageStatsReport
Report.report2 = ContactTimesReport

ProphetRouter.secondsInTimeUnit = 30

Optimization.cellSizeMult = 5
Optimization.randomizeUpdateOrder = true
\end{lstlisting}

% SprayAndWaitRouter.nrofCopies = 6
% SprayAndWaitRouter.binaryMode = true

% GUI.UnderlayImage.fileName = data/helsinki_underlay.png
% GUI.UnderlayImage.offset = 64, 20
% GUI.UnderlayImage.scale = 4.75
% GUI.UnderlayImage.rotate = -0.015
% GUI.EventLogPanel.nrofEvents = 100

%------------------------------------------------------------

%%%% TEMPLATE OF MessageStatsReport

% Message stats for scenario Epidemic1001
% sim_time: 43200.1000
% created: 1463
% started: 59515
% relayed: 31962
% aborted: 27553
% dropped: 31879
% removed: 0
% delivered: 369
% delivery_prob: 0.2522
% response_prob: 0.0000
% overhead_ratio: 85.6179
% latency_avg: 4690.3407
% latency_med: 3250.0000
% hopcount_avg: 4.6829
% hopcount_med: 4
% buffertime_avg: 1385.0640
% buffertime_med: 894.2000
% rtt_avg: NaN
% rtt_med: NaN

% -------------------------------------------------------------------------
% -------------------------------------------------------------------------
% -------------------------------------------------------------------------

\end{document}